\documentclass[apj,twocolumn,numberedappendix]{emulateapj}
\usepackage{epsfig}
\usepackage{amsmath}
\usepackage{natbib}
\usepackage{subfigure}
\usepackage{longtable}
\usepackage{color}

\def\spose#1{\hbox to 0pt{#1\hss}}
\def\simlt{\mathrel{\spose{\lower 3pt\hbox{$\mathchar"218$}}
     \raise 2.0pt\hbox{$\mathchar"13C$}}}
\def\simgt{\mathrel{\spose{\lower 3pt\hbox{$\mathchar"218$}}
     \raise 2.0pt\hbox{$\mathchar"13E$}}}

\usepackage{graphicx}

\usepackage{graphicx}
\begin{document}

\title{A new method to search for high redshift clusters using photometric redshifts}
\author{G. Castignani\altaffilmark{1}, M. Chiaberge\altaffilmark{2,3,4}, A. Celotti\altaffilmark{1,5,6},
 C. Norman\altaffilmark{2,7}} 
\altaffiltext{1}{SISSA, Via Bonomea 265, 34136, Trieste, Italy\\
                 e-mail:castigna@sissa.it }
\altaffiltext{2}{Space Telescope Science Institute, 3700 San Martin Drive, Baltimore, MD 21218}
\altaffiltext{3}{INAF - IRA, Via P. Gobetti 101, Bologna, I-40129}
\altaffiltext{4}{Center for Astrophysical Sciences, Johns Hopkins University, 3400 N. Charles Street, Baltimore, MD 21218, USA}
\altaffiltext{5}{INAF - Oservatorio Astronomico di Brera, via Bianchi 46,  23807, Merate, Italy}
\altaffiltext{6}{INFN- Sezione di Trieste, via Valerio 2, 34127, Trieste, Italy}
\altaffiltext{7}{Department of Physics and Astronomy, Johns Hopkins University, Baltimore, MD 21218, USA}
%\altaffiltext{4}{Department of Physics and Astronomy, Johns Hopkins University, Baltimore, MD, USA}

\begin{abstract}
%NEW ABSTRACT (reduced down to 250 words)
{We describe a new method (Poisson Probability Method, PPM) to search for high redshift galaxy clusters and groups by using photometric redshift information and 
galaxy number counts. The method relies on Poisson statistics and is primarily introduced to search for Mpc-scale environments around a specific beacon.
The PPM is tailored to both the properties of the FR~I radio galaxies in the \citet{chiaberge2009} sample, that are selected within the COSMOS survey, and on the specific dataset used. We test the efficiency of our method 
of searching for cluster candidates against simulations. Two different approaches are adopted. 
i) We use two $z\sim1$ X-ray detected cluster candidates found in the COSMOS survey and we shift them to higher redshift up to $z = 2$. We find that the PPM detects the cluster candidates up to $z = 1.5$, and it correctly estimates 
both the redshift and size of the two clusters. ii) We simulate spherically symmetric clusters of different size and richness, and we locate them at different redshifts (i.e. $z = 1.0$~,1.5, and 2.0) in the COSMOS field. We find that the PPM detects the simulated clusters within the considered redshift range with a statistical 1-$\sigma$ redshift accuracy of $\sim0.05$. 
The PPM is an efficient alternative method for high-redshift cluster searches that
may also be applied to both present and future wide field surveys such as SDSS Stripe 82, LSST, and Euclid. Accurate photometric redshifts and a survey depth similar or better than that
of COSMOS (e.g. I$<25$) are required.}

\end{abstract}


%\maketitle

\section{Introduction}\label{sec:intro}
Cluster of galaxies are among the most massive large scale structures in the Universe.
They form from gravitational collapse of matter concentrations induced by perturbations of the primordial density field 
\citep{peebles1993,peacock1999}. 
Galaxy clusters have been extensively studied to understand how large scale structures form and evolve during cosmic time, 
from galactic to cluster scales \citep[see][for a review]{kravtsov_borgani2012}.

Despite this,
the properties of the cluster galaxy population and their changes with redshift 
in terms of galaxy morphologies, types, masses, colors \citep[e.g.][]{bassett2013,mcintosh2013},
and star formation content \citep[e.g.][]{zeimann2012,santos2013,strazzullo2013,gobat2013,casasola2013,brodwin2013,zeimann2013,alberts2013} are still debated, 
especially at redshifts $z\gtrsim1.5$.

It is also unknown when the Intra-Cluster Medium (ICM) 
virializes and starts emitting in X-rays 
and upscattering the CMB through the Sunyaev - Zel'dovich (SZ) effect \citep{sunyaev_zeldovich1972}.
See \citet{rosati2002} for a review.
In general, the formation history of the large scale structures and the halo assembly
history \citep[e.g.][]{sheth2004,dalal2008} are not fully understood.

High redshift clusters counts are used to constrain cosmological parameters \citep[e.g.][]{planckXX_2013}, to test the
validity of the $\Lambda$CDM scenario and quintessence models \citep{jee2011,mortonson2011,benson2013}.
Cluster counts are strongly sensitive to the equation of state of the Universe, especially at $z\gtrsim1$ \citep{mohr2005}, when the
Universe starts accelerating and the dark energy component starts becoming dominant.
The Sunyaev-Zel'dovic (SZ) effect, weak lensing measurements \citep{rozo2010}, X-ray scaling relations and data \citep{vikhlinin2009,mantz2010}
are used to evaluate the mass, the redshift of the clusters, and their mass function.
Moreover, high redshift cluster samples might be used to test 
the  (non-)Gaussianity of the primordial density field and to test alternative 
theories beyond General Relativity \citep[see][and references therein for a review]{allen2011,weinberg2012}.

Searching for high redshift $z\gtrsim1$ galaxy clusters is therefore a fundamental issue of modern astrophysics 
to assist our understanding of open problems of extra-galactic astrophysics and cosmology
from both observational and theoretical perspectives.

An increasing number of high redshift $z\gtrsim1$ spectroscopic confirmations 
of cluster candidates have been obtained in the last years. 
To the best of our knowledge, there are in the literature only 11 spectroscopically confirmed 
$z\gtrsim1.5$ clusters 
\citep{papovich2010,fassbender2011,nastasi2011,santos2011,gobat2011,brodwin2011,brodwin2012,zeimann2012,stanford2012,muzzin2013,newman2013}.
Only some of them have estimated masses greater than $10^{14}$~M$_\odot$.
In addition, \citet{tanaka2013} spectroscopically confirmed a $z=1.6$
X-ray emitting group, whose estimated mass is $3.2\times10^{13}~M_\odot$.
A $z\sim1.7$ group associated with a $z\sim8$ lensed background galaxy was found by \citet{barone_nugent2013}.

High redshift clusters have been searched for using several independent techniques; such as e.g.
those that use X-ray emission \citep[e.g.][]{cruddace2002,bohringer2004,henry2006,suhada2012} or 
the SZ effect \citep[e.g.][]{planckXXIX_2013,hasselfield2013,reichardt2013}.
However, such methods require a minimum mass and are rapidly insensitive 
to detecting $z\gtrsim1.2$ clusters \citep[see  e.g. discussion in][]{zeimann2012}.
This seems to be true also for the SZ effect.
%, even if it
%is supposed to be independent of the redshift \citep[see  e.g. discussion in][]{zeimann2012}.

It is commonly accepted that  
early-type passively evolving galaxies segregate within the cluster core  
and represent the majority among the galaxy population, at least at redshifts $z\lesssim1.4$ 
\citep[e.g.][]{menci2008,tozzi2013}.

Various methods search for distant clusters
taking advantage of this segregation. 
Its evidence is observationally
suggested by a tight red-sequence relation
which early-type galaxies exhibit in the color magnitude plots. 
Such high-$z$ searches are therefore based on overdensities of red objects and are 
commonly performed adopting  either optical \citep{gladders2005} or 
infrared \citep{papovich2008} color selection criteria. 
They find a great number of cluster candidates, even at $z\sim2$ \citep[e.g.][]{spitler2012}. However, all these
methods seem to be
less efficient at redshifts $z\gtrsim1.6$. Moreover, such methods require a significant presence of red galaxies.
There might be a bias in excluding clusters with a significant amount of star forming galaxies or, at least,  
such searches  might be biased towards large scale structures with specific properties in terms of galaxy colors
\citep{scoville07b,george2011}. 

The red-sequence has been confirmed in clusters and observed to be fairly
unevolved up to $z=1.4$ \citep{mei2009,rosati2009,strazzullo2010}.
However, its presence and evolution
are still debated at higher redshift.
Recent work showed evidence of
increasing star formation activity in cluster cores at $z\gtrsim1.5$ \citep{hilton2010,fassbender2011,santos2011}.  

Several methods use photometric and/or spectroscopic redshifts 
to search for high redshift overdensities \citep{eisenhardt2008,knobel2009,knobel2012,adami2010,adami2011,george2011,wen_han2011,jian2013}.
Similarly to the methods outlined above, they are generally less efficient at $z\gtrsim1.5$.
This is due to the difficulty of obtaining spectroscopic redshift information for a sufficient 
number of sources at $z>1$, to the significant photometric redshift uncertainties, and
to the small number density of objects.

 In fact, typical 1-$\sigma$ statistical photometric redshift uncertainties are $\sim$0.15 at redshifts $z=1.5$, while the mean number of galaxies within a
redshift bin $\Delta z = 0.3$ and a circle of 1~Mpc diameter is $\lesssim 9$ and $\lesssim 3$, at $z= 1.5$ and $z =2.0$, respectively.

Powerful radio galaxies \citep[i.e. FR~IIs,][]{fr74} have been  extensively used for high redshift cluster searches \citep[e.g.][]{rigby2013}.
High redshift (i.e. $z\gtrsim2$) high power radio galaxies
are frequently hosted in Lyman-$\alpha$ emitting protoclusters \citep[see][for a review]{miley_debreuck2008}.
Recently \citet{galametz2012} and \citet{wylezalek2013} searched for Mpc-scale 
structures around high redshift (i.e. $z\gtrsim1.2)$  high power radio galaxies 
using a IR color selection \citep{papovich2008}.
%They found evidence that radio galaxies of FR~II type reside preferentially
%in dense environments.
%These studies were originally encouraged by the evidence that radio galaxies are hosted in massive elliptical galaxies (REF)
%and reside locally in cluster environment (REF).

FR~I radio galaxies \citep{fr74} are 
intrinsically dim and are more difficult to find 
at high redshifts than the higher power FR~IIs.
This has so far 
limited the environmental study of the high redshift ($z\gtrsim1$) radio galaxy population to the FR~II class only.

However, due to the steepness of the luminosity function, FR~I radio galaxies represent the great majority among
the radio galaxy population.
Furthermore, on the basis of the radio luminosity function,
hints of strong evolution have been observationally suggested by previous work \citep{sadler2007,donoso2009}. 
Their comoving density is expected to reach a maximum around $z\sim1.0-1.5$ according to some 
theoretical model \citep[e.g.][]{massardi2010}.

%In addition to this, the non-thermal emission
%of low power FR~I radio galaxies  contaminates less
%that coming from the cluster environment, at each wavelength,  
%than high power FR~IIs radio galaxies do.

At variance with FR~II radio galaxies or other types of AGNs,
low-redshift FR~Is are typically hosted by undisturbed ellipticals or cD galaxies \citep{zirbel96}, which 
are often associated with the 
Brightest Cluster Galaxies \citep[BCGs,][]{vonderlinden2007}.
FR~Is are preferentially found locally in dense environments at least at low redshifts
\citep{hill1991,zirbel1997,wing2011}. 
This suggests that FR~I radio galaxies might be more effective 
for high redshift cluster searches than FR~IIs.

\citet[][hereinafter C09]{chiaberge2009} derived the first sample of $z\sim1-2$ FR~Is 
within the COSMOS field \citep{scoville07}.
\citet{chiaberge2010} suggested the presence of overdensities around three of their 
highest redshift sources. 
Based on galaxy
number counts, the authors found that the Mpc-scale environments of these sources are 4$\sigma$ denser than the mean COSMOS density. 
\citet{tundo2012} searched for X-ray emission in the fields of the radio galaxies of the C09 sample.
They took advantage of the Chandra COSMOS field (C-COSMOS).
They did not find any evidence for clear diffuse X-ray emission from the surroundings of the radio galaxies.
Their stacking analysis suggests that, if present, any X-ray emitting hot gas would have temperatures lower than $\sim$2-3~keV.
Furthermore, \citet{baldi2013} derived accurate photometric redshifts for each of the sources in the \citet{chiaberge2009} sample.

The goal of this project is to search for high redshift clusters or groups using FR~I radio galaxies  as beacons.
In this paper we introduce our newly developed method and we test it against simulations.
The method is tailored to the specific properties of the sample (C09) we consider, and it uses 
photometric redshifts. 
In a companion paper \citep[][hereafter Paper~II]{castignani2013} we apply our method 
to the C09 sample. 
We will refer to the sources in the sample using the ID number only, as
opposed to the complete name COSMOS-FR~I~nnn.  

In Sect.~\ref{sec:method_motivations} we outline the motivations for introducing our new method,
in Sect.~\ref{sec:method_PPM} we briefly describe our newly developed method. 
Then we test our method against simulations.
In Sect.~\ref{sec:cluster_shifting_sim} we consider two $z\sim1$ clusters,
 and we test the efficiency of our method to detect them when they are located at different redshifts.
In Sect.~\ref{sec:simulated_clusters} we perform similar tests on simulated clusters.
In Sect.~\ref{sec:conclusions} we summarize our results and we draw conclusions.
In Appendix~\ref{app:PPM} we fully describe the details of our method.

Throughout this work we adopt a standard flat $\Lambda$CDM cosmology with 
matter density $\Omega_m=0.27$ and Hubble constant $H_0=71~\hbox{km}\,\hbox{s}^{-1}\,\hbox{Mpc}^{-1}$ \citep{Hinshaw2009}.
The physical projected distance is fairly constant within the redshift range of our interest, and it varies by $\sim5\%$ from $z=1$ to $z=2$. At redshift $z=1.5$, a projected separation of 60~arcsec corresponds to 512~kpc (physical units).

\section{Motivations for a new method}\label{sec:method_motivations}
%We have searched for overdensities in the fields of the radio galaxies in the sample as follows.
%(i) A similar but more accurate analysis
%as the one adopted in \citet{chiaberge2010} has been attempted.
%(ii) Angular cumulative number count distributions of the sources in the fields have been inspected. 
%The angular Kolmogorov - Smirnov test is not conclusive and do not add further information about the environment of the radio galaxies.
%(iii) The possible presence of the red sequence has been investigated. The inspection of the color - magnitude plots does not lead to 
%a better comprehension of the environment. However, we will discuss the possible presence of the red sequences in the
%fields in a forthcoming paper.

%We take advantage of both the several studies to search for high redshift cluster candidates, and the work has been done
%to search for evidence of dense environments around the radio galaxies in our sample.
As outlined in Sect.~\ref{sec:intro}, the goal of this work is to introduce a new method to search for high redshift ($z\gtrsim1$)
Mpc-scale overdensities on the basis of photometric redshifts only.
We primarily introduce the method in order to search for groups and clusters 
in the COSMOS field \citep{scoville07}
around the FR~I radio galaxies in the C09 sample.
Due to the COSMOS multiwavelength coverage, increasingly accurate 
photometric redshift determinations have been derived \citep{mobasher2007,ilbert2009}.

Our method is tailored to the specific properties of both the sample and the survey adopted,
but it can be also applied to other multiwavelength surveys and samples,
if accurate photometric redshift information is available.

Furthermore, the method requires the projected coordinates of fiducial beacons
(e.g. in our project we adopt the sample of $z\sim1-2$ radio galaxies). This implies that
our method relies on a positional prior, i.e. it 
is introduced to search for a cluster or group environment around assigned locations in the projected sky.
Therefore, it is not properly a method to search blindly for 
clusters and groups within a given survey, even if it can be possibly applied for such purposes. 
This strategy is similar to that adopted in \citet{george2011}, 
who associated galaxies with previously selected groups and that adopted in \citet{hao2010} who
searched for clusters around the BCGs.

In this section we briefly discuss the problems that affect methods that search for high redshift clusters 
on the basis of number densities, with particular attention to those that use photometric redshifts.
Then, we focus on the peculiarities of our sample and the resulting need for introducing 
a new method to search for high redshift Mpc-scale overdensities.

\begin{enumerate}
\item As pointed out by \citet{scoville07b}, methods that identify high redshift structures
 on  the basis of the observed surface densities
have to discriminate galaxies at different redshifts, to avoid projection effects. 
As noted in \citet{eisenhardt2008}, galaxy number counts are more susceptible to projection 
effects than, for example, the detection of X-ray emission from the Intra-Cluster Medium (ICM).
This makes problematic the identification of the structures at different distances along the line of sight. 

\item Number densities are increasingly small for increasing redshifts, at $z\gtrsim1$. This affects
also the deepest sky surveys. For example, the COSMOS field survey has, on averages,
number densities per unit redshift of $\sim$25, 10, and 3 counts~arcmin$^{-2}$~dz$^{-1}$
 at redshift $z\sim$1, 1.5, and 2.0, respectively \citep{ilbert2009}, where
only those galaxies with $i^{+}$ AB magnitudes in the range $21.5<i^{+}<24.5$
are considered.
These low number counts imply  that shot-noise strongly affects any $z\sim1-2$ cluster search based on galaxy number counts and
photometric redshifts, since
Mpc-scale overdensities are extended and  detected over scales of $\sim$1~arcmin \citep[e.g.][]{santos2009},
typical of those of cluster cores.
In fact, 1~arcmin corresponds to $\sim480$~kpc at $z=1$.

\item Typical statistical photometric redshift uncertainties 
are $\sigma_z\sim0.1-0.2$ at redshifts $z\sim1-2$. This applies to
surveys such as COSMOS \citep{ilbert2009} and CFHTLS \citep{coupon2009}.
Note that a distance of  $\sigma_z=0.1$ along the line of sight
corresponds to more than 100~Mpc, which is significantly more than the typical size
of large scale structures in the Universe.
Therefore, these uncertainties highly affect the line of sight discrimination of real cluster members from the foreground
and any attempts to determine cluster membership 
on the basis of photometric redshifts only.

Furthermore, the typical statistical photometric redshift uncertainty increases 
 within the redshift interval of our interest and undergoes a catastrophic failure at $z\gtrsim1.5$ \citep{ilbert2009}.
In fact, photometric and spectroscopic redshift information cannot 
be easily obtained between $z\sim1-2$ because most of the relevant spectral features fall 
outside of the instrumental wavelength bands in that redshift range, which is therefore 
called {\it redshift desert}
\citep{steidel2004,banerji2011}.

\item Mpc-scale overdensities might undergo significant evolution between $z\sim1-2$.
Their structure and number density might significantly
change with cosmic time.
In fact, diffuse protoclusters with star-forming
galaxies have been in fact found at redshifts higher than
$z \sim2.0$ \citep{steidel2000,venemans2007,capak2011}.
\end{enumerate}

Methods that search for high redshift groups or galaxy clusters that are based 
on optical number counts and photometric redshifts 
have to carefully identify the different Mpc-scale structures 
that are present along the line of sight, in order to avoid projection effects.

Most of the existing methods such as those that are based on wavelets, Friend-of-Friends algorithms, 
peak finding methods, Delaunay, Voronoi tessellations, adaptive kernel 
\citep[see e.g.][]{ebeling1993,postman1996,scoville07b,eisenhardt2008}
that search for high redshift clusters on the basis of number counts and redshifts 
are very efficient at $z\lesssim1.5$, but show reduced efficiency at higher redshifts because of
the above mentioned problems.

All these methods are based on the 2-d surface density more than on the 3-d number density. 
As noted in \citet{scoville2013}, considering the 3-d number density would require a more accurate 
photometric redshift information.
All the above mentioned methods characterize the projected space with a high accuracy, in order to identify Mpc-scale structures
of different scales.
However, such a detailed multi-scale projected space analysis implies that involves establishing whether multiple 
overdensity peaks in the 2-d projected density field are part of a single larger structure
in practice becomes extremely difficult and subjective \citep{scoville2013}.
For this reason, previous studies are not always able to provide galaxy cluster and group candidate catalogs \citep{scoville2013}.
Therefore, we will introduce a less sophisticated but flexible method to overcome
to these limitations. 

Furthermore, high photometric redshift uncertainties do not allow 
us to consider the 3-d number density. Therefore, we consider the 
 2-d surface density and the redshift information separately.
In order to overcome the problems listed above,
a detailed distance discrimination based on photometric redshifts is therefore required. 
As we show in the following this can be achieved to the detriment of a less 
detailed tessellation of the projected space.

%However, trying to detect large scale structures in the projected space with a flexible and not sophisticated tasselation of the sky 
%is a simplistic approach.
%This turns out to be a strong constraint, unless the location
%in the projected space around which we search for overdensities is known.
%In our case we want to test if radio galaxies are good beacons for dense environments.

\section{The Poisson Probability Method (PPM)} \label{sec:method_PPM}
Our method is based on galaxy number counts and photometric redshifts.  It
consists in searching for a dense environment around a given location in the sky.
Concerning our specific goal to search for cluster environments around the FR~Is
in the C09 sample, we will adopt the photometric redshift information for the galaxies in the 
COSMOS field as given in the \citet{ilbert2009} catalog. Limiting the sample to only FR~Is,
 we consider their recently estimated photometric redshifts
from \citet{baldi2013}, when spectroscopic redshifts from  the zCOSMOS-bright \citep{lilly2007}
and MAGELLAN \citep{trump2007} catalogs are not available (see also Paper~II).
Note that this applies to any catalog and dataset with characteristics similar to the those we adopted. 

The Poisson Probability Method (PPM), is adapted from that
proposed by \citet{gomez1997} to search for X-ray emitting substructures within clusters. 
The authors note how their method naturally overcomes the inconvenience of
dealing with low number counts per pixel ($\gtrsim 4$), which prevents them from
applying the standard techniques based on $\chi^2$-fitting, e.g.  \citet{davis1993}, see \citet{gomez1997}.
We are similarly dealing with the problem of small number counts. 
Therefore standard methods might not be appropriate, as discussed above.  We refer to Appendix \ref{app:PPM} for a
comprehensive description of the PPM.  Here we briefly summarize the
basic steps of the procedure:

\begin{itemize}
\item We tessellate the projected space with 
a circle centered at the coordinates of the beacon and a number of consecutive adjacent annuli.
The regions are concentric and have the same area (2.18~arcmin$^{2}$). 

\item For each region, we count galaxies with photometric redshifts
within a given interval $\Delta z$ centered at the centroid redshift $z_{\rm centroid}$,
for different values of $\Delta z$ and $z_{\rm centroid}$. The values of $\Delta z$ and  $z_{\rm centroid}$
densely span between $0.02-0.4$ and $0.4-4.0$, respectively.

\item  For each area and for a given redshift bin
we calculate the probability of the null hypothesis (i.e. no clustering) to have more than the observed number of galaxies, assuming
Poisson statistics and the average number density estimated from the COSMOS 
field.\footnote{We test 
if cosmic variance affects our analysis selecting 
four disjoint quadrants in the COSMOS survey to estimate the field density separately from each quadrant.
We verify that the results are independent of the particular choice of the field.}
Starting from the coordinates of the beacon we select only
the first consecutive overdense regions
for which  the probability of the null hypothesis
is $\leq30\%$. 
We merge the selected regions and we compute the probability, separately, as done for each of them.
We estimate the detection significance of the number count excess as the complementary probability.
We do not consider overdensities that start to be detected at ${\rm r}\gtrsim130$~arcsec, corresponding to $\gtrsim1$~Mpc, from the location of the beacon.

%----------------------- Figure: PPM plot -------------------------------
\begin{figure*} \centering
%\subfigure[]{\includegraphics[width=0.45\textwidth]{001.jpg}}\qquad
\subfigure[]{\includegraphics[width=0.45\textwidth,natwidth=610,natheight=642]{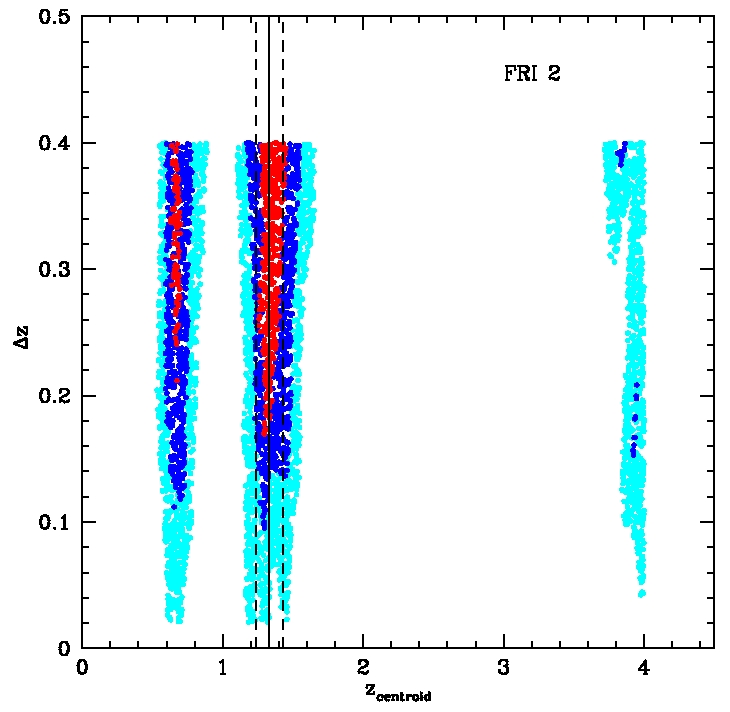}}\qquad
\subfigure[]{\includegraphics[width=0.45\textwidth,natwidth=610,natheight=642]{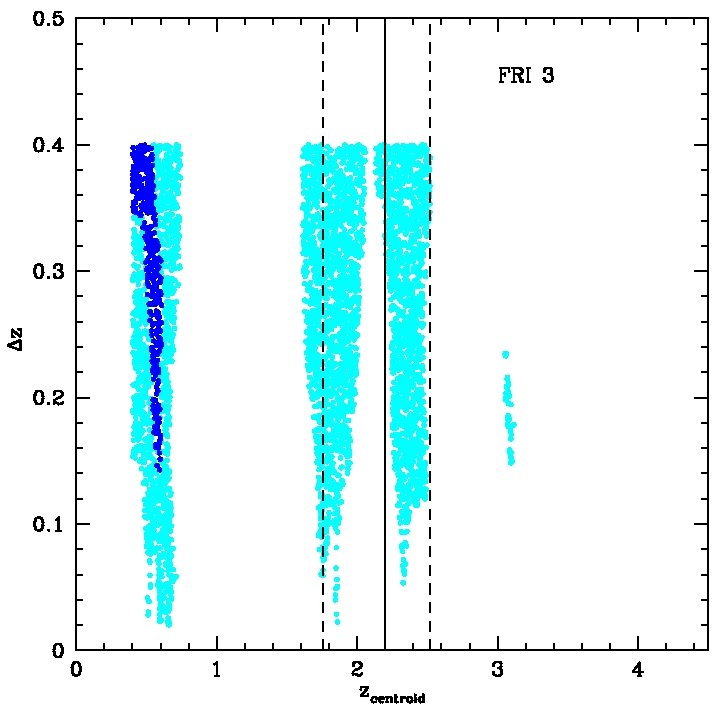}}\qquad
\subfigure[]{\includegraphics[width=0.45\textwidth,natwidth=610,natheight=642]{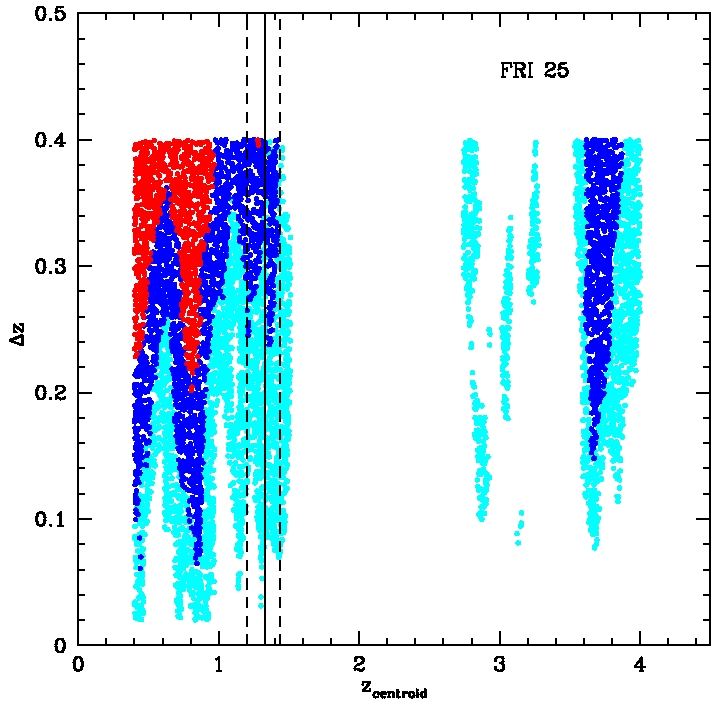}}\qquad
\subfigure[]{\includegraphics[width=0.45\textwidth,natwidth=610,natheight=642]{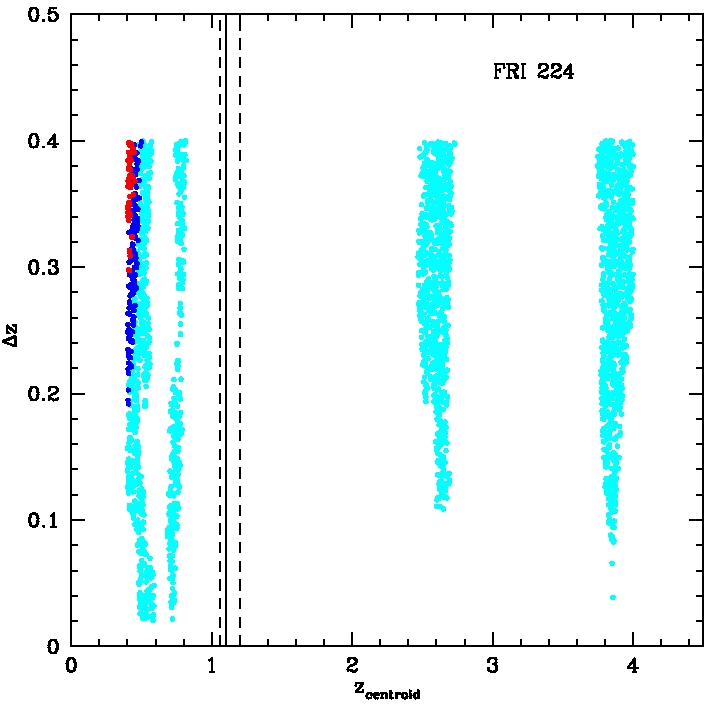}}\qquad
\caption{PPM plots for the fields of sources 02 (a), 03 (b), 25 (c), and  224 (d) of the \citet{chiaberge2009} sample.
The abscissa of the vertical solid line is at the redshift of the source. The vertical dashed
lines show its uncertainties as given in \citet{baldi2013}. 
Each point represents the detection significance of the number count excess for a specific choice of the values of the redshift bin $\Delta z$ (within which we perform the number count) and its centroid $z_{\rm centroid}$. The detection significance is estimated as the complementary probability of the null hypothesis (i.e. no clustering) to have more than the observed number of galaxies in the field of the beacon (i.e. the FRI radio galaxy in our case), 
assuming Poisson statistics and the average number density estimated from the COSMOS field. 
We plot only the points corresponding to overdensities with a $\geq2\sigma$ detection significance. Color code: $\geq2\sigma$ (cyan points), $\geq3\sigma$ (blue points),
$\geq4\sigma$ (red points). A Gaussian filter which eliminates high frequency noisy patterns is applied.}
\label{fig:ppm_plots}
\end{figure*}
%--------------------------------------------------------------------------------

\item In Figure~\ref{fig:ppm_plots}  we report the PPM plots for the fields of some of the sources in the \citet{chiaberge2009} sample.
For each choice of the parameters $z_{\rm centroid}$ and $\Delta z$ we plot the detection significance defined in the previous step.
We adopt the following color code: $\geq2\sigma$, $\geq3\sigma$, and $\geq4\sigma$  points 
are plotted in cyan, blue, and red, respectively.
We do not plot $<2\sigma$ points.
The abscissa of the vertical solid line is at the redshift of corresponding source. The vertical dashed
lines show the redshift uncertainties as given in B13.
We apply a Gaussian filter to eliminate high frequency noisy patterns. 
Figure~\ref{fig:ppm_plots} shows the plot where the filter is applied.

\item We define as overdensities only those regions in the filtered plot for which consecutive $\geq2\sigma$ points are present 
in a region of the PPM plot at least $\delta z _{\rm centroid} = 0.1$ long on the redshift axis $z_{\rm centroid}$ 
and defined within a tiny $\delta(\Delta z) = 0.01$ wide interval centered at $\Delta z=0.28$.
These values are chosen because of the properties of the errors of the photometric redshifts of our sample and of
the size of the Gaussian filter we apply.
In particular the redshift bin ($\Delta z=0.28$) corresponds to the estimated statistical 2-$\sigma$ photometric redshift uncertainty 
at $z\sim1.5$ for dim galaxies \citep[i.e. with AB magnitude ${\rm i}^{+}\sim24$,][]{ilbert2009}.
These magnitudes are typical of the galaxies we expect to find in clusters in the redshift range of our interest.
We verified that the results are stable with respect to a sightly different choice of the redshift bin $\Delta z$.

\item In order to estimate the actual significance of each Mpc-scale overdensity 
we apply the same procedure outlined in the previous step, but progressively increasing
the significance threshold until no overdensity is found.
We assign to each overdensity a significance equal to the maximum significance threshold 
at which the overdensity is still detected. 
Note that in case the overdensity displays multiple 
local peaks we do not exclude the lower significance ones.
%At this point we are able to redefine as overdensity each separate peak.
%This is the case of source 25 where 4 overdensities are detected in the PPM plot within the same  $\geq2\sigma$
%region \citep[see discussion in Sect. ?? in ][]{PPMmethod}.

\item We estimate the redshift of each overdensity as the centroid redshift $z_{\rm centroid}$
at which the overdensity is selected in the PPM plot.

\item We also estimate the size of each overdensity in terms of the minimum and maximum distances from the FR~I beacon at 
which the overdensity is detected. In order to do so  we consider all the points in the PPM plot within 
the region centered around $\Delta z=0.28$ and at least $\delta z _{\rm centroid} = 0.1$ long on the redshift axis $z_{\rm centroid}$ 
which defines the overdensity.
For each of these points the overdensity is detected within certain minimum and maximum distances.  
We estimate the minimum and maximum distances of the overdensity as the average (and the median) of the minimum and maximum distances 
associated with all of these points, respectively.
We also compute the rms dispersion of the distances as an estimate for the uncertainty.

\item In order to estimate the fiducial uncertainty for the redshift of the overdensity we consider all sources located
between the median value of the minimum distance and the median value of the maximum distance from the coordinates 
of the source within which the overdensity is detected in 
the projected space.
We also limit our analysis to the sources that have photometric redshifts within a redshift bin $\Delta z=0.28$ centered at the 
estimated redshift of the overdensity. This value is chosen to ensure consistency with the value
used for our detection procedure (see above).
We estimate the overdensity redshift uncertainty as 
the rms dispersion around the average of the photometric redshifts of
the sources that are selected in the field of the radio galaxy.

\item We associate with each radio galaxy any overdensity in its field that is located 
at a redshift compatible to that of the radio source itself (see Appendix~\ref{app:PPM}).
Note that multiple overdensity associations are not excluded.
\end{itemize}

Our approach implicitly assumes azimuthal symmetry around the axis oriented at the coordinates of the beacon.
Since we extend the tessellation up to $\sim$6~arcmin 
(i.e. $\sim$3~Mpc at $z=1.5$) from the coordinates of the beacon, 
we do not exclude the possibility 
of detecting non circularly symmetric systems.
\citep[see][for a similar methodology]{postman1996}.
Furthermore, our method is also flexible enough to find clusters even if the coordinates
of the cluster center are known with $\sim100$~arcsec
accuracy (as tested with simulations, see Sect.~\ref{sec:shifting_results} and Sect.~\ref{sec:increasing_offset}). 

We note that the great majority of low-power radio sources in clusters or groups
are found within $\sim200$~kpc from the core center up to $z\simeq1.3$ \citep{ledlow_owen1995,smolcic2011}.
Furthermore, FR~Is are typically hosted by undisturbed ellipticals or cD galaxies \citep{zirbel96},
which are often associated with the Brightest Cluster Galaxies \citep[BCGs,][]{vonderlinden2007}.
Similarly to the FR~Is, BCGs are preferentially found within $\sim41$~kpc from the cluster centers \citep{zitrin2012,semler2012}.
Therefore, this suggests that FR~Is in cluster environments are preferentially hosted 
within the central regions of the core, at least at low redshifts.
The results presented in the companion paper (see discussion in Sect.~7.10 of Paper~II).
for the $z\sim1-2$ cluster candidates associated with  the \citet{chiaberge2009} sample suggest that this is generally true also at higher redshifts.
This motivates 
the peculiar projected space tessellation 
described in this section and adopted for our cluster search (see Sect.~\ref{app:PPM} for further discussion) 
\\

In the next section we test the PPM against simulations. 
We use the COSMOS survey and the photometric redshift catalog of \citet{ilbert2009}.
We follow two different approaches: i) we use two
clusters discovered in the COSMOS field at $z\sim1$ and then we shift them
to higher redshifts in order to assess
the PPM efficiency to detect Mpc-scale structures at progressively high
redshifts. ii) We simulate spherically symmetric clusters of
different size and richness, and we locate them at different redshifts (i.e. $z=1.0$, 1.5, and 2.0)
in the COSMOS field. Then, we 
apply the PPM and we test if we can detect the simulated clusters.

Note that we do not test our method adopting mock catalogs derived from N-body numerical simulations  
to simulate the COSMOS density field shown in previous work for groups in COSMOS up to
to $z\simeq1$ \citep[e.g.][]{george2011,jian2013}.
This test omission is motivated by the fact that we lack of sufficient spectroscopic redshift information. 
We also have both smaller number count statistics and larger photometric redshift uncertainties
both of which strongly affect these studies at higher redshifts (i.e. $z\gtrsim1$).

\vspace{0.1cm}
\subsection{The PPM theory}
 In this section we report a sequential list of logical statements that clarify the theory the PPM procedure is based on.
We refer to Sect.~\ref{app:PPM} for the proofs.
\begin{itemize}
 
\item Since high photometric redshift uncertainties affect any high-z cluster search, the redshifts and 
projected coordinates of the galaxies are considered separately.
The field is tessellated with concentric regions of equal area centered at the 
projected coordinates of the beacon, i.e. the radio galaxy in our case (see Sect.~\ref{sec:projected_space}).
  
\item Sources with photometric redshifts within the redshift bin $\Delta z$ centered at the redshift centroid $z_{\rm centroid}$ are
selected. The values of $\Delta z$ and $z_{\rm centroid}$ densely span the ranges of our interest 
(see Sect.~\ref{sec:distance_discrimination}).
  
\item The probability of the null hypothesis (i.e. no clustering) is calculated for each region, given the values of $\Delta z$ and 
$z_{\rm centroid}$ (see Sect.~\ref{app:the_method}).

\item Starting from the projected coordinates of the beacon, the first consecutive regions for which the null hypothesis is rejected 
at a level $\geq70\%$ are selected. Then, the regions are merged to form a new one (see Sect.~\ref{app:the_method}). This procedure aims at
selecting the region in the projected space where the overdensity is present.

\item The probability $1-\mathcal{P}$ of null hypothesis is calculated for the new region, 
for each value of $\Delta z$ and $z_{\rm centroid}$. 
The null hypothesis is rejected with a probability $\mathcal{P}$.
Photometric redshift uncertainties are implicitly neglected (see Sect.~\ref{sec:theoretical_framework} and Sect.~\ref{app:the_method}).

\item Fluctuations of $\mathcal{P}$ on scales $\delta\Delta z$ and $\delta z_{\rm centroid}$ 
smaller than the typical statistical photometric redshift uncertainties are not physical and ultimately due to noise.
They are locally removed by convolving $\mathcal{P}$ with
a Gaussian filter, i.e. $\overline{\mathcal{P}} = \mathcal{W}\ast\mathcal{P}$.\footnote{The function $h = f\ast g$
is the convolution of the function $f$ with the function $g$.} 
$\overline{\mathcal{P}}$ is an effective mean field defined on the space of $(z_{\rm centroid} ; \Delta z)$, 
see Sect.~\ref{sec:noise_subtraction}.
%, while 
%$\mathcal{P}$ is also defined on the ensemble of all the possible redshift realizations of the galaxies in the field

\item We apply a variational approach and show that 
the filter $\mathcal{W}$ simultaneously suppresses (in linear approximation)
 the variations of $\mathcal{P}$ both in the space of $(z_{\rm centroid} ; \Delta z)$
and in the ensemble of all the possible redshift realizations of the galaxies in the field
(see Sect.~\ref{sec:noise_subtraction}). The variations of $\mathcal{P}$ in the ensemble  are originated by the fact that 
photometric redshift uncertainties are neglected when calculating $\mathcal{P}$ (see Sect.~\ref{sec:theoretical_framework}).

\item $\overline{\mathcal{P}}$ is a good estimate for (i) the probability that the null hypothesis is rejected, where photometric 
redshift uncertainties are not neglected, and (ii) the significance of the number count excess in the field.
In fact, the significance of the number count excess is 
decreased by the filtering procedure to take the additional variance due to the photometric redshift uncertainties into account
(see Sect.~\ref{sec:theoretical_framework} and Sect.~\ref{sec:noise_subtraction}).

\item We conservatively fix a 2-$\sigma$ wide redshift bin $\Delta z = 0.28$
and we apply 
the peak finding algorithm we developed for our discrete case. Such a procedure belongs to
a more general context known as Morse theory (see Sect.~\ref{sec:peak_finding_alg}).
The algorithm allows to select the cluster candidates in the field within the redshift range of our interest and, for each overdensity, it
provides (i) an estimate for its redshift,
(ii) an estimate for the cluster core size, and (iii) a rough estimate for the cluster richness 
(see Sect.~\ref{sec:cluster_candidates_selection}).

\item The association of the cluster candidates detected in the field with the beacon (i.e. in our case the radio galaxy) 
is performed by using
the cluster redshift estimate and the redshift of the radio galaxy, as well as the corresponding uncertainties (see Sect.~\ref{sec:cl_ff1_ass}).

\item A generalization of the method to other datasets and surveys is provided in Sect.~\ref{sec:generalization}.

\end{itemize}

\section{Clusters at \scriptsize{z}\normalsize$\simeq1$ shifted to higher redshifts.}
\label{sec:cluster_shifting_sim} 

In this section we test the effectiveness of the PPM in detecting
overdensities as a function of redshift.
We consider the $z\sim1$ cluster candidates with id numbers 62 and 126 (hereafter  F062 and F126, respectively)
in the \citet[][hereinafter FGH07]{finoguenov2007} group COSMOS catalog,  
selected by using XMM-{\it Newton} observations \citep{hasinger2007}.

F062 is in the field of the source COSMOS-FRI~01. This source is
part of the C09 sample and it is found in rich Mpc-scale environment by the PPM 
(see Paper~II for further discussion).
%F062 is the only cluster candidate in the FGH07 catalog that is also found with the PPM within the C09 sample.
The offset between the X-ray centroid of F062 (estimated in FGH07) and the projected coordinates of COSMOS-FRI~01
is about $\sim$10~arcsec.
This corresponds to 78~kpc at the spectroscopic redshift of the radio galaxy, i.e. $z_{spec} =0.88$ \citep{lilly2007,trump2007}.
The redshift of F126 as estimated in FGH07 is $z=1.0$.
FGH07 also estimated masses $M_{500}=(5.65\pm0.37)\times10^{13}~M_\odot$ and $(6.87\pm0.69)\times10^{13}~M_\odot$ for F062 and F126, 
respectively.\footnote{Here M$_{500}$ is the mass enclosed within the radius 
encompassing the matter density 500 times the critical.}

%-------------------- Figure: RGB images of F062, F126 (Finoguenov et al., 2007)----------
\begin{figure*}[htbc] \centering
\subfigure{\includegraphics[width=0.4\textwidth,natwidth=610,natheight=642]{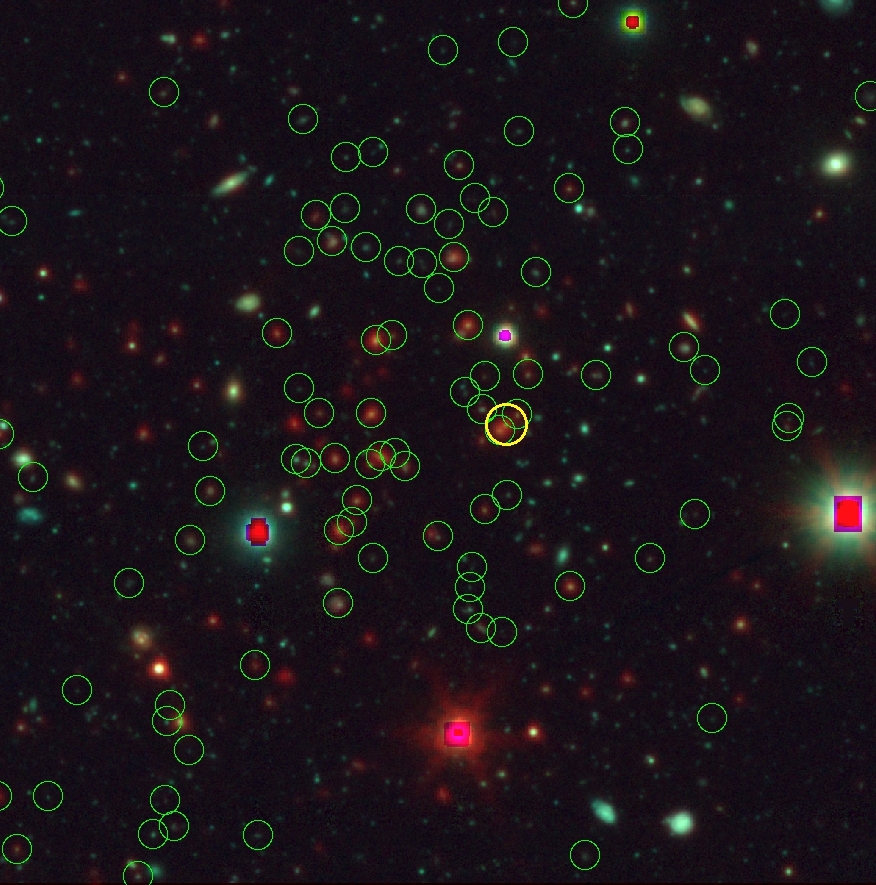}}\qquad
\subfigure{\includegraphics[width=0.4\textwidth,natwidth=610,natheight=642]{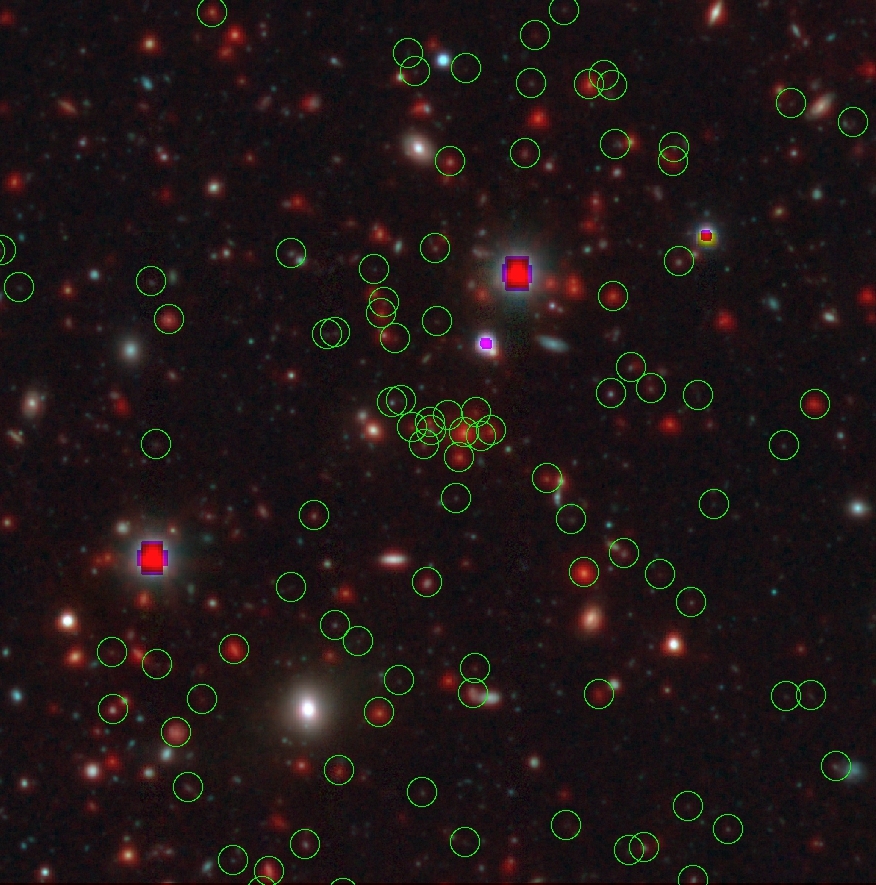}}
\caption{RGB images of F062 (left) and F126 (right) centered at the X-ray coordinates of the clusters, as in FGH07. The images are obtained using Spitzer 3.6$\mu$m,
 Subaru r$^+$- and V-band images for the R, G, and B channels, respectively. Green circles indicate objects with $0.78<z_{\rm phot}<0.98$ (left) and 
$0.9<z_{\rm phot}<1.1$ (right). The yellow circle in the left panel shows the location of COSMOS-FR~I~01. The projected sizes of the fields are 180''$\times$180''. North is up.}
\label{fig:RGBimages_F62_F126}
\end{figure*}
%-------------------------------------------------------------------------------------------

In Figure~\ref{fig:RGBimages_F62_F126} we plot the RGB images of F062 (left panel) and F126 (right panel) centered at the X-ray coordinates of the clusters, as in FGH07. We plot as green circles the locations of the galaxies in the \citet{ilbert2009} 
catalog with photometric redshifts within a $\Delta z = 0.2$ long 
redshift bin centered at the redshift of the cluster.
Concerning F062, we show as a yellow circle the location of COSMOS-FR~I~01.
The images are obtained using Spitzer 3.6$\mu$m, Subaru r- and V-band images for the R, G, and B channels, respectively.
As clear from visual inspection, both F062 and F126 exhibit a clear segregation of red objects 
within their Mpc-scale core.
Note that the brightest cluster member of F126 is associated with a radio source that is 
below the flux threshold of the C09 catalog. 
The two groups considered here also have comparable core sizes, X-ray fluxes, luminosities and temperatures (see Table~1 in  FGH07 for further details).
They have similar X-ray properties, but
F126 seems significantly richer than F062 (see Figure~\ref{fig:RGBimages_F62_F126}). 
Hence, we prefer to consider both of them, instead of one only.
This is in order to make our conclusions more robust.
In fact, if we adopted one single cluster candidate, our simulations might be biased by the specific properties
of that overdensity and our results might not be valid in a more general sense.
In the following we outline the different tests we perform. In Sect.~\ref{sec:shifting_results} we will describe the results 
in detail.
\begin{enumerate}
\item Firstly, we apply the PPM and we test if it detects 
 F062 and F126 (at their actual redshift).
\item We apply the PPM using increasing offsets between coordinates of the center of the PPM tessellation and the X-ray coordinates of the cluster center. 
This is done to estimate the required accuracy in the projected coordinates of the cluster center 
in order to detect cluster candidates with the PPM.
\item We shift both F062 and F126 to higher redshifts and we test whether the PPM is able to detect them. The procedure is quite complex and we describe it in the following. 
We select the fiducial cluster members adopting a color
$(I-K)_{AB}$ selection criterion to identify the redder sources. 
A cluster membership is required since we want to shift the cluster members to increasing redshifts.
The cluster members are selected within a redshift bin centered at the redshift of 
the cluster candidate and within a projected area centered at the X-ray cluster coordinates. 
Both the redshift bin and the projected area are selected accordingly to the PPM, as we will discuss in details.  
A color selection is preferred to a cluster membership assigned on the basis of the photometric redshift information. Our choice is motivated by the fact that we select galaxies that are in the field and at the redshift of F062 and F126, until the mean COSMOS density is reached. A selection based on the photometric redshift information \citep[e.g.][]{papovich2010} 
might be biased towards selecting cluster galaxies as well as field galaxies. This would imply an overestimation of the number of the 
cluster members as well as of number count excess associated with F062 and F126. Conversely, our color criterion avoids it, since we select galaxies starting from the reddest ones, that are most likely the elliptical galaxies of the cluster.

We subtract the cluster members from the fields of F062 and F126. 
We apply the PPM to see if any residual structure is detected to see if the cluster membership has been correctly assigned. 
Some of the cluster members may have not been identified. 
If this is the case, the PPM might still detect an overdensity in the field, once the cluster members are subtracted.
However, the opposite case in which too many sources are selected as cluster members is
not tested with this approach. This is because the PPM is not used to detect the presence of underdense regions.

We add the fiducial cluster members to the fields of two sources of the C09 sample, namely COSMOS-FRI~70 and COSMOS-FRI~66, where no overdensity is detected by the PPM in the redshift range $z\sim1-2$ 
of our interest (see also Paper~II). 
This is done applying a rigid rotation to the projected coordinates of the selected cluster members.
Two fields are used because weak overdensities not detected by the PPM procedure might be present in the redshift range of our interest.
Therefore, the clusters would be more easily detected if their members are shifted to the redshifts of these non-detected overdensities. This might imply that the cluster detection significance is overestimated. The choice of two fields reduce the possibility that this bias occurs.
Then we apply the PPM in the new field to test if Mpc-scale overdensities is still detected.

We shift the fiducial cluster members to $z_{\rm c,sim}=$~1.5, 2.0. We firstly estimate the AB I-band magnitude I$_{\rm sim}$ that
each of the cluster member would have if located at 
higher redshift $z_{\rm c,sim}$. Then we reject all of the cluster members with 
I$_{\rm sim}\geq$25. This is the same selection criterion applied 
in \citet{ilbert2009} in estimating the photometric redshifts. This is done to simulate the COSMOS sensitivity and to 
properly reject the faintest galaxies that would not be detected 
when shifted to a redshift higher than their own.

We assign a photometric redshift to each of the cluster members selected 
with the previous procedure, according to a Gaussian probability distribution. 
The average is set equal to the redshift of the simulated cluster $z_{\rm c,sim}$. 
We adopt a variance equal to the square of the typical statistical 1-$\sigma$ accuracy 
$\sigma_z(z_{\rm c,sim}) = 0.054(1+z_{\rm c,sim})$
of the photometric redshifts around $z_{\rm c,sim}$ for sources
with $i^+\sim24$ and redshifts within $1.5<z<3$ \citep[see Table~3 of][]{ilbert2009},
typical of the cluster galaxies we consider.
This is done in order to assign properly a photometric redshift to each 
of the cluster members once they are shifted to a redshift higher than the true redshift of the cluster.

We finally apply the PPM to see if the clusters are still detected 
by the PPM at $z=1.5$ and 2.0. 
\end{enumerate}
\subsection{Results}\label{sec:shifting_results}
%-------------------- Figure: PPM plots for F062, F126 (Finoguenov et al., 2007)----------
\begin{figure*} \centering
\subfigure[]{\includegraphics[width=0.45\textwidth,natwidth=610,natheight=642]{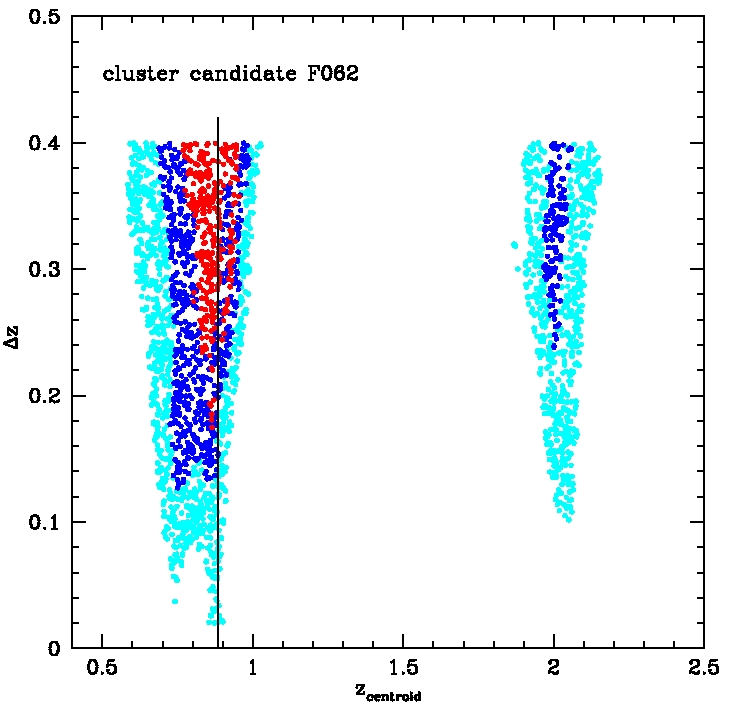}}\qquad
\subfigure[]{\includegraphics[width=0.45\textwidth,natwidth=610,natheight=642]{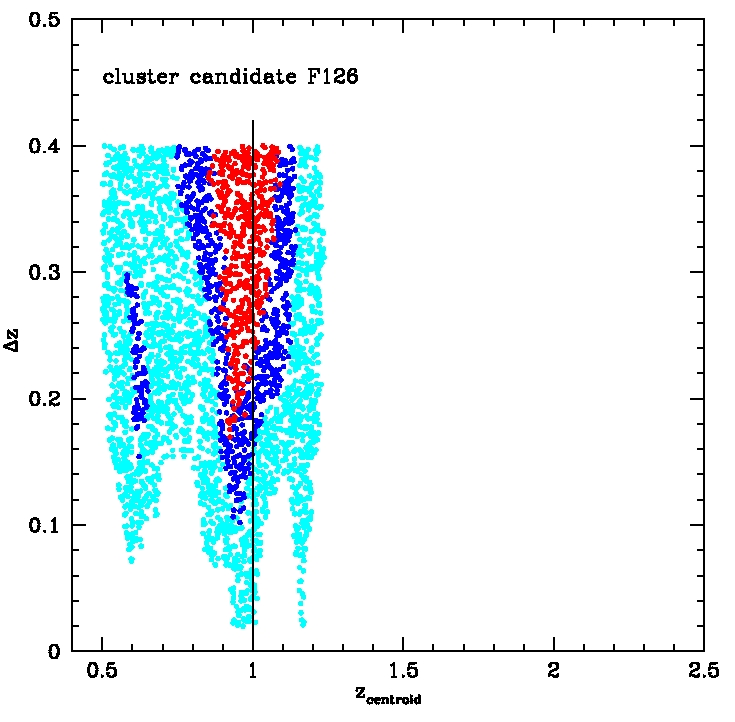}}
\caption{PPM plots for the cluster candidates F062 (left), F126
(right), as given in FGH07.  Overdensities:
$\geq2\sigma$ (cyan points), $\geq3\sigma$ (blue points),
$\geq4\sigma$ (red points). The vertical solid lines indicate the
redshift of the cluster candidate.}
\label{fig:PPMplots_F62_F126}
\end{figure*}
%-------------------------------------------------------------------------------------------
1.\hspace{0.4cm} In Figure~\ref{fig:PPMplots_F62_F126} we show the PPM plots (as in Figure~\ref{fig:ppm_plots}, see Sect.~\ref{sec:method_PPM}) for F062 (left) and F126 (right).
We adopt the following color code: $\geq2\sigma$,
$\geq3\sigma$, and $\geq4\sigma$ points are plotted in cyan, blue, and red,
respectively. The abscissa of the vertical solid line indicates the redshift of the cluster candidate.

In Table~\ref{tab:PPM_results_F062_F126} we report the PPM results for F062 and F126 (top table).
We also report the PPM results of our simulations, when these two clusters are added to the fields of COSMOS-FR~I~66 and  COSMOS-FR~I~70 of the \citet{chiaberge2009} sample (middle table). In the bottom table we report the PPM results where the clusters 
are shifted to $z=1.5$.
In the first four columns we list the cluster ID number (i.e. F062, F126), the cluster redshift, the cluster redshift as estimated by the PPM, and the cluster detection significance.
In the fifth column we report the distance ${\rm r_{max}}$ from the location of the radio galaxy in the projected space at which the overdensity formally ends.
For the above quantity, the average, the rms dispersion around the average and the median value (between parenthesis) in units of arcsec are reported, as estimated by the PPM procedure. 
The rms dispersion and the median value are not reported where the former is null, i.e. where the estimated ${\rm r_{max}}$ is maximally stable with respect to $z_{\rm centroid}$, i.e. where the rms dispersion is null. 
In the sixth column we report the field to which the cluster is added; 66 and 70 denote that the
cluster members are added to the fields of COSMOS FR I 66 and COSMOS FR I 70, respectively. The symbol --- denotes that the PPM is applied to the fields of F062 and F126, where the cluster
members are not subtracted.

Concerning the PPM results for F062 and F126 (top table), they are detected with significance levels
of 3.8$\sigma$ and 4.3$\sigma$, respectively.
The estimated redshifts are $z=0.86$ and $z=0.96$, respectively.
In addition to the cluster candidate at $z\sim1$, the PPM detects another 2.7$\sigma$ overdensity in the field of F126  at $z=0.64$.
This is a clear example of a projection effect.

Note that our redshift estimates fully agree with the actual redshifts of the two cluster candidates (i.e. $z = 0.88$ and $z=1.0$ for F062 and F126, respectively)
 and that the PPM effectively finds systems whose masses are compatible to 
those of rich groups and, therefore, they are even below the typical cluster mass cutoff $\sim1\times10^{14}$~M$_\odot$, 
as it is the case for F062 and F126.
Hereinafter we do not estimate the redshift uncertainties following the PPM procedure prescription.
This is mainly because we know the redshift of the cluster in our simulations. Therefore, we can 
directly compare our estimates with the original cluster redshifts to derive the statistical uncertainties.
Conversely, in Paper~II we estimate redshift uncertainties (following the procedure described above)
for the overdensities we find within the C09 sample.
\\

%  -----------------------    TABLE: PPM results for F062 and F126  -----------------------
\begin{table}
\caption{} 
\label{tab:PPM_results_F062_F126} \centering                       

\begin{tabular}{c}
 PPM results for F062 and F126
\end{tabular}

\begin{tabular}{cccccc}       
\hline\hline 
ID & $z_{\rm cluster}$ & $z_{\rm PPM}$ & significance & r$_{\rm max}$ (arcsec) & field \\ 
\hline                 
F062 &  0.88  & 0.86 &  3.8$\sigma$     & 72.6$\pm$5.1     (70.7)    & ---  \\
F126 &  1.00  & 0.96 &  4.3$\sigma$     & 181.3$\pm$33.4   (165.8)   & --- \\
\hline\hline
\vspace{0.01cm}
\end{tabular}
\\

\begin{tabular}{c}
 F062 and F126 added to the ECFs
\end{tabular}

\begin{tabular}{cccccc}       
\hline\hline 
ID & $z_{\rm cluster}$ & $z_{\rm PPM}$ & significance & r$_{\rm max}$ (arcsec) & field \\ 
\hline                 
F062 &  0.88  & 0.86 &  3.5$\sigma$     & 71.6$\pm$3.6     (70.7)     & 66 \\
F062 &  0.88  & 0.86 &  4.0$\sigma$     & 74.3$\pm$6.7     (70.7)     & 70 \\

F126 &  1.00  & 0.94 &  4.9$\sigma$     & 109.9$\pm$6.1    (111.8)    & 66 \\
F126 &  1.00  & 1.00 &  4.1$\sigma$     & 92.0$\pm$10.0    (86.6)     & 70 \\
\hline\hline
\vspace{0.01cm}
\end{tabular}
\\

\begin{tabular}{c}
 F062 and F126 added to the ECFs and shifted to higher redshift.
\end{tabular}

\begin{tabular}{cccccc}       
\hline\hline 
ID & $z_{\rm cluster}$ & $z_{\rm PPM}$ & significance & r$_{\rm max}$ (arcsec) & field \\ 
\hline                 
F062 &  1.50  & 1.61 &  2.6$\sigma$     & 50.0 ------                 & 66 \\
F062 &  1.50  & 1.56 &  2.9$\sigma$     & 50.0 ------                 & 70 \\

F126 &  1.50  & 1.51 &  3.1$\sigma$     & 111.3$\pm$2.4    (111.8)    & 66 \\
F126 &  1.50  & 1.59 &  2.5$\sigma$     & 85.4$\pm$4.2     (86.6)     & 70 \\
\hline\hline
\vspace{0.01cm}
\end{tabular}
\\

\tablecomments{PPM results for F062 and F126 where the cluster members are not removed (top table).
PPM results where the cluster members are added to the ECFs (middle table) and where they are also shifted to $z = 1.5$ (bottom table).
\\
Column description: (1) cluster ID number; (2) cluster redshift; (3) cluster redshift estimated with the PPM; (4) significance of the overdensity estimated by the PPM in terms of $\sigma$; (5) average maximum radius [arcsec] of the overdensity along with the rms dispersion around the average (both estimated with the PPM). The median value [arcsec] is written between the parenthesis;
(6) field to which the cluster is added; 66 and 70 denote that the cluster members are added to the fields of COSMOS-FR~I~66 and COSMOS-FR~I~70, respectively. The symbol ---  denotes that the PPM is applied to the fields of F062 and F126, where the cluster members are not subtracted.}
\end{table}
%----------------------------------------------------------------------------------------

2.\hspace{0.4cm} We then apply the PPM using increasing offsets $\theta_{off}$
 between the cluster center of the PPM tessellation and the actual center as measured from the X-ray emission.
This is done to find the required accuracy in the coordinates of the cluster center in order to detect Mpc-scale overdensities
with the PPM. 
We keep the right ascension of the center of the PPM tessellation fixed and we change its declination from $\theta_{off}=$10 up to 500~arcsec.

We find that F062 and F126 are detected up to $\theta_{off}=150$ and 500~arcsec, respectively.
%, at their true redshifts, i.e.
%between $z=0.82-0.91$ and $z=0.95-1.02$, respectively.
The clusters are detected with a fairly constant significance (between $\sim3.2-3.8\sigma$ and $\sim3.7-4.8\sigma$ for F062 and F126, respectively). 
However a mild trend 
of decreasing significance for increasing offsets is observed.
F062 is detected with significances of 3.8, 3.2, and 3.2$\sigma$ at $\theta_{off}=0$, 75, and 150~arcsec, respectively.
 In fact, F126 is detected with significances of 4.3, 3.9, 3.7, and 3.7$\sigma$ at $\theta_{off}=0$, 100, 300, and 500~arcsec, respectively.

A clear trend between $r_{\rm max}$ and $\theta_{off}$ is observed for F062. The estimated size increases up to $r_{\rm max}\simeq150$~arcsec
for $\theta_{off}=125$~arcsec. While the estimated size for $\theta_{off} = 0$~arcsec is 
$r_{\rm max}=72.6 \pm 5.1$~arcsec. 
Conversely, no trend is observed for F126.
%, for which the estimate cluster sizes vary between r$_{\rm max}\simeq140-200$~arcsec 
%at increasing the offset up to $\theta = 500$~arcsec, that are $\sim$4~Mpc at $z=1$.
%This, in addition to the fact that F126 is always detected starting from the coordinates of the center of the PPM
%tessellation (i.e. $r_{\rm min}=0$~arcsec) at varying the offset $\theta$, 
%suggests that F126 might be part of a more extendend $z\sim1$ structure.
%Similarly, F062 is also detected starting from the coordinates of the center of the PPM
%tessellation (i.e. $r_{\rm min}=0$~arcsec) at varying the offset up to $\theta=125$~arcsec. 
%It is instead detected starting from a positive projected separation $r_{\rm min}=50$~arcsec 
%at the maximum offset $\theta=150$~arcsec, at which F062 is still detected by the PPM.

These results suggest that the PPM is effective to detect Mpc-scale overdensities
even if the projected coordinates of the cluster center are known 
with an accuracy of only $\sim100$~arcsec. This implies that the PPM can be efficiently applied even if the cluster center coordinates are not accurately known.
\\

3.\hspace{0.4cm} We want to shift these two groups to redshifts higher than $z\sim1$, thus we select the fiducial cluster members of both F062 and F126. 
We select those sources that fall within circular regions of radius 70.7 and 165.8~arcsec centered at the coordinates of the cluster center, for F062 and F126, respectively.
These are the regions in the projected space within which the clusters are detected by the PPM.

We conservatively select sources with photometric redshifts 
within a redshift slice $\Delta z=4\sigma_z(z_{c})$ centered around the redshift $z_{c}$ of the cluster,
where $\sigma_z(z_c) = 0.054(1+z_c)$ is the 1-$\sigma$ statistical photometric redshift uncertainty
of faint galaxies with $i^+\sim24$ and $1.5<z<3$ sources \citep[see Table~3 of][]{ilbert2009},
typical of the cluster galaxies we consider. 
The redshift slice considered here is higher than that adopted throughout the PPM procedure (i.e. $\Delta z=0.28$) 
to make sure that the large majority of the sources at the redshift of the cluster are included in the bin,
 even if the accuracy of their photometric redshifts is poor. This is not a cluster membership assignment. 
In fact, the cluster members will be selected among these sources by using an $(I-K)_{AB}$ color criterion, as we will describe in the following.\footnote{We denote as I the 
{\it Subaru} $i^+$ magnitude and, if absent, the CFHT $i'$ magnitude. We denote as K the CFHT K-band magnitude. All these magnitudes are from the I09 catalog, they are in AB system, 
and they are measured within an aperture of 3~arcsec diameter.}

Since red and passively evolving galaxies constitute the majority among the cluster core galaxies
at $z\sim1$ we also adopt a color selection criterion to define the fiducial cluster members.
We sort the selected galaxies according to their $(I-K)_{AB}$ color, from the redder to the bluer.
This specific color criterion has been chosen because the rest frame $\sim$4000~\AA~absorption feature
typical of the spectra of early type galaxies
falls just between the $K$- and $I$- bands at redshift $z\sim1$.

The cluster members are then removed from the field starting from the reddest source until the 
average COSMOS number density within the selected $4\sigma_z$ bin is reached.

According to the outlined procedure we select as cluster members 57 and 249 galaxies down to $(I-K)_{AB}=1.12$ and 1.30 magnitudes
for F062 and F126, respectively. 
As expected, these cluster members are faint, since their I-band magnitudes are between $I\sim21.9-25.3$ and $I\sim21.1-25.7$, for 
F062 and F126, respectively.

%------------------------- Figure: radial profiles cluster F062, F126 -----------------------
\begin{figure*} \centering
\subfigure{\includegraphics[width=0.4\textwidth,natwidth=610,natheight=642]{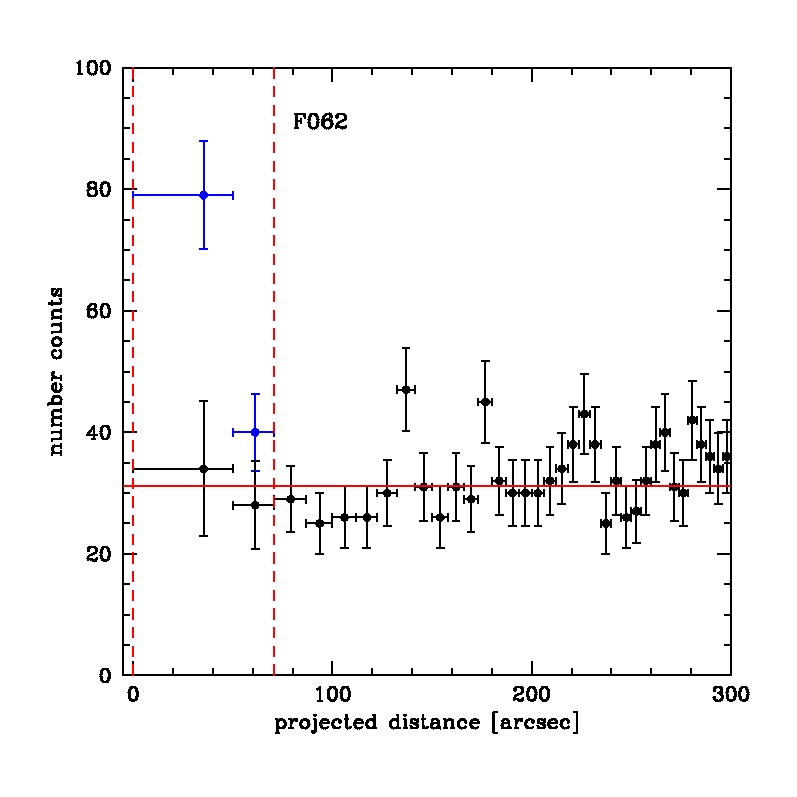}}\qquad
\subfigure{\includegraphics[width=0.4\textwidth,natwidth=610,natheight=642]{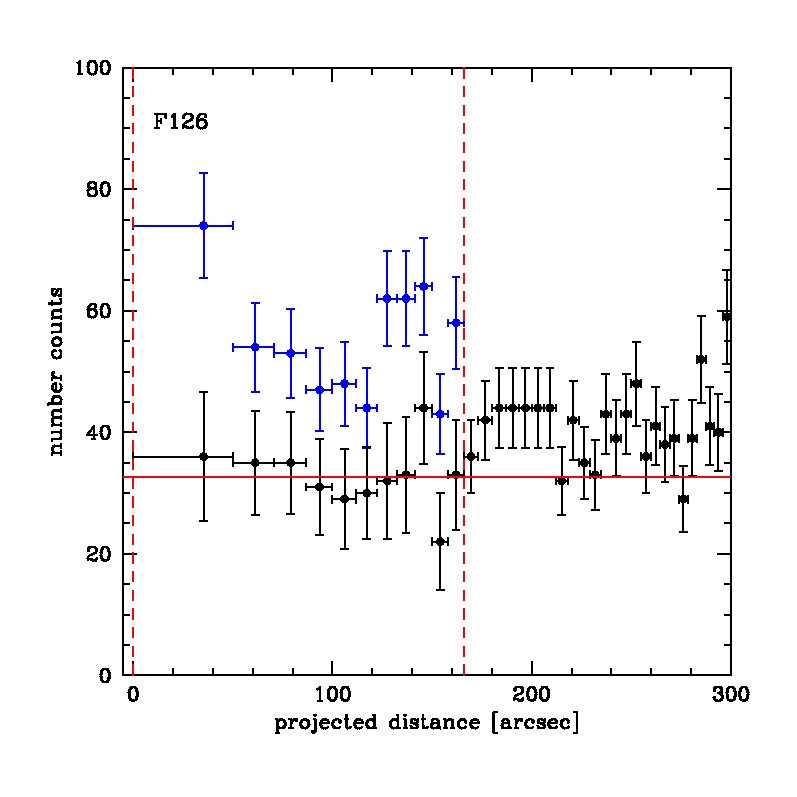}}
\caption{Blue points: differential number counts of the sources in the fields of F062 (left panel) and F126 (right panel), as a function of the distance from the cluster center coordinates. Sources are counted within a $\Delta z = 4\sigma_z$ redshift bin centered at the redshift of the cluster (i.e. $\Delta z = 0.406$ and 0.432 for F062 and F126, respectively). 
Black points: differential number counts, as for blue points, where the cluster members are subtracted from the field of the cluster. Number count 1-$\sigma$ uncertainties are plotted along the y-axis. The uncertainty in the radial coordinate is the half-width of each region within 
which the number counts are performed.
Vertical dashed lines show the region where the cluster members are selected.
The horizontal solid red line show the mean COSMOS number count per area.}

\label{fig:radial_profiles_fino}
\end{figure*}
%---------------------------------------------------------------------------------------------

We want to make sure that the cluster membership is not biased towards preferentially selecting sources that are located in certain regions in the projected space around the cluster center. 
In order to do this we verify whether the differential
 radial number counts are consistent with a constant -- no clustering -- once the cluster members are removed from the field.

In Figure~\ref{fig:radial_profiles_fino} we plot the differential number counts of the sources in the fields of F062 (left panel) and F126 (right panel), as a function of the distance from the projected cluster center coordinates. 
Sources are counted within the $4\sigma_z$ redshift slice adopted throughout the cluster membership procedure and within regions of equal areas 
(i.e. 2.18~arcmin$^2$), analogously to what done for the PPM. The areas are chosen equal among each other
in order to have a constant mean field density per region.

The galaxy number counts along with the corresponding 1-$\sigma$ Poisson uncertainties for the error are plotted as blue symbols. 
Number counts, once the cluster members are subtracted from the fields of both F062 and F126, are plotted as black points, along with the 1-$\sigma$ uncertainties estimated according to the Skellam distribution.\footnote{The Skellam distribution is the discrete probability distribution of the difference of two statistically independent random variables each having Poisson distributions.
In this case, the Skellam probability is chosen because we subtract the cluster members from the field.}
The uncertainty in the radial coordinate corresponds to the radial half-width of each region.

The vertical red dashed lines
show the radial interval within which the cluster members are selected.
By construction, according to the cluster membership procedure, black and blue points coincide outside of this interval.
The horizontal line shows 
the mean COSMOS number counts of $\sim30$ galaxies associated with a 2.18~arcmin$^2$ area around which the black points are scattered.

The radial profiles of both F062 and F126 clearly show that the number count excess (blue points) is limited within the projected area defined within the vertical dashed lines. Furthermore, once the cluster members are subtracted, such a number count excess disappears. In fact, the values associated with the black points are consistent with the mean COSMOS number density within the reported 1-$\sigma$ uncertainties.

\begin{figure*} \centering
\subfigure{\includegraphics[width=0.45\textwidth,natwidth=610,natheight=642]{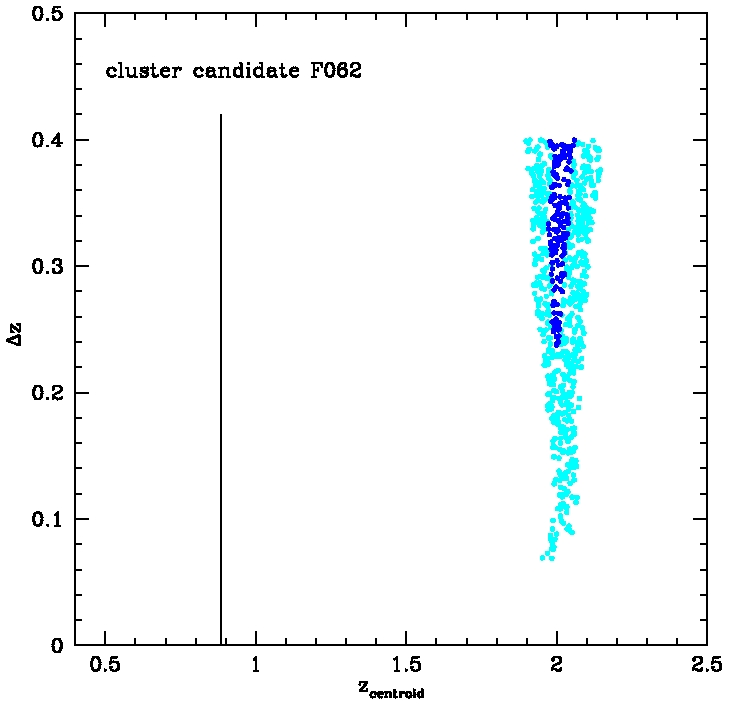}}\qquad
\subfigure{\includegraphics[width=0.45\textwidth,natwidth=610,natheight=642]{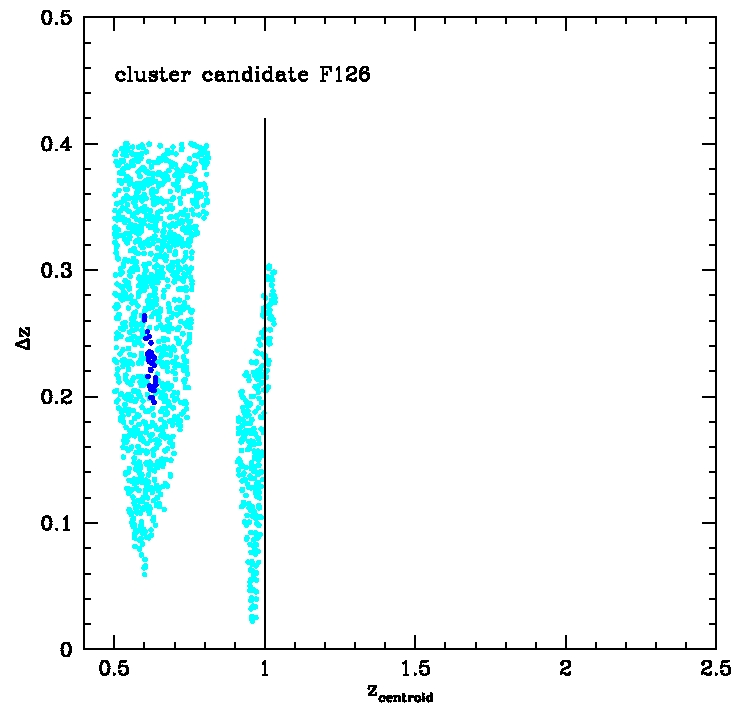}}
\caption{PPM plots where the clusters F062 (left) and F126 (right), in the FGH07 catalog, are subtracted. Overdensities: $\geq2\sigma$ (cyan points), $\geq3\sigma$
(blue points), $\geq4\sigma$ (red points). The vertical solid line indicates the redshift of the cluster candidate.}
\label{fig:PPMcluster_sutracted_F62_F126}
\end{figure*}

In Figure~\ref{fig:PPMcluster_sutracted_F62_F126} we report the PPM plots
of the fields of F062 and F126, where the cluster members are
subtracted. The adopted color code is analogous to that of Figure~\ref{fig:PPMplots_F62_F126}.
We apply the PPM and we verify that neither F062 nor F126 are now detected.
In fact, as is clear from visual inspection, the high significance pattern at the redshift of the cluster completely disappears in the case of F062, while a residual $\gtrsim2\sigma$ feature is still present in the case of F126 at its redshift.
According to the PPM procedure, such a feature is interpreted as noise because it is not enough extended to be detected as overdensity, 
i.e. it is less than $\delta z_{\rm centroid} = 0.1$ long on the redshift 
axis $z_{\rm centroid}$ at fixed $\Delta z = 0.28$.

The other Mpc scale overdensity that was previously detected by the PPM in the field of F126 at $z=0.64$ with a significance of 2.7$\sigma$ is still detected with similar significance (2.6$\sigma$) and redshift ($z=0.61$). This confirms that the specific cluster membership assigned 
here combined with the PPM  is efficient at removing the degeneracy resulting from the projection effect.

%-------------------------Figure: PPM plots, cluster subtracted, F062, F126 -----------------------
%---------------------------------------------------------------------------------------------

%-------------------- Figure: radial PPM plots for F062, F126 (Finoguenov et al., 2007)------
\begin{figure*} \centering
%\subfigure[]{\includegraphics[width=0.45\textwidth,natwidth=610,natheight=642]{rmin_PPM_finoguenov62_offset0arcsec.jpg}}\qquad
\subfigure{\includegraphics[width=0.45\textwidth,natwidth=610,natheight=642]{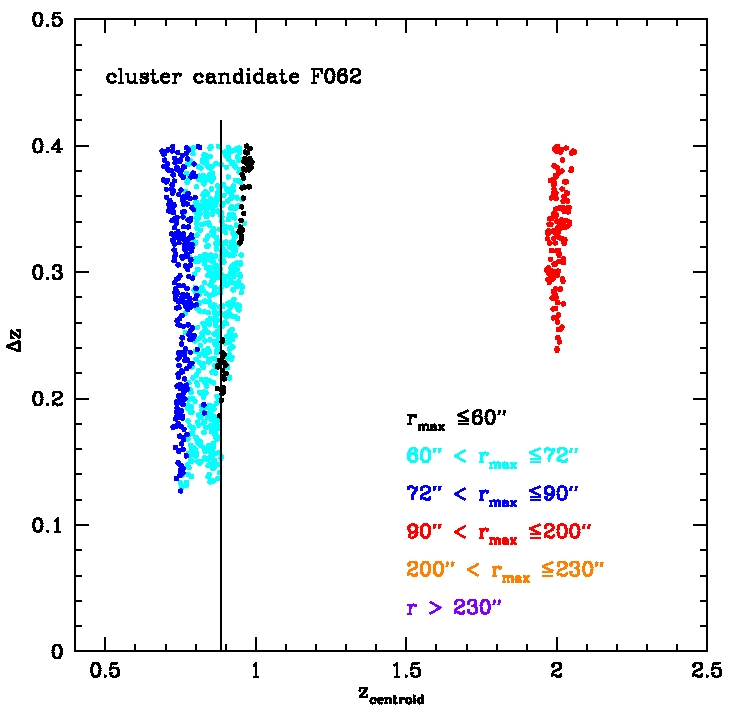}}
%\subfigure[]{\includegraphics[width=0.45\textwidth,natwidth=610,natheight=642]{rmin_PPM_finoguenov126_offset0arcsec.jpg}}
\subfigure{\includegraphics[width=0.45\textwidth,natwidth=610,natheight=642]{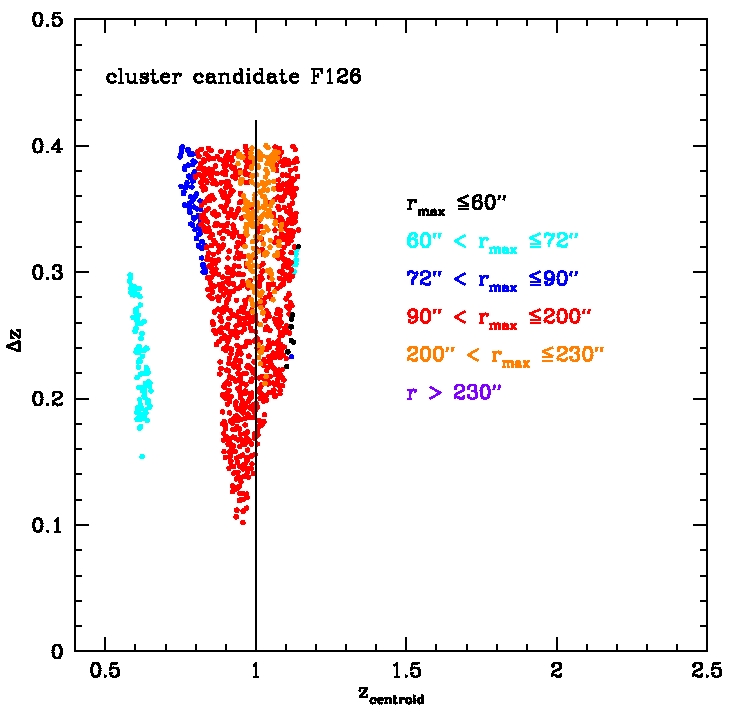}}
\caption{PPM plots for F062 (left) and F126 (right). We plot only those points that correspond to $\geq3\sigma$ overdensities for a specific choice of the redshift bin $\Delta z$ and its centroid $z_{\rm centroid}$. 
Different colors correspond to different values of the
cluster size $r_{\rm max}$ associated with each point and estimated with the PPM.
See the legend in the plots for further information about the color code adopted.}
\label{fig:radialPPMplots_F62_F126}
\end{figure*}
%-------------------------------------------------------------------------------------------

As explained above, for our simulations we perform the cluster membership by selecting fiducial cluster members within a circular region centered at the X-ray coordinates of F062 and F126. The radius of the region corresponds to the projected size of the cluster, as estimated by the PPM. In the following we reconsider our estimates by using the PPM plots and we compare the fiducial sizes estimated with the PPM with those obtained by previous work.

In Figure~\ref{fig:radialPPMplots_F62_F126} we report the PPM plots for F062 (left panel) and F126 (right panel), where only the points corresponding to $\geq$3-$\sigma$ overdensities are plotted.
This is because the two clusters are detected with significance higher than 3-$\sigma$.
Since we are interested in the cluster size, we plot the cluster size (${\rm r_{max}}$) associated with each point in the plot, as estimated by the PPM for specific $z_{centroid}$ and $\Delta z$.
We refer to the legend in Figure~\ref{fig:radialPPMplots_F62_F126} for
the specific color code adopted. As clear from visual inspection of the plots, the values of ${\rm r_{max}}$ are stable with respect to the $\Delta z$ parameter. 

By averaging among the values associated with the points at $\Delta z \simeq 0.28$ that
define the overdensities we estimate the sizes r$_{\rm max}=72.6\pm5.1$~arcsec and r$_{\rm max}=181.3\pm33.4$~arcsec for  F062 and F126, respectively, 
according to the PPM procedure. Here we report the average value and the rms dispersion around the average. The cluster sizes 70.7~arcsec and 165.8~arcsec that are assumed when performing the cluster membership correspond to the median values of r$_{\rm max}$ for F062 and F126, respectively. 

On the basis of the X-ray surface brightness FGH07 estimated a core size r$_{500}$~=~48~arcsec for both F062 and F126.
By assuming spherical symmetry and a $\beta$-model density profile for the cluster matter distribution \citep{cavaliere_fusco1978}
we estimate r$_{200}=76$~arcsec for both F062 and F126\footnote{Here r$_{500}$ (r$_{200}$) is the radius at which the enclosed mass  
encompasses the matter density 500 (200) times the critical. 
In estimating r$_{200}$ for both F062 and F126 we also assume hydrostatic equilibrium. We use Eq.~(3) 
of \citet{reiprich_bohringer99} and the core radius estimates as in Eq.~(4) of FGH07.}.

\citet{george2011} estimated core sizes  r$_{200}$~=~73~arcsec and 81~arcsec 
and core masses  $M_{200}=5.25\times10^{13}M_{\odot}$ and $8.32\times10^{13}M_{\odot}$, for F062 and F126
\footnote{Here M$_{200}$ is the mass enclosed within the radius 
encompassing the matter density 200 times the critical.}, respectively, on the basis of  the mass vs. X-ray luminosity relation given in \citet{leauthaud2010}.
By using virial assumption and spectroscopic redshift information \citet{knobel2012} estimated a size of 659~kpc (i.e. $\sim$84~arcsec at the redshift of the cluster) for F062.

We find that our size estimates are in good agreement with those reported by previous work.
However, for F126 our estimate is higher than previous work.

Since we want to shift the cluster members of both F062 and F126 to higher redshifts
we select two fields where no overdensity is detected with the PPM in the redshift range $z\sim1-2$.
We prefer to shift the cluster members of both F062 and F126 into other fields because we want to make sure than no overdensity is detected by the PPM in the redshift range $z\sim1-2$ for the considered field. 
We note in fact that this is not the case of F062, i.e. a 2.4$\sigma$ overdensity is detected at $z=2.00$. 
Furthermore, the choice of the same field for both  F062 and F126 allows us to directly compare the results we obtain with the PPM for the two clusters, once the cluster members are added to such a field.

\begin{figure*} \centering
\subfigure{\includegraphics[width=0.45\textwidth,natwidth=610,natheight=642]{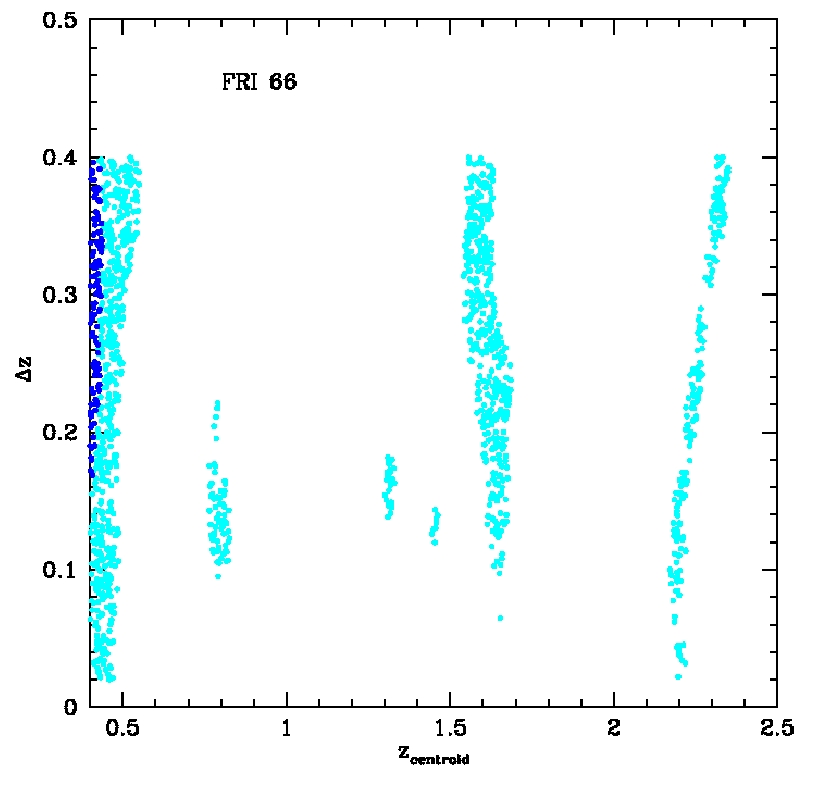}}\qquad
\subfigure{\includegraphics[width=0.45\textwidth,natwidth=610,natheight=642]{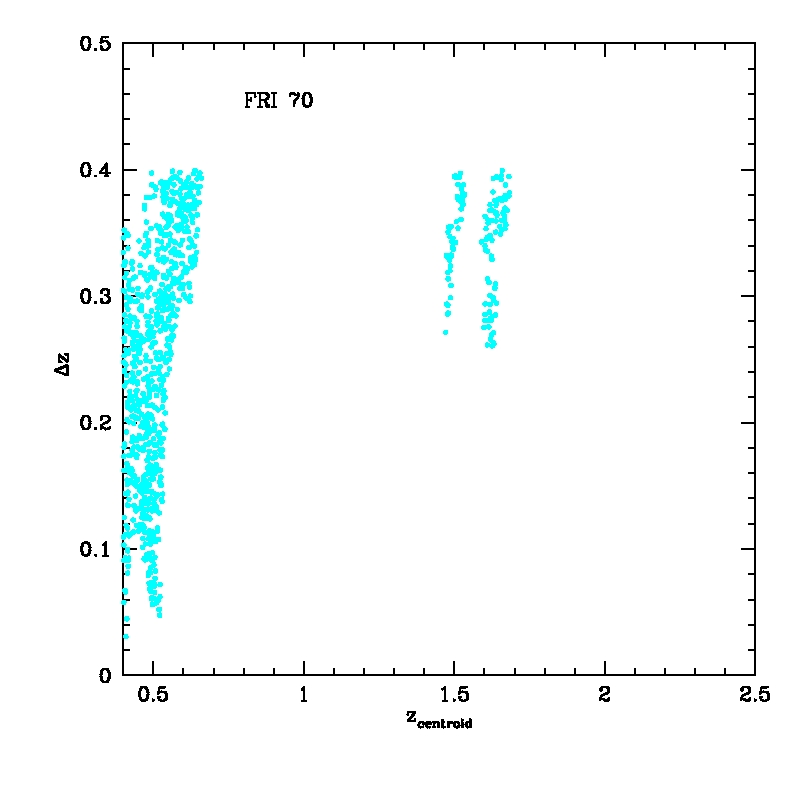}}
\caption{PPM plots for the fields of COSMOS-FR~I~66 (left) and COSMOS-FR~I~70 (right) in the \citet{chiaberge2009} sample.
Overdensities: $\geq2\sigma$ (cyan points), $\geq3\sigma$ (blue points), 
$\geq4\sigma$ (red points).}
\label{fig:PPMplot_066_070}
\end{figure*}

In Figure~\ref{fig:PPMplot_066_070} we report the PPM plots for the fields of COSMOS-FR~I~66 (left panel) and COSMOS-FR~I~70 (right panel). 
As clear for visual inspection of the plots, no high significance pattern is detected
in these plots within the redshift range $z\sim1-2$.
A weak 2$\sigma$ overdensity is detected by the PPM at redshift $z=1.60$ 
in the field of COSMOS-FR~I~66. However such a feature is not detected if a slightly different redshift bin (i.e. $\Delta z =0.24$) is adopted 
throughout the PPM procedure.
All of the other isolated $\gtrsim2\sigma$ patterns clearly visible in the plots are interpreted as noise.
This is because either they are not located around the y-axis value $\Delta z =0.28$ that is relevant for the overdensity detection or they are not enough extended to be detected as overdensity 
(i.e. they are less than $\delta z_{\rm centroid} = 0.1$  long on the redshift axis $z_{\rm centroid}$ at fixed $\Delta z =0.28$), according to the PPM procedure.

Since no clear overdensity is detected in the fields of COSMOS-FR~I~66 and COSMOS-FR~I~70 we  
use them as empty control fields (ECFs).
Note that we cannot exclude the presence of underdense or dense regions 
that are not detected by the PPM, but still present in these two ECFs at the redshifts of our interest. 

If a cluster is superimposed on an underdense region the PPM might underestimate the detection significance or it might detect no overdensity. Conversely, if the cluster is added to an overdense region, the PPM tends to overestimate the overdensity significance.
The reason to choose two ECFs instead of one is to see whether these two scenarios occur.
In particular, we will compare our results obtained from each ECF separately to look for a possible mismatch.

%------------------------- Figure: PPM plots for 70, 66 -----------------------
%---------------------------------------------------------------------------------------------

\begin{figure*} \centering
\subfigure{\includegraphics[width=0.45\textwidth,natwidth=610,natheight=642]{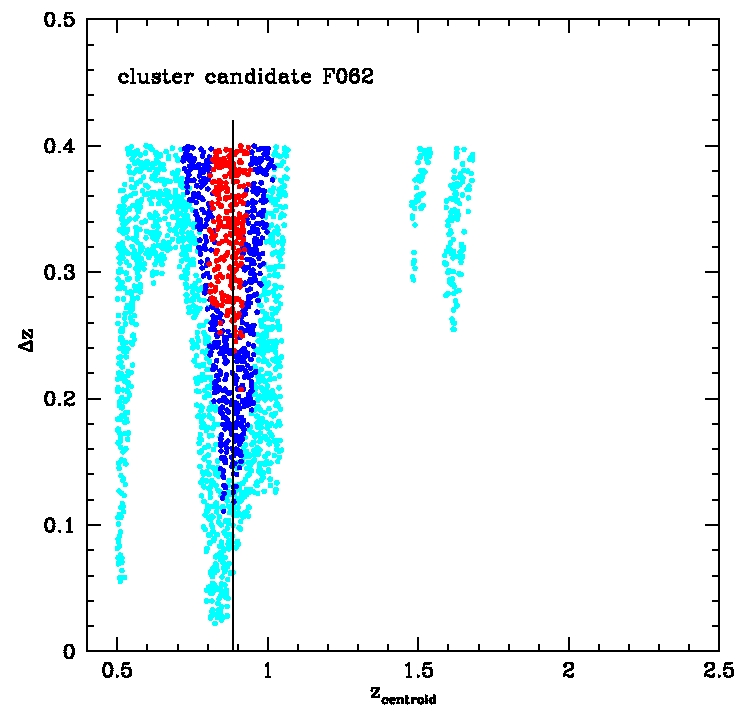}}\qquad
\subfigure{\includegraphics[width=0.45\textwidth,natwidth=610,natheight=642]{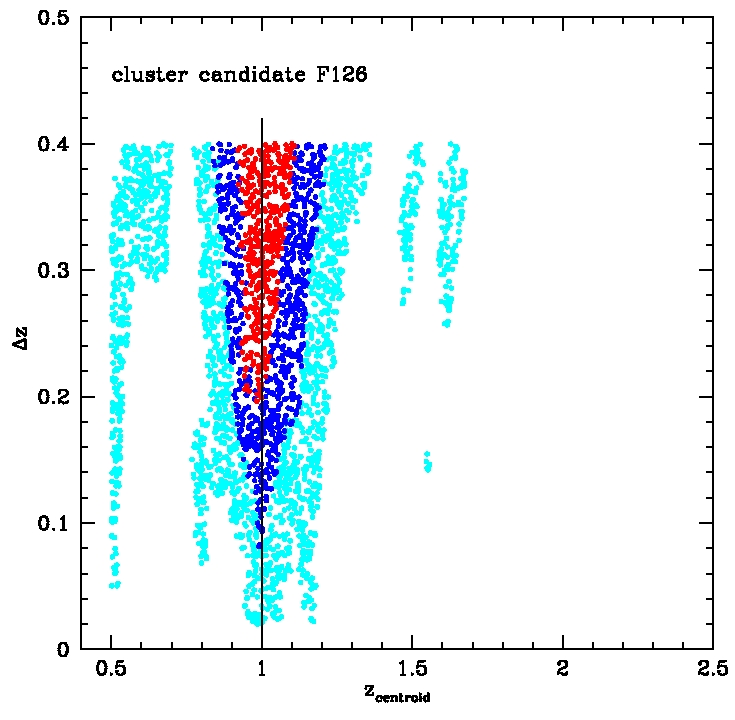}}
\caption{PPM plots of F062 (left) and F126 (right), where their cluster members are added to the field of COSMOS-FR~I~70. The vertical solid  line in each panel is located at the redshift of the cluster. Overdensities: $\geq2\sigma$ (cyan points), $\geq3\sigma$ (blue points), 
$\geq4\sigma$ (red points).}
\label{fig:PPMplot_F62_F126_in_the_field_of070}
\end{figure*}

We add to each ECF the fiducial cluster members of F062 and F126, separately, 
and we apply the PPM at the coordinates of COSMOS-FR~I~66 and 
COSMOS-FR~I~70.\footnote{We add the cluster members into each ECF by applying a rigid rotation to all the projected coordinates of the cluster members.}
In Figure~\ref{fig:PPMplot_F62_F126_in_the_field_of070} we report the
resulting PPM plots, where the clusters F062 (left panel) and F126 (right panel) are in the field of COSMOS-FR~I~70. 
As clear from visual inspection of the plots, both F062 and F126 are still detected at their true redshift with significances between $\sim$3-4$\sigma$.
In Table~\ref{tab:PPM_results_F062_F126} (middle table) we report the PPM results of our simulations.
The estimated redshifts for F062 and F126 are $z=0.86$ and $z=1.0$, respectively, where the cluster members are added to the fields of COSMOS-FR~I~70.
Therefore, the estimated redshifts fully agree with those of the clusters.
The estimated sizes of F062 and F126 are $r_{\rm max}= 74.3\pm6.7$~arcsec and $r_{\rm max}= 92.0\pm10.0$~arcsec, respectively, where the cluster members are added to the field of COSMOS-FR~I~70.
These size estimates agree, indepentently of the ECF adopted (within the errors), with those previously obtained where the cluster members are not subtracted from their own fields (see Table~\ref{tab:PPM_results_F062_F126}, top table).
These results suggest that the cluster properties estimated by the PPM are not affected by the applied cluster membership and by the fact that the cluster members are added to a field different from that original of the cluster.

%-------------- Figure: PPM plots for F062, F126 rigidly rotated in the field of 70--------
%---------------------------------------------------------------------------------------------

In order to shift the cluster members of both F062 and F126 to higher
redshifts (i.e. $z_c=1.5$ and 2.0) we need to address the problem of the detection limit.
The COSMOS number density drops off rapidly with increasing redshift.
In fact, the number density per unit redshift is, on average,
${\rm dn}/{\rm d}z/{\rm d}\Omega \simeq 25$, 10, and 3~arcmin$^{-2}$
 at redshift $z\sim$1, 1.5, and 2.0, respectively \citep[see][]{ilbert2009}. 
Therefore we expect some of the selected cluster members 
would not be detected if they were located at higher redshifts, since
their flux would be lower than the survey threshold.

In order to address this problem we estimate the I band magnitude each cluster member would have if located at higher redshift. 
Then we  reject all the sources with I$\geq25$, that is the
magnitude cut applied to the \citet{ilbert2009} catalog.

We assume that each cluster member is located at the redshift of the clusters F062 and F126, i.e. $z_{c} = 0.88$ and 1.0, respectively. Then, we estimate the simulated I-band magnitude each cluster member would have if shifted to $z_{c,sim}=1.5$ and $2.0$. Practically, we perform the K-correction by using the SED of each object, i.e. we linearly interpolate the flux measurements reported in the I09 catalog, and we correct the apparent magnitude for the luminosity distance.

Then, as outlined above, we reject all the members for which I$_{\rm AB}\geq25$. 
This procedure reduces the number of the cluster
members from 57 to 9 sources ($z_{c,sim}=1.5$) and one source  ($z_{c,sim}=2.0$) in the case of F062 and from 249 to 58
($z_{c,sim}=1.5$) and 9 galaxies ($z_{c,sim}=2.0$) for F126. 
We note that the magnitude cut (I$<25$) is applied in the \citet{ilbert2009} to the I(auto) magnitude, that 
corresponds to the Subaru $i^+$ band magnitude obtained with SExtractor \citep{bertin_arnouts1996}. 
Therefore, the I(auto) magnitudes should be considered instead of the I (Subaru or CFHT) magnitudes. However, we prefer to adopt the I magnitude instead of the I(auto) magnitude because the latter is automatically estimated by SExtractor and, therefore, the former is more reliable for our simulations.
However, we verify that the I(auto) magnitudes of the selected cluster members are, on average only 0.3$\pm$0.4 and 0.3$\pm$0.2 lower than the corresponding I magnitudes, for F062 and F126, respectively. The reported uncertainty is the rms dispersion around the average.
Therefore, the I magnitudes are consistent within $\sim$1-$\sigma$ 
with the I(auto) magnitudes for the selected cluster members. 
This suggests that the results of our simulations would not change if we chose the I(auto) instead of the I magnitudes.

In the following we will address the problem of assigning coordinates to the cluster members of both F062 and F126, when they are located at  $z_{c,sim}\geq1.5$. 
The K-correction applied here neglects any contribution from possible evolution.
\\
%We note that the re-selection of cluster members performed here is independent of
%the particular assignment for their coordinates.

%In particular, we would be underestimating the number of cluster members at redshift $z_{c,sim}\geq1.5$
%if the cluster members we are considering experienced previous epochs of
%star formation activity at $z_{c,sim}\geq1.5$. 
%On the contrary, an increasing of the star formation might also imply an increasing of the dust absorption 
%from the host galaxy. If this is the case each galaxy would be redder than estimated.
%This would result in an overestimation of the number of the cluster members.

4.\hspace{0.4cm} Having addressed the problem of cluster membership, we assign 
fiducial coordinates to each of the cluster members,
when the overdensity is shifted to a higher redshift.
We assume that the coordinates in the projected space of
each galaxy remain unchanged when the overdensity is
shifted to higher redshift. Therefore, projection effects and the peculiar motions of the  galaxies are neglected.
This approximation is good enough because a high accuracy of the projected coordinates of the cluster members is not required in order to apply the PPM. In fact, each area of the PPM tessellation has a projected size of a few $\sim100$~kpc. 
Such a size is much larger than the projected positional uncertainty resulting from our approximation.
%This is an approximation because i) galaxies that lie along the same line of sight do not remain aligned if the
%cluster is shifted to higher redshift. ii) The cluster might be 
%still forming and assembling at redshift $z\gtrsim1.5$. 

Concerning the galaxy redshifts we assume that all the selected members are at the same distance to the observer, corresponding to redshift $z_{c,sim}=1.5$,~$2.0$, equal to 
that of the simulated cluster.
Then, we assign to each cluster member a photometric redshift
accordingly to a Gaussian probability distribution centered at the redshift of the cluster $z_{c,sim}$
and a standard
deviation $\sigma_{c,sim}=0.054(1+z_{c,sim})$. 

This value corresponds to the 1-$\sigma$ statistical photometric redshift uncertainty
of $i^+\sim24$ and $1.5<z<3$ sources \citep[see Table~3 of][]{ilbert2009},
typical of the cluster galaxies we consider. 
\\

%----------------------- Figure: PPM plots of F062 and F126 at z = 1.5 ---------------------------
\begin{figure*}[hctb] \centering
\subfigure[]{\includegraphics[width=0.45\textwidth,natwidth=610,natheight=642]{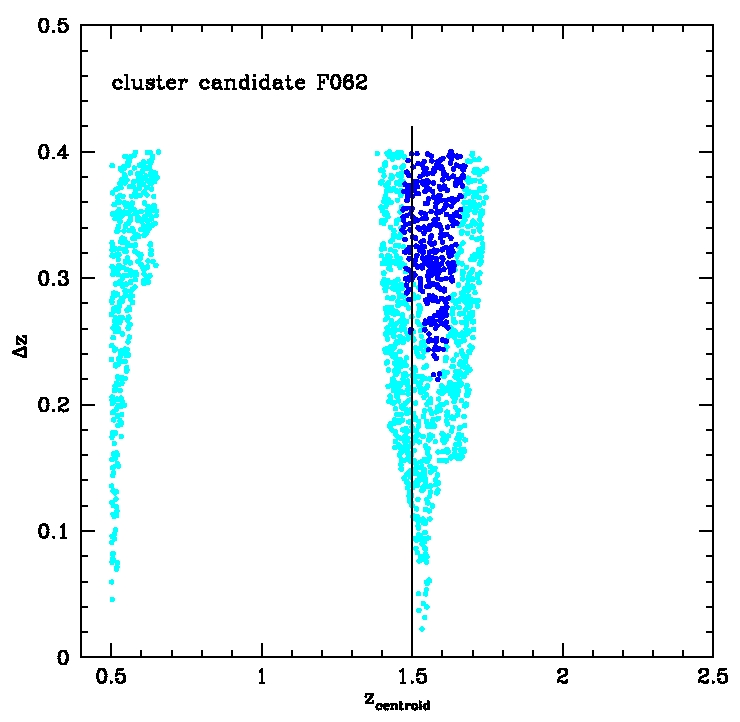}}\qquad
\subfigure[]{\includegraphics[width=0.45\textwidth,natwidth=610,natheight=642]{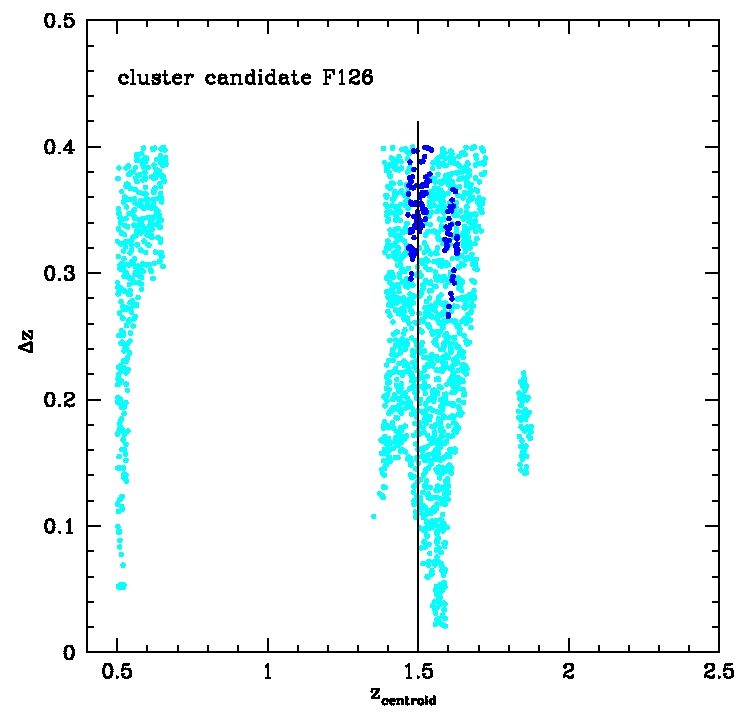}}\qquad
\caption{PPM plots for the F062 (left panel) and F126 (right panel), shifted at $z_{c,sim}=1.5$ and located 
in the field of COSMOS-FR~I~70.
The abscissa of the vertical solid line is at the redshift of the overdensity ($z_{c,sim}=1.5$). 
We plot only the points corresponding to detected overdensities for 
different values of $\Delta z$ and $z_{\rm centroid}$. Color code: $\geq2\sigma$ (cyan points), $\geq3\sigma$ (blue points),
$\geq4\sigma$ (red points). The Gaussian filter which eliminates high frequency noisy patterns is applied.}
\label{fig:ppm_plots_F062_and_F126_z15}
\end{figure*}
%--------------------------------------------------------------------------------

5.\hspace{0.4cm} We shift both F062 and F126 to higher redshift, i.e. $z_{c,sim}=1.5$, where the cluster members are added to the fields of COSMOS-FR~I~66 and COSMOS-FR~I~70, separately.
In Figure~\ref{fig:ppm_plots_F062_and_F126_z15} we report the corresponding PPM plots for both F062 (left panel) and F126 (right panel). The vertical solid line is located at the redshift of the overdensity.
As clear from visual inspection of the PPM plots, both F062 and F126 are still detected, if they are located at $z_{c,sim}=1.5$. 
In Table~\ref{tab:PPM_results_F062_F126} (bottom table) we report the PPM results for these simulations.
F062 and F126 are detected with 2.9$\sigma$ and 2.5$\sigma$ significance levels; the estimated redshifts are $z=1.56$ and $z=1.59$, respectively, if the cluster members are added to the field of COSMOS-FR~I~70.
This suggests that the PPM is effective in finding high redshift 
groups at $z\simeq1.5$, albeit with lower significance than at $z\sim1$ (i.e. $\sim$2.5-3$\sigma$).
The estimated sizes for both F062 and F126 are consistent, within the reported errors, with those previously obtained for these two clusters
at their true redshift (see Table~\ref{tab:PPM_results_F062_F126}, top table) The results are quite independent of the ECF considered. 
Neither F062 nor F126 is detected at $z_{c,sim}=2.0$.

\section{Simulated clusters.}
\label{sec:simulated_clusters} 

We now perform another set of simulations, by creating
simulated clusters with different richness and size. 
Then, we apply the PPM to
test if they are detected at different redshifts.

We consider as cluster
members $N_c$ sources uniformly distributed within a sphere of
comoving radius $R_c$ centered at the redshift $z_c$.

\begin{figure*} \centering
\subfigure{\includegraphics[width=0.45\textwidth,natwidth=610,natheight=642]{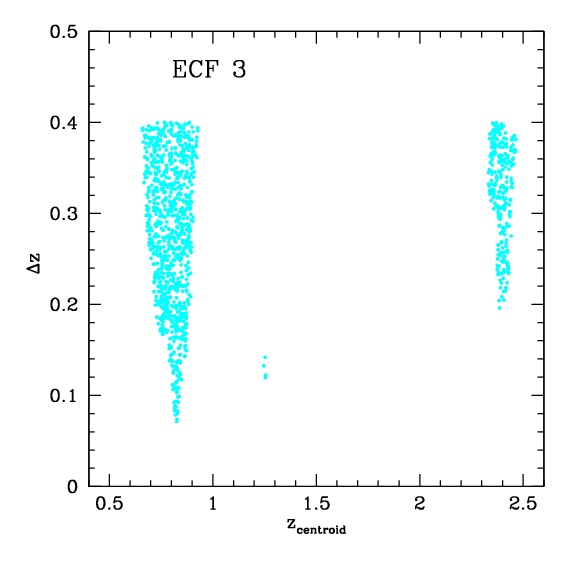}}\qquad
\subfigure{\includegraphics[width=0.45\textwidth,natwidth=610,natheight=642]{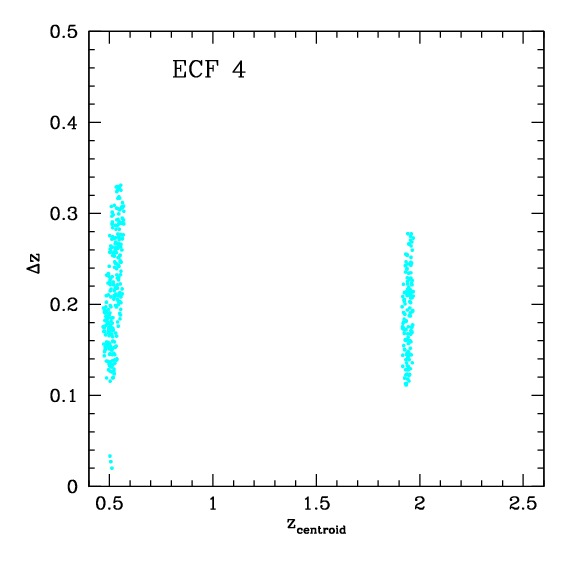}}\qquad
\subfigure{\includegraphics[width=0.45\textwidth,natwidth=610,natheight=642]{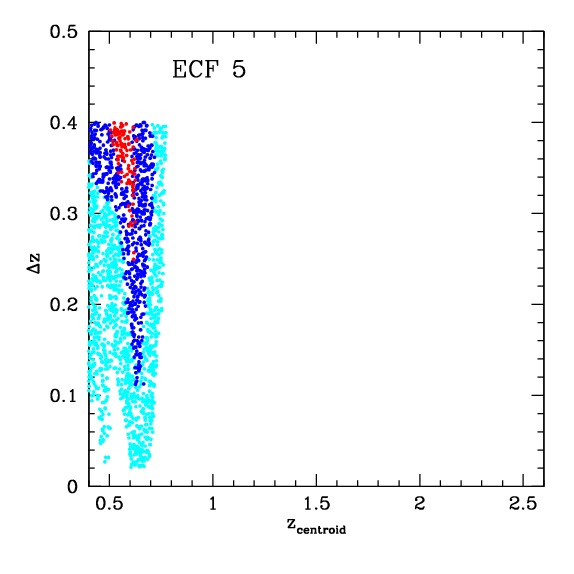}}\qquad
\subfigure{\includegraphics[width=0.45\textwidth,natwidth=610,natheight=642]{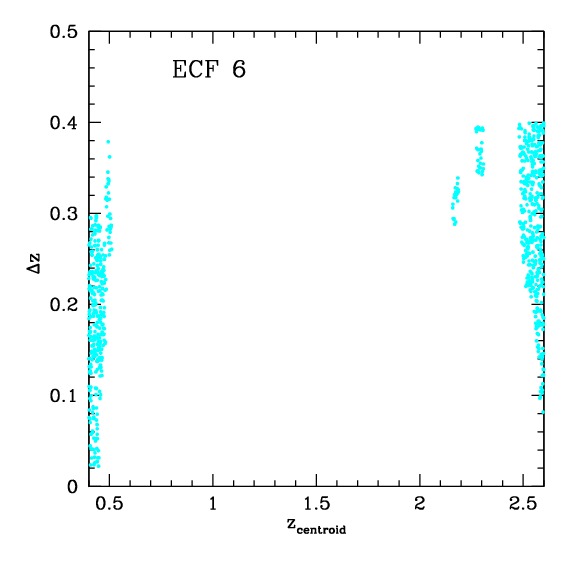}}
\caption{PPM plots for the four additional Empty Control Fields.
Overdensities: $\geq2\sigma$ (cyan points), $\geq3\sigma$ (blue points), 
$\geq4\sigma$ (red points).}
\label{fig:PPMplots_additionalECFs}
\end{figure*}

We consider both the two ECFs used in Sect.~\ref{sec:cluster_shifting_sim}
and four additional ECFs (denoted as ECF 3, 4, 5, and 6) where no overdensity is detected 
by the PPM within the redshift range $z\sim1-2$. We increase the number of ECFs with respect to our previous analysis because the ECFs 
might host some overdensities that are just slightly below the $2\sigma$ PPM detection threshold,
but they might be detected once other galaxies are added to the same field.
The effect of overdensities and underdensities in the location of our simulated clusters should be marginalized with the increased number 
of random fields.
In Figure~\ref{fig:PPMplots_additionalECFs} we report the PPM plots for the four additional ECFs.
Concerning the cluster sizes, in this section we only refer to comoving
sizes, unless otherwise specified.

%------------------------- Figure: PPM plots for the four additional ECFs -----------------------
%---------------------------------------------------------------------------------------------

We choose the following parameters: $N_c$~=~10, 30, 60, 100, 150, and 200; $z_c$~=~1.0, 1.5, and 2.0;
$R_c$~= 1.0, 2.0, and 3.0 Mpc. This results in 54 simulated clusters obtained by considering all the possible 
combinations of the values of $N_c$, $z_c$, and $R_c$.
In particular, the redshifts are chosen in the range of our interest, while
the adopted comoving sizes and the considered values for the richness are typical of clusters and groups we expect to find
in the COSMOS surveys adopting our method.

In fact, in Paper~II we estimate cluster core sizes for the 
$z\sim1-2$ cluster candidates found in the fields of the \citet{chiaberge2009} sample. Average physical and comoving 
core sizes $r_{max} = ( 772 \pm 213 )$~kpc and $r_{max} = ( 1762 \pm 602 )$~kpc are obtained, respectively.    
The average is performed using all the cluster candidates and the reported uncertainty is the 1-$\sigma$ rms dispersion around the average.

As discussed in Paper~II the estimated number of the fiducial cluster members varies with the cluster detection significance
from $\sim$10 for our cluster candidates at the highest redshifts ($z\sim2$) to more than $\sim$200 for our $z\sim1$ clusters
candidates.

Note that clusters of galaxies usually include up to thousands of galaxies. 
Here we adopt smaller values for $N_c$ because the \citet{ilbert2009} catalog lacks of faint 
I$>25$ galaxies that still constitute a significant fraction of the cluster galaxies at redshifts $z\gtrsim1$ \citep{rudnick2012}.

As discussed in Paper~II,
mass estimates are found in the literature for some of our cluster candidates at redshift $z\sim1$. In particular, FGH07 estimated a cluster 
mass $M_{500}= 5.65\times10^{13}~M_\odot$ for the rich group associated with the source 01, for which the PPM selects $\sim100$ 
cluster members within a circle of $r_{max}=70.7$~arcsec radius and a redshift bin $\Delta z =0.28$ centered at the spectroscopic redshift $z = 0.88$
of the cluster. \citet{knobel2009,knobel2012} reported masses within $M\sim 1.4-2.2\times10^{13} M_\odot$  for the cluster candidates 
in the fields of sources 16, 18, and 20, for which $\sim100$, $\sim200$, and $\sim100$  fiducial
cluster members are selected by the PPM, respectively.
Source 16 has a spectroscopic redshift $z = 0.97$, while the photometric redshifts of sources 18 and 20 are $z = 0.92$ and $z = 0.88$, respectively.
As pointed out in Sect.~\ref{sec:shifting_results}, such mass estimates and cluster detections 
further suggest that PPM effectively finds systems whose masses are compatible to 
those of rich groups. Therefore, the PPM is able to detect structures whose mass is 
even below the typical cluster mass cutoff $\sim1\times10^{14}$~M$_\odot$.

For a given simulated cluster we change the exact redshift of
each of the $N_c$ members to account for the observational
uncertainties. Conversely, we do not change their projected coordinates of the cluster members 
because  the angular positional uncertainties are negligible with respect to the
photometric redshift uncertainties \citep{ilbert2009}.
For the same reason we also neglect the galaxy peculiar velocities and, therefore, 
all of the cluster members are assumed to be at the same redshift $z_c$. 
We assign to each of the $N_c$ sources a photometric redshift drawn from a
Gaussian probability distribution centered at the mean $z_{c}$ and whose square standard deviation is $\sigma_c=0.054(1+z_c)$. This is the 1-$\sigma$ statistical photometric redshift uncertainty
of $i^+\sim24$ and $1.5<z<3$ sources \citep[see Table~3 of][]{ilbert2009},
typical of the cluster galaxies we consider, consistently with what done throughout this work 
(see e.g. Sect.~\ref{sec:cluster_shifting_sim}).

We consider the case where 
(i) the equatorial coordinates at which we choose to center the tessellation of the PPM, 
(ii) the equatorial coordinates of the adopted ECF and (iii) the  
equatorial coordinates of the center of the spherically symmetric simulated cluster all coincide. 
In particular, for our simulations we keep (i) the equatorial coordinates at which we choose to center the tessellation of the PPM
and (ii) the equatorial coordinates of the adopted ECF unchanged, i.e. (i) and (ii) will always coincide.
The ECFs are in fact chosen because the PPM does not detect any overdensity in these fields at the redshift range ($z\sim1-2$)
of our interest. Conversely, some overdensities might be present 
at a certain offset from the equatorial coordinates of the adopted ECF.
In this case the PPM might detect these overdensities if the 
the equatorial coordinates at which we choose to center the PPM tessellation do not coincide with those of the center of the adopted ECF.

However, in order to test the efficiency of the PPM
to detect clusters if the cluster center coordinates is not accurately known,
in Sect.~\ref{sec:increasing_offset} we will offset (by an angle $\theta$) the 
input PPM coordinates with respect to the center of the spherically symmetric simulated cluster.

\subsection{General results and trends}
In Table~\ref{table:sim_clusters_null_offset} we summarize the results for all the 54 simulated clusters.
Each entry of the table shows the fraction of ECFs in which the cluster with specific values of 
${\rm N_c}$, ${\rm R_c}$,  and $z_c$ is detected.
For example, the fraction 4/6 means that the cluster is detected in four out of the six ECFs.

Our simulations suggest that the majority, i.e. 47, 44, and 41 out of the 54 simulated clusters
are detected at least in three, four, and five ECFs. For the 41 clusters that are detected in at least five out of the six ECFs
the redshift $z_{\rm PPM}$ estimated 
by the PPM is fully consistent with the input simulated cluster redshift $z_{c}$.
In fact, the average difference for the 41 clusters is 
$\langle z_{\rm PPM}-z_{c}\rangle=0.02\pm0.05$, where all the detections for the 41 clusters are considered
and the reported uncertainty is the rms dispersion around the average.
Therefore, the statistical 1$\sigma$ uncertainty for our redshift estimates is $\sim0.05$. It is estimated 
by Gaussian propagation of the mean offset with the rms dispersion.
Interestingly, this is fully consistent with the independently estimated redshift uncertainties ($\sim0.06-0.09$) described 
throughout the PPM procedure 

The 41 overdensities that are found in at least five ECFs are 
detected with significances spanning from $\gtrsim2.7\sigma$ up to $\sim12\sigma$, depending on 
the adopted parameters, and a median value of $5.2\sigma$. 
At a fixed richness ($N_c$) and size ($R_c$), the clusters are more easily detected for increasing redshifts.
This is because the mean COSMOS number density rapidly drops down for increasing redshifts.
At fixed richness ($N_c$) and redshift ($z_c$), more compact clusters are more easily detected with higher significance
than more extended overdensities. This is because compact clusters have higher number densities than more extended overdensities.

Furthermore, clusters with low values for $N_c$ are
more easily detected at redshifts higher than at $z_c=1$. This is due to
the decreasing mean number density for increasing redshifts.
In fact, at redshifts $z\geq1.5$, only 10 cluster members seem to be sufficient 
(see also the results outlined in Paper~II).
In fact, among the six clusters with $N_c=10$ and  $z\geq1.5$, five are detected in at least three ECFs.
However, only one of them is detected in at least five ECFs.

The reported trends are clearly due the fact that we consider the cluster parameters $N_c$, $R_c$, and $z_c$ as independent.
In fact, we do not change the cluster parameters $N_c$ and $R_c$ when we shift the cluster to higher redshift. 
This is motivated by the fact that the statistics at $z\gtrsim1$ is poor and we prefer to investigate whether the PPM 
is able to detect overdensities over a wide range of adopted parameters.
However, the results of these simulations are clearly dependent on all the simplifications we made
(e.g. spherical symmetry, $N_c$, $R_c$, and $z_c$ are considered independent).
We note that this is a different approach to that adopted for the simulations in Sect.~\ref{sec:cluster_shifting_sim},
 where we simulate how the cluster would be observed if it were located at higher redshift.
Given all the assumptions we make, the accuracy of our simulations is reasonably good 
for our purpose to detect high-redshift overdensities on the basis of number counts and photometric 
redshifts, given the specific properties of both real clusters and the adopted survey. 

The general relationship between the richness parameter $N_c$, the size $R_c$ of the cluster, the cluster mass, and the 
significance of the cluster detection is complex 
(i.e. it depends on the depth of the photometric 
catalog, the redshifts, the evolution of luminosity function), especially at the redshift of our interest ($z\sim1-2$),
where the properties of cluster galaxy population in terms of luminosity and segregation within the cluster
are expected to evolve and are not fully understood.
 
We will address problems 
of completeness and purity of the cluster catalogs 
derived with the PPM in a forthcoming paper (Castignani et al., in prep).

%  --------------------    TABLE 1: simulated clusters results null offset ---------------------

\begin{table*}[hbct]
\caption{Simulated cluster detections (null offset)} 
\label{table:sim_clusters_null_offset} \centering                       

\begin{tabular}{|c|c|c|c|c|c|c|c|c|c|c|c|c|c|c|c|c|c|}
 \hline
  \multicolumn{1}{|c|}{R$_c$} &
  \multicolumn{3}{|c|}{$N_c =10$} & 
  \multicolumn{3}{|c|}{$N_c =30$} &
  \multicolumn{3}{|c|}{$N_c =60$} & 
  \multicolumn{3}{|c|}{$N_c=100$} & 
  \multicolumn{3}{|c|}{$N_c\geq150$} \\

(Mpc)  &$z_c=1$ & 1.5 & 2 & 1 & 1.5 & 2 & 1 & 1.5 & 2 & 1 & 1.5 & 2 & 1 & 1.5 & 2  \\
\hline
 1.0 & $0/6$ & $4/6$   & $5/6$ & $6/6$ & $6/6$ & $6/6$ & $6/6$  & $6/6$ & $6/6$ & $6/6$ & $6/6$ & $6/6$ & $6/6$ & $6/6$ &$6/6$   \\[0.1cm]
 2.0 & $0/6$ & $3/6$   & $3/6$&  $1/6$ & $5/6$ & $6/6$  & $5/6$ & $6/6$ & $6/6$ & $6/6$ & $6/6$ & $6/6$ & $6/6$& $6/6$ & $6/6$   \\[0.1cm]
 3.0 & $0/6$ & $2/6$ & $3/6$ &   $0/6$  & $4/6$ & $6/6$  & $0/6$ & $6/6$ & $6/6$ & $4/6$ & $6/6$ & $6/6$ & $6/6$ & $6/6$ & $6/6$  \\ [0.1cm]
\hline
\end{tabular}

\tablecomments{Detection results for the simulated clusters with different input richness (${\rm N_c}$), redshift ($z_{\rm c}$),
and size (${\rm R_c}$), in the case where 
(i) the equatorial coordinates at which we choose to center the tessellation of the PPM, 
(ii) the equatorial coordinates of the adopted ECF and (iii) the  
equatorial coordinates of the center of the spherically symmetric simulated cluster all coincide.\\
Column description: (1) comoving size (Mpc) of the simulated cluster; (2-13) detection rates for
simulated clusters of different richness ${\rm N_c}$ and redshift $z_c$. 
Each fraction $n/6$ denotes that the cluster is detected in $n$ out of the six adopted ECFs.}
\end{table*}
%-----------------------------------------------------------------------------------------

\subsubsection{Projected cluster sizes}
We find that the comoving sizes, estimated by our method, are 
consistent with the comoving cluster sizes ($R_c$) of our simulations within $\sim30\%$.

\subsubsection{The case of a 2~Mpc comoving size cluster }
As outlined in Table~\ref{table:sim_clusters_null_offset}, at   $z_c = 1$, 11 out of 18 simulated clusters
are detected in at least five ECFs. Among the 18 simulated clusters, 
apart for the clusters with  $R_c\geq2$~Mpc and $N_c = 10$,
the remaining 16 simulated clusters are all detected at 
$z_c \geq 1.5$ in at least four ECFs.

The simulated cluster with $N_c=30$ cluster members and $R_c=2.0$~Mpc is one of the 
four clusters that are detected in at least four ECFs only if located at $z_c \geq 1.5$

\begin{figure*} \centering
\subfigure[]{\includegraphics[width=0.31\textwidth,natwidth=610,natheight=642]{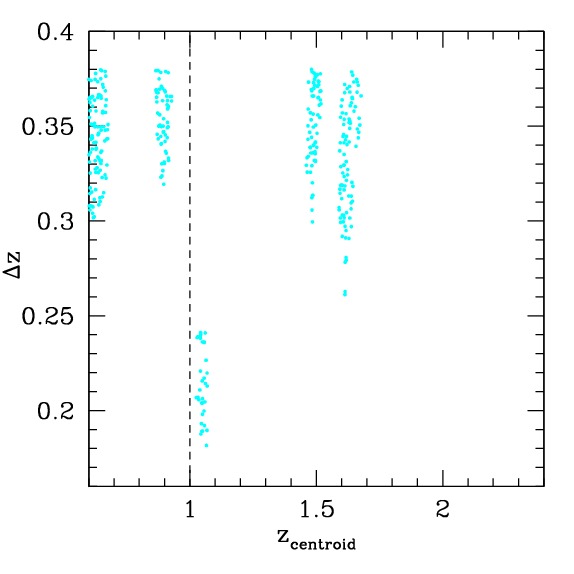}}\qquad
\subfigure[]{\includegraphics[width=0.31\textwidth,natwidth=610,natheight=642]{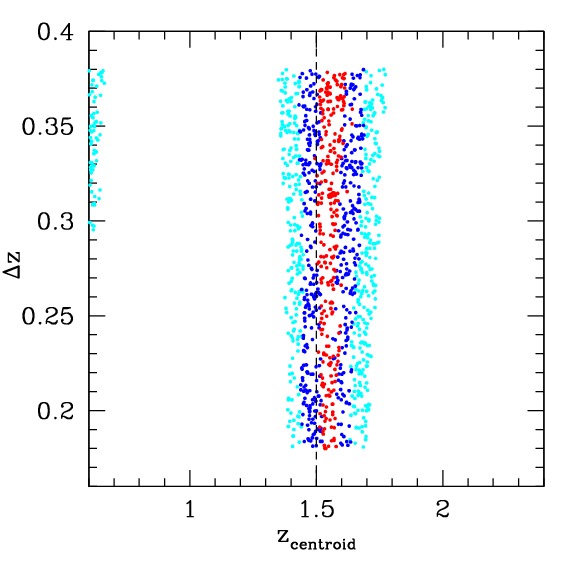}}\qquad
\subfigure[]{\includegraphics[width=0.31\textwidth,natwidth=610,natheight=642]{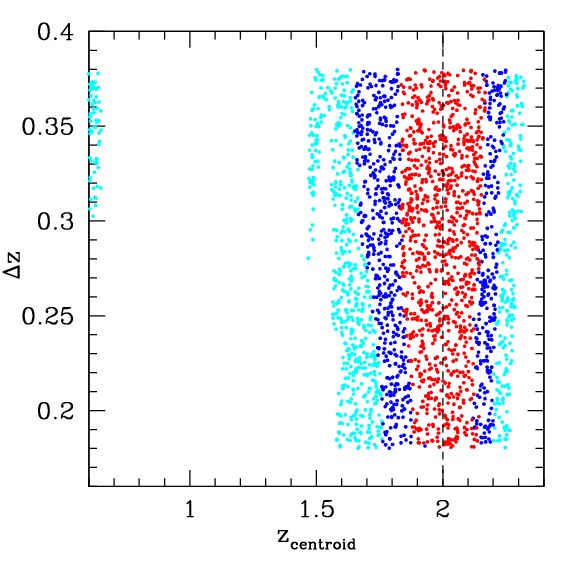}}\qquad
\caption{PPM plots for the simulated cluster located in the field of COSMOS-FR~I~70, with $\theta=0$~arcsec, and at different redshifts: $z=1.0$ (left), $z=1.5$ (middle), and $z=2$ (right).
Dashed vertical line corresponds to the input redshift of the simulated cluster. 
Overdensities: $\geq2\sigma$ (cyan points), $\geq3\sigma$ (blue points),
$\geq4\sigma$ (red points). A Gaussian filter to eliminate high frequency noisy patterns is applied.}
\label{fig:ppm_plots_simulated_cluster}
\end{figure*}

In Figure~\ref{fig:ppm_plots_simulated_cluster} we report the PPM plots for this specific cluster in the case where 
it is located in the field of COSMOS-FR~I~70 and where the offset $\theta=0$~arcsec.
The abscissa of the vertical dashed line is equal to the input redshift of the simulated cluster: 
$z_c=1.0$ (left panel), $z_c=1.5$ (middle panel), and $z_c=2.0$ (right panel).
We adopt the same color code as in Figure~\ref{fig:ppm_plots}.

In particular, the cluster at $z_c=1.5$ and $z_c=2.0$ is detected with significances of $3.7\sigma$ and  $4.7\sigma$, respectively.
The estimated redshifts are $z= 1.56$ and $z= 2.00$, respectively.
The estimated sizes are $r_{max}= 86.6$~arcsec and $r_{max}= 70.7$~arcsec, that correspond to 1.9~Mpc and 1.8~Mpc (comoving), 
at the estimated cluster redshifts, respectively.

Interestingly, both the redshift and size of these overdensities fully agree with 
the input parameters of our simulations.
All these results further confirm that the PPM is very effective in finding clusters and rich groups and 
in estimating their properties such as redshift and size, 
if the projected cluster coordinates are known. 
In the following we will test the PPM against simulations in the case where the projected cluster coordinates are known with an
accuracy of 100 and 200~arcsec.

%----------------------- Figure: PPM plot simulated clusters -------------------------------
%--------------------------------------------------------------------------------

\subsection{Increasing the offset $\theta$}\label{sec:increasing_offset}
We repeat all the 54 simulations for increasing offsets $\theta$ between the 
input coordinates of the PPM
and the center of the simulated cluster. 
In order to do so, we shift the coordinates of cluster members by $\theta= 100$~arcsec. 
We keep unchanged both the PPM input cluster
equatorial coordinates and the equatorial coordinates of
the adopted ECF.
As explained above, this is because the surroundings of the ECFs might host dense regions in the redshift range of our interest.
In Table~\ref{table:sim_clusters_100arcsec_offset} we summarize our results, where $\theta= 100$~arcsec,
analogously to what is  reported in Table~\ref{table:sim_clusters_null_offset} for the case of null offset $\theta$.

For $\theta>100$~arcsec the PPM becomes highly inefficient mainly because of the constraint applied to 
an angular separation of $\sim$2~arcmin from the coordinates at which the PPM tessellation is centered.
In fact, for $\theta = 200$~arcsec, the simulated clusters are all detected in less than five ECFs.
Only four high redshift ($z_c = 1.5$), rich ($N_c \geq 60$) and extended ($R_c \geq 2$~Mpc)  clusters are detected in at least three ECFs.
Among the four, only the cluster with  with $z_c = 1.5$, $N_c = 200$, and $R_c = 3$~Mpc is detected in four ECFs.
This is not surprinsing. In fact, for such a high value of $\theta$ very extended and rich structures have more chances to be detected. 

Note that by considering all the possible combinations of the parameters (i.e. the different empty control fields, 
the offsets, and the different values for $N_c$, $R_c$, and $z_c$) 972 clusters are simulated as part of this work.

%  -----    TABLE 2: simulated clusters results offset = 100 arcsec -----------------------------
\begin{table*}[hbct]
\caption{Simulated cluster detections ($\theta =100''$ offset)} 
\label{table:sim_clusters_100arcsec_offset} \centering                       

\begin{tabular}{|c|c|c|c|c|c|c|c|c|c|c|c|c|c|c|c|c|c|c|c|c|}
 \hline
  \multicolumn{1}{|c|}{R$_c$} &
  \multicolumn{3}{|c|}{$N_c =10$} & 
  \multicolumn{3}{|c|}{$N_c =30$} &
  \multicolumn{3}{|c|}{$N_c =60$} & 
  \multicolumn{3}{|c|}{$N_c=100$} & 
  \multicolumn{3}{|c|}{$N_c=150$} &
  \multicolumn{3}{|c|}{$N_c=200$}  \\

(Mpc)  &$z_c=1$ & 1.5 & 2 & 1 & 1.5 & 2 & 1 & 1.5 & 2 & 1 & 1.5 & 2 & 1 & 1.5 & 2 &  1 & 1.5 & 2 \\
\hline
 1.0 & $0/6$ & $1/6$   & $1/6$ & $0/6$ & $3/6$ & $6/6$ & $2/6$  & $5/6$ & $5/6$ & $4/6$ & $5/6$ & $6/6$ & $6/6$ & $6/6$ &$5/6$  & $6/6$ & $6/6$ & $6/6$  \\[0.1cm]
 2.0 & $1/6$ & $1/6$   & $0/6$&  $0/6$ & $3/6$ & $5/6$  & $2/6$ & $5/6$ & $6/6$ & $2/6$ & $5/6$ & $5/6$ & $6/6$& $6/6$ & $6/6$ &$6/6$  & $6/6$ & $6/6$   \\[0.1cm]
 3.0 & $0/6$ & $0/6$ & $0/6$ &   $0/6$  & $2/6$ & $4/6$  & $1/6$ & $3/6$ & $5/6$ & $3/6$ & $6/6$ & $6/6$ & $4/6$ & $6/6$ & $6/6$ & $6/6$ & $6/6$ &  $6/6$ \\ [0.1cm]
\hline
\end{tabular}

\tablecomments{Detection results for the simulated clusters with different input richness (${\rm N_c}$), redshift ($z_{\rm c}$),
and size (${\rm R_c}$), in the case where 
(i) the equatorial coordinates at which we choose to center the tessellation of the PPM, 
(ii) the equatorial coordinates of the adopted ECF and (iii) the  
equatorial coordinates of the center of the spherically symmetric simulated cluster do not coincide.
We fix the offset $\theta=100$~arcsec between (i) and (iii), changing 
the equatorial coordinates of the center of the spherically symmetric simulated cluster,
while (i) and (ii) still coincide. \\
Column description: (1) comoving size (Mpc) of the simulated cluster; (2-13) detection rates for
simulated clusters of different richness ${\rm N_c}$ and redshift $z_c$. 
Each fraction $n/6$ denotes that the cluster is detected in $n$ out of the six adopted ECFs.}
\end{table*}

%  -------------------------------------------------------------------------------------------------

\subsubsection{General results and trends}
Our simulations suggest that the great majority, i.e. 
30 out of the 41 clusters that are detected 
in at least five ECFs in the case where $\theta=0$~arcsec are 
also found in at least five ECFs if $\theta=100$~arcsec.
Note that 23 out of the 30 clusters have $N_c \geq60$ and $z_c>1$ (see Table~\ref{table:sim_clusters_100arcsec_offset}).

For the 30 simulated clusters the redshift $z_{\rm PPM}$ estimated 
by the PPM is fully consistent with the input simulated cluster redshift $z_{c}$.
In fact, the average mismatch for the 30 clusters is 
$\langle z_{\rm PPM}-z_{c}\rangle=0.02\pm0.07$, where all the detections for the 30 clusters are considered 
and the reported uncertainty is the rms dispersion around the average.
Therefore, the statistical 1$\sigma$ uncertainty for our redshift estimates is $\sim0.07$,
estimated with Gaussian propagation of the mean offset with the rms dispersion.
Interestingly, this is again fully consistent with the independent redshift estimate uncertainties ($\sim0.06-0.09$) described 
throughout the PPM procedure. 

Furthermore, note that poorer overdensities or clusters of intermediate richness (i.e. $N_c\leq60$) 
are more difficult to detect than richer clusters. This is especially true at redshift $z_c=1.0$ and in the case of 
$\theta=100$~arcsec, where $N_c = 100$  cluster members or more
are required, unsurprisingly.
%None of the simulated cluster candidates with $N_c=10$ are detected in the case where $\theta=100$~arcsec, independently of the ECF adopted.
%This also holds for all the $z_c=1.0$ clusters with $N_c\leq60$.
At variance with the case of null offset (i.e. $\theta=0$~arcsec),
poor and intermediate richness clusters are more difficult to detect  by the PPM at such a high offset 
$\theta=100$~arcsec. This in fact 
corresponds to a comoving distance of 1.6~Mpc, at redshift $z_c=1.0$, that is comparable to the input 
$\sim$Mpc size of the simulated clusters. 

Among the 18 clusters with intermediate richness, i.e. $N_c=30$ and 60, seven are detected in at least five ECFs.
Six among the seven have $z_c\geq1.5$ and $R_c\leq2$~Mpc.

Rich simulated clusters with $N_c\geq100$  are always detected at $z_c\geq1.5$ 
in at least five ECFs.
At $z_c=1.0$, these rich clusters are detected with more difficulty, especially in the case $N_c=100$, while 
they are detected in at least four ECFs if they have $N_c\geq150$.

\subsubsection{Detection significances}
The overdensities found in the case of $\theta=100$~arcsec are detected, by construction, with significances 
$\geq2\sigma$, and a median value
of $3.9\sigma$.
Therefore, the overdensities tend to be detected with lower significances than in the case of $\theta=0$~arcsec.

In particular, for $\theta=100$~arcsec and similarly to what is found in the case of null offset,
at fixed size and richness, the clusters are detected with
increasing significances for increasing redshifts.
However, at variance with the case $\theta=0$~arcsec, at each fixed
richness and redshift, no specific trend is observed for 
the detection significances, for increasing sizes $R_c$.
This is because of two competing effects:
a larger sized cluster is more easily detected than more compact overdensities, at a
larger offset $\theta$.
On the contrary, similarly to what discusses in the case of null offset $\theta=0$~arcsec,
a larger size implies a lower projected number density that makes
the cluster detection more difficult.

%These two competing effects might explain the fact that, as outlined in the previous section, 
%the rich cluster with $N_c=100$ cluster members, at $z_c=1.0$, is detected in at least one ECF if its comoving size 
%is $R_c=1$~Mpc or $R_c=3$~Mpc, while is detected in neither of the ECF if it has $R_c=2$~Mpc.

\subsubsection{Projected cluster sizes}
The comoving cluster size estimated by the PPM as $r_{\rm max}$ in the case where $\theta=100$~arcsec is $\sim$1.6 times the input comoving cluster size $R_c$
(and up to a factor of $\sim$2.6 at 1-$\sigma$). 
The reason of this mismatch is due to the fact that, by construction,
the simulated cluster formally ends at a larger distance from the input PPM coordinates
than in the case of $\theta=0$~arcsec.

These aspects suggest that, if the coordinates of the cluster center are not accurately known,
the cluster sizes might be overestimated up to a factor of $\sim2.6$, that corresponds to the extremal case where we are looking for a cluster environment around a radio source that resides in the outskirts of the cluster core.

\section{Summary and conclusions}\label{sec:conclusions}
The goal of this project is to search for high redshift $z\gtrsim1$ clusters or groups using FR~I radio galaxies as beacons.
In this paper we have introduced a new method we developed to achieve such a goal.
The method is tailored to the specific properties of the $z\sim1-2$ FR~I radio galaxy sample we consider \citep{chiaberge2009},
selected within the COSMOS survey \citep{scoville07}, and to the specific dataset used. 

The Poisson Probability Method (PPM), is adapted from the method proposed by
\citet{gomez1997} to search for X-ray emitting substructures within clusters in the low number count regime. 
Here we are similarly dealing with the problem of small number densities. 

We test the efficiency of the PPM in searching for cluster candidates against simulations. Two different approaches are adopted.
i) We use two $z\sim1$ X-ray detected clusters found in the COSMOS survey within
the \citet{finoguenov2007} catalog. We shift them to higher redshift up to $z=2$.
We find that the PPM detects both clusters up to $z=1.5$ and it correctly estimates both 
the redshift and the size of the two clusters. 
ii) We simulate spherically symmetric clusters of
different size and richness, and we locate them at different redshifts (i.e. $z = 1.0$, 1.5, and 2.0) in the COSMOS
field. We find that the PPM detects the simulated clusters within the entire redshift range considered 
with a statistical 1$\sigma$ redshift accuracy of $\sim0.05$. 
This is remarkably comparable to the statistical photometric redshift uncertainty of photometric redshift catalogs over the same redshift 
range \citep{mobasher2007,ilbert2009}.

Our results suggest 
that almost all of our  simulated clusters are detected.
Compact clusters (i.e. 1~Mpc comoving size) and rich clusters are more easily detected than lower richness clusters, 
when the cluster center coordinates are accurately known.
The majority of these clusters are also detected even if the coordinates of the cluster center are known with poor accuracy of $\sim100$~arcsec. 
Furthermore, poor overdensities and clusters of intermediate richness are more difficult to detect
in the case where the cluster coordinates are known with an accuracy of $\sim$100~arcsec.
Concerning cluster sizes, we found that the PPM provides estimates with a 33$\%$ rms fractional accuracy, if the cluster center coordinates are known.

We applied the Kolmogorov-Smirnov (KS) test to the cumulative number distributions of galaxies in the fields of the \citet{chiaberge2009}
sample similarly to what done by previous work on COSMOS \citep[e.g.][]{harris2012}.
We checked that the KS test is ineffective in dealing with these types of cluster searches and
the results are not conclusive. This is because  
shot noise fluctuations affect the results of the KS test.

We found that our method is effective in finding clusters up to high redshift.
We believe that the PPM is a valuable alternative to previously considered methods to search for high-redshift clusters
based on photometric redshifts. In fact, with the inclusion of a solid positional prior and an accurate
redshift sampling we overcome, at least in part, the problem of
establishing whether multiple overdensity peaks in the 2-d projected 
density field are part of a single larger structure \citep{scoville2013} and, thus, 
identifying different structures at different redshifts.

Although the PPM is primarily introduced for the COSMOS survey
(see Paper~II), it may be applied to
wide field surveys to blindly search for cluster candidates.
Accurate photometric redshifts and a survey depth similar or better than that of COSMOS (e.g. I$<25$) are required. 

However, our method is less effective for those surveys that will provide sufficient spectroscopic redshift information, where standard 3-d methods such as correlation functions might be more successfully applied.
Conversely, the PPM might be also applied to SDSS Stripe 82 and
future wide field surveys such as LSST and Euclid that will provide accurate photometric redshift information.
Another possible use of the PPM is a search for (proto-)clusters at $z\gtrsim 2$, by adopting radio galaxies or other sources such as Lyman break galaxies as beacons.

\appendix

\section{The Poisson Probability Method (PPM)}
\label{app:PPM}

As mentioned in Sect.~\ref{sec:method_PPM}, the PPM is adapted from the method proposed by \citet[][see their Appendix~A]{gomez1997} 
to search for X-ray emitting substructures within clusters.
We introduce such a method to search
for the presence of clusters in a given field, around specific projected coordinates, by using photometric redshift information.
Therefore, our method is not properly a method to search for cluster candidates, but to verify their presence around a given beacon
\citep[this is an approach similar to that adopted by previous work, e.g.][]{george2011}.

The method we describe in the following is tailored to the specific properties of the sample and dataset used, to which 
we refer throughout the method description.
More specifically, the PPM has been introduced to search for cluster candidates around $z\sim1-2$ FR~Is  
selected in the COSMOS field \citep{scoville07} by \citet{chiaberge2009}.
We also use photometric redshifts from the \citet{ilbert2009} catalog.

The PPM is based on photometric redshifts and galaxy number counts.
Similarly to other methods that use photometric redshift information \citep[e.g.][]{eisenhardt2008},
we consider the redshift information and the coordinates in the projected space separately.
This is because the photometric redshift uncertainties are much larger 
than the typical scale of clusters. Therefore, such uncertainties are  
significantly dominant with respect to any other observable uncertainty (e.g., flux uncertainties, 
projected space coordinate uncertainties). 
In the following we will focus first on the
projected space and then on the redshifts.

%Approaches similar to the PPM were adopted by e.g. \citet{steidel1998} used the Erlang distribution \citep{eadie1971} to establish the significance of a concentration of galaxies at $z\sim3$.

\subsection{The projected space}\label{sec:projected_space} 

We tessellate the projected space with a circle centered at the coordinates of the beacon
(in our specific case this is the location of the FR~I radio  galaxy)
 and 49 consecutive adjacent
annuli. These regions are concentric and have the same area, i.e. 2.18~arcmin$^{2}$. This is done in order to have the 
same average field density for each of the regions.
The inner radius of the $i$-th annulus is equal to $\arccos[1+i(-1+\cos 50~ {\rm arcsec})] \simeq 50\times\sqrt{i}$~arcsec. 
This means that the radius of the circle centered at the coordinates of the beacon
is equal to 50~arcsec. Such an angular separation 
is consistent with that adopted by previous work 
focused on high-$z$ clusters \citep[e.g.][]{santos2009,adami2010,adami2011, durret2011, galametz2012, spitler2012}.
In fact, it corresponds to 427~kpc at redshift $z=1.5$, that is typical of the cluster core size at redshift $z\sim1$. 
If we chose a smaller scale we would be highly affected by shot noise. 
In fact, on average, for a fixed area of 2.18~arcmin$^{2}$,  the differential number counts 
(${\rm d}N/{\rm d}z$) per unit redshift in the COSMOS field are quite small and equal to $\sim55$, 22, and 7,
at redshifts $z\simeq1$, 1.5, and 2.0, respectively \citep{ilbert2009}.
Conversely, if we chose a greater scale we would characterize the cluster environment of the FR~Is in our sample
with a rough (Mpc-scale) accuracy only.

In general, due the specific tessellation of the PPM, the method
is effective to detect Mpc-scale rich groups and clusters (as also discussed in Paper~II).
Conversely, it might be less efficient in 
finding poor groups and larger, i.e. a few Mpc-scale, diffuse structures.
Therefore, a more detailed treatment of the projected space typical of 
sophisticated tessellations such as Voronoi, Delaunay  \citep[see e.g.][]{ebeling1993} 
correlation estimators \citep[e.g.][]{adami2011},
wavelet analysis \citep[e.g.][]{eisenhardt2008}, filter techniques, and adaptive kernels \citep[e.g.][]{scoville07b} 
might be ineffective and difficult to
apply for cluster searches with the use of photometric redshifts only \citep[see also][for further discussion]{scoville2013}.
On the other hand, such methods might be more useful to study
spectroscopically confirmed clusters or groups \citep[e.g.][]{jelic2012}.

The typical size inspected by
our tessellation changes at most $\sim6\%$ within the entire redshift range of our interest.
In fact, 1~arcmin corresponds to a physical size of 482, 512, and 509~kpc, at redshift $z=1.0$, 1.5, and 2.0, respectively. 
Depending on the adopted cosmology, the 
angular distance assumes a maximum between $z\sim1-2$. 
%Therefore, we keep this projected space tessellation fixed, even if we are interested
%in clusters or groups within the entire the redshift range $z\sim1-2$.
These considerations imply that our tessellation is effective to characterize
Mpc-scale overdensities with the required accuracy, independently of redshift, under the assumption that cluster core size does not dramatically change for increasing redshifts.

Furthermore, our approach implicitly assumes azimuthal symmetry around the axis oriented at the coordinates of the beacon.
We do not exclude the possibility  
to detect non circularly symmetric systems,
since we extend the tessellation up to $\sim$6~arcmin 
(i.e. $\sim$3~Mpc at $z=1.5$) from the coordinates of the beacon \citep{postman1996}. 
Moreover, our method is also flexible enough to find clusters even if the coordinates of the 
cluster center are known within $\sim100$~arcsec only (as tested with simulations in Sect.~\ref{sec:increasing_offset}). 

However, we note that the great majority of of low-power radio sources in clusters or groups
are found within $\sim200$~kpc from the core center up to $z\simeq1.3$ \citep{ledlow_owen1995,smolcic2011}.
Therefore, this suggests that low-power radio galaxies in cluster environments are preferentially hosted 
within the central regions of the core, at least at low or intermediate redshifts.
The results presented in the companion paper (see discussion in Sect.~7.10 of Paper~II)
for the $z\sim1-2$ cluster candidates within the \citet{chiaberge2009} sample suggest that this is generally true also at higher redshifts .
Therefore, all these results support 
the specific projected space tessellation method 
described in this section and adopted for our cluster search.

\subsection{Redshift information}\label{sec:distance_discrimination} 
As discussed in \citet{scoville07b}, identifying large scale structures on the basis of 2-d number densities requires 
a careful selection of those galaxies that are at the redshift of the structure. This is because, especially in the 
case of high-$z$ clusters, foreground galaxies contaminate the field. 
Despite this, the contamination from foreground sources is limited by the smaller angular size of high-$z$ clusters with respect to that
of those at lower redshifts.

As pointed out in \citet{scoville07b}, three different criteria are commonly adopted to discriminate among the galaxies at different
distances by using (i) color selections \citep[e.g.][]{papovich2008,gladders2005}, (ii) spectroscopic redshifts
\citep[e.g.][]{knobel2009,knobel2012}, or (iii) photometric redshifts \citep[e.g.][]{adami2010,durret2011}.

(i) Color selection is not used here, since it might be biased towards large scale structures with specific properties in terms
of galaxy colors. 
This is particularly important especially at redshift $z\gtrsim1.5$, where 
  the properties of the cluster galaxy population and their changes with redshift 
in terms of galaxy morphologies, types, masses, colors \citep[e.g.][]{bassett2013,mcintosh2013},
and star formation content \citep[e.g.][]{zeimann2012,santos2013,strazzullo2013,gobat2013} are still debated;
(ii) spectroscopic redshifts are preferred to the photometric redshifts. However, spectroscopic redshift 
catalogs \citep[e.g.][]{lilly2007} are limited to a small fraction of the galaxies in the COSMOS field;
(iii) we adopt the photometric redshift catalog of \citet{ilbert2009},
that was obtained considering sources with AB magnitude I$<25$.

We consider $M\gg1$ redshift bins, i.e. the closed intervals $[z_l^i, z_r^i]$, with $z_r^i>z_l^i$, $i = 1$, 2, ..., $M$, and $M = 22,500$.
We define the length and the centroid of each interval as $\Delta z^i = z_r^i - z_l^i$ and $z_{\rm centroid}^i = (z_r^i+z_l^i)/2$, 
respectively. The subscripts $l$ and $r$ stand for left and right, respectively.
Both the redshift lengths $\Delta z^i\in[0.02;0.4]$ and the redshift centroids $z_{\rm centroid}^i\in[0.4;4.0]$
 are randomly and independently chosen assuming a uniform distribution.
Since $M\gg1$, the considered ranges are densely spanned concerning both the redshift bin and the redshift centroid.
The redshift range of our interest is $z\sim1-2$, while the typical statistical photometric redshift uncertainties at those redshifts are
$\sigma_z\sim0.1-0.2$ \citep{ilbert2009}. 
Therefore, both $\Delta z^i$ and $z_{\rm centroid}^i$ are conservatively selected over wider intervals than 
those of our interest. This is done in order to avoid spurious boundary effects that might derive from our selection. 
Before describing in detail our method in the following section we will discuss its theoretical framework.

\subsection{Theoretical framework}\label{sec:theoretical_framework}
We denote as $n$ the galaxy number density within a given projected area of the sky subtended by a solid angle $\Omega$.
The total variance in the number counts $n$ is given by \citet{peebles1980}

\begin{equation} 
\label{eq:clustering_variance}
{\Biggl\langle}\left(\frac{n-\langle n \rangle}{\langle n \rangle}\right)^2{\Biggr\rangle} = 
\frac{1}{\langle n \rangle}+\sigma_v^2~,
\end{equation} 

where 

\begin{equation} 
\label{eq:clustering_term}
\sigma_v^2 = \frac{1}{\Omega^2}\int\int \omega(\theta)\hspace{0.1cm}d\Omega_1d\Omega_2~,
\end{equation} 

is the sampling variance due to source clustering. As pointed out e.g. in \citet{massardi2010}, 
$\sigma_v^2$ adds a significant contribution to the uncertainties in the case of small-area fields.
As pointed out in \citet{peebles1980}, assuming ergodicity, the average denoted by the brackets $\langle\hspace{0.2cm}\rangle$ is
either the ensemble average (i.e. the average among all 
the field realizations)  or the volume average (i.e. the average among different areas in the survey, each of them
is subtended by a solid angle $\Omega$).
The clustering term in Eq.~\ref{eq:clustering_term} is expressed as the integral over the field of the 
projected two-point correlation function $\omega(\theta)$,
where $\theta$ is the angular separation between the solid angle elements  d$\Omega_1$, d$\Omega_2$.

%It is not straightforward to derive an explicit form of the distribution for the variable $n$, whose
%second-order moment is expressed in Equation~\ref{eq:clustering_variance}.
%This is ultimately due to the presence of the clustering term, $\sigma_v^2$.
We note that $n$ would be Poisson distributed if the clustering term were not present.  
Then, we ask what is the probability that the null hypothesis (i.e. no clustering)
occurs for the given field. This is equivalent to set $\sigma_v^2=0$ and to assume that  $n$ is Poisson distributed.
According to our formalism, we estimate the probability of not clustering as the 
probability to have a number density $n'$ equal or higher than the observed value $n$:

\begin{equation}
 \label{eq:prob_poisson}
 \mathcal{P}_{Poisson}(n' \ge n ) = \sum_{k= n\times\Omega}^{\infty}\frac{(\langle n\rangle\times\Omega)^k}{k!}e^{-\langle n\rangle\times\Omega}~.
\end{equation}

As for Eq.~\ref{eq:clustering_variance}, $\langle n\rangle$ and $\Omega$ are the average number density for the survey considered
and the solid angle subtended by the selected field for which we estimate the null hypothesis probability, respectively.
Therefore, the null hypothesis is rejected with a probability

\begin{equation}
 \label{eq:prob_poisson2}
 \mathcal{P} = 1 -\mathcal{P}_{Poisson}(n' \ge n ) = \sum_{k= 0}^{(n-1)\times\Omega}\frac{(\langle n\rangle\times\Omega)^k}{k!}e^{-\langle n\rangle\times\Omega}~.
\end{equation}

 The probability $\mathcal{P}$ is higher in those fields where the sources are more clustered, i.e. where
$\sigma_v^2$ is non negligible 
with respect to the shot noise term $1/\langle n \rangle$.
Therefore, $\mathcal{P}$  can be independently considered in this paper as
the probability that  an overdensity is present in the field.
% or, alternatively, (ii) 
%of the significance level of the number count excess with respect to the mean field density.}

So far, our formalism implicitly assumes that the sample selection is a negligible source of uncertainty 
(i.e. it does not contribute to the total variance of Eq.~\ref{eq:clustering_variance}). 
However, the cluster membership selection (based on e.g. fluxes, colors, or redshifts)
always contributes to the total number count variance because of the observable uncertainties.
Equivalently, observable uncertainties imply that the total 
 variance in Eq.~\ref{eq:clustering_variance} is underestimated.
Consequently, Eq.~\ref{eq:prob_poisson2} overestimates the probability $\mathcal{P}$ that null hypothesis (i.e. no overdensity) 
is rejected.

Limiting the analysis to the PPM, our method is based on number counts and photometric redshifts.
According to the PPM procedure, for each redshift bin $\Delta z^i$, only those sources 
within the redshift interval $[z_l^i, z_r^i]$ are considered (see below, Sect.~\ref{app:the_method}).
Photometric redshift uncertainties are significantly higher than any other observable (e.g. flux, 
projected coordinates) associated with our sample.

This implies that we can estimate the additional term to the number count variance in Eq.~\ref{eq:clustering_variance} due to 
observable uncertainties by considering photometric redshift uncertainties only.
We denote such a term as $\sigma^2_{ph}$, where the notation $ph$ stands for photometric redshifts.
This term can be  independently estimated as the number count variance obtained by averaging over
 all the possible realizations  
%$\{z_1, z_2, z_3,..., z_k,...\}_i$ 
of the photometric redshifts 
of the galaxies in the selected field. These realizations are ideally drawn
from the redshift probability distributions of the galaxies in the field.
Hence, we have:

\begin{equation} 
\label{eq:sigma_ph}
\sigma_{ph}^2 = {\Biggl\langle}\left(\frac{n-\langle n \rangle_{ph}}{\langle n \rangle_{ph}}\right)^2{\Biggr\rangle}_{ph}~,
\end{equation}

where $\langle\hspace{0.2cm}\rangle_{ph}$ denotes the average over the ensemble constituted by 
all the possible redshift realizations.
The net effect  to increase the number count variance in Eq.~\ref{eq:clustering_variance}
by the amount $\sigma_{ph}^2$ is similar to that described in e.g. \citet{sheth2007} in the context of
luminosity functions estimated by adopting photometric redshifts.
\citet{sheth2007} showed that photometric redshift uncertainties (i.e. distance errors) have the effect of scattering objects to the 
low-luminosity and high-luminosity ends of the luminosity function (i.e. towards higher and lower luminosities).
Similarly, in our case photometric redshift uncertainties have the effect of scattering the number counts
over a wider range.

Given the additional term $\sigma_{ph}^2$ in the right hand side of Eq.~\ref{eq:clustering_variance},
a rigorous calculation of the probability of the null hypothesis (i.e. no clustering, $\sigma_{v}^2=0$)
would require to consider the  redshift probability distribution associated with each galaxy in the field.
Estimating the correct expression of the null hypothesis probability
might be done with simulations, i.e. adopting different realizations of the redshifts, 
that are drawn from the redshift probability distribution associated with each galaxy in the field. 
However, this procedure would be enormously demanding in terms of time and would not add a significant contribution to the PPM. 
We prefer to adopt a different approach. We neglect any additional terms in the number count variance due to observational uncertainties.
We prefer to take account for
such an approximation, by correcting our final estimate (see Sect.~\ref{sec:noise_subtraction}). 
Our correction will decrease the probability $\mathcal{P}$.
In the next sections we will describe how our method works.

\subsection{The method}\label{app:the_method}

Firstly, we consider each pair defined by the redshifts $z_l^i$ and $z_r^i$,
as in Sect.~\ref{sec:distance_discrimination}, and the sources 
 in the \citet{ilbert2009} catalog that have redshifts 
within the interval $z_l^i\leq z < z_r^i$ and that fall within the largest circle that can be inscribed in the COSMOS survey.
Such a circle subtends a $\sim1.25$~deg$^2$ solid angle.
Then, we estimate the average mean density
$\langle n \rangle_{i}$ as the ratio of the number of these sources to the solid angle subtended by the circle.
We test if cosmic variance affects our analysis by adopting different choices in estimating such an average number density (i.e.
we also selected 
four disjoint quadrants in the COSMOS survey and we estimated the average number density for each quadrant, separately).
We verified that the results are independent of the choice adopted. 
%Concerning this average number count density estimate, for sake of semplicity we do not consider all the COSMOS survey because
%the number counts are sufficient to avoid cosmic variance.

For a given beacon, i.e. in our case the projected coordinates of the radio galaxy,
 we consider each of the 50 projected regions defined in Sect.~\ref{sec:projected_space}.
We denote them with the index $t = 1, ..., 50$, where $t=1$ corresponds to the central 
circle, while the adjacent annuli are denoted with progressively higher indexes.
We denote as $N_t^i$ the observed number of  galaxies within the chosen redshift
interval $z_{l}^i\leq z< z_{r}^i$ that fall within the $t$-th area.
By construction, each region subtends a solid angle $\Delta\Omega_t = 2.18$~arcmin$^2$.
We also define the observed number density for the selected region and redshift bin as $n_t^i = N_t^i/\Delta\Omega_t$.
The probability of the null hypothesis (i.e. no clustering) for the $t$-th region, consistently 
with Eq.~\ref{eq:prob_poisson}, is
\begin{equation}
 \label{eq:prob_poisson2bis}
  \mathcal{P}_{Poisson}^{i,t}({n'}_t^i\geq n_t^i) = \sum_{k= N_t^i}^{\infty}\frac{(\langle n\rangle_i\times\Delta\Omega_t)^k}{k!}e^{-\langle n\rangle_i\times\Delta\Omega_t}~,
\end{equation}
that corresponds to the probability of having a number density ${n'}_t^i$ higher or equal than the observed one,
 according to the Poisson statistics.
Note that, because of the low number counts, the Gaussian statistics is not a good approximation of the
Poisson statistics, therefore the latter is required.

Starting from the central region corresponding to $t=1$ we select the first adjacent regions, denoted with
the indexes  $\{\overline{t}_i,\overline{t}_i+1, ..., \overline{t}_i + h_i\}$,
 for which that the probability of the null hypothesis in 
Eq.~\ref{eq:prob_poisson2bis} is $\leq30\%$. Here $\overline{t}_i\geq1$ and $h_i\geq0$ are both integer numbers.
Note that we do not exclude the possibility to select one single region, i.e. $h_i=0$.
The central circle may be selected or not, depending on whether the threshold criterion is satisfied or not 
by that region. 

According to our prescription, we have selected only those regions for which 
the null hypothesis is rejected at a level of $\geq70\%$, i.e. down to about 1$\sigma$.
Note that the adopted $\sim1\sigma$ threshold this is not a tight constraint. 
A $70\%$ threshold is also adopted by \citet{gomez1997} and is similar to the values 
adopted for other selection criteria applied by previous papers 
that were focused on high-$z$ clusters and that used photometric redshift information
 \citep[e.g.][]{papovich2010,finkelstein2010}.
%(GUARDA NOTE 24 FEBBRAIO 2012).

%However, as pointed out in Sect.~\ref{sec:projected_space}, small number counts,
%i.e. between $\sim$2-10 within a redshift bin $\Delta z=0.2$,
%are expected, on average, at redshifts $z\sim1-2$, in each of the 50 areas.
%Therefore, therefore a weak constraint is required.

According to the procedure, the probabilities are always estimated according to Poisson statistics. 
Even if the Gaussian statistics is not adopted we often refer to the probability in terms of $\sigma$ 
(e.g. we refer to 68.27\%, 95.45\%, and 99.73\%
probabilities as 1-$\sigma$, 2-$\sigma$, and 3-$\sigma$ significances, respectively). 
We adopt this notation for practical reasons, for sake of convenience.

Similarly to what done in \citet{gomez1997}, in order to determine the true significance of the number count excess in the field,
we merge together all the $h_i+1$ adjacent regions to form a (larger) circle  or a annulus, depending whether the central circle is included or 
not, respectively.
Then, we define the total observed number count $N_i$ for the new region as
\begin{equation}
 N_i = \sum_{t=\overline{t}_i}^{\overline{t}_i+h_i}N^i_t\hspace{0.1cm}\rm{,}
\end{equation}
and the corresponding number density $n_i$ as:
\begin{equation}
 n_i=\frac{N_i}{(h_i+1)\Delta\Omega_t}\hspace{0.1cm}\rm{.}
\end{equation}

We stress that the goal of the procedure described so far is not to quantify the number 
count excess associated with the field.
Merging the $h_i+1$ regions aims at selecting 
an area in the projected space for which there is indication  
of a number count excess and that is most likely associated with the projected coordinates of the beacon.

Note that if we chose a more constraining (i.e. $>1\sigma$) threshold criterion we would select only those regions that 
show a significant number count excess. Therefore, we might be biased towards selecting those regions (i) that are 
associated with highly overdense substructures in 
the cluster or (ii) show high shot noise fluctuations.
These scenarios might occur because of 
the small area subtended by each region 
(i.e. $\sim2~$arcmin$^2$) and the small number densities that occur at the redshifts of our interest.
This discussion suggests that a more constraining (i.e. $>1\sigma$) criterion
might not be effective in selecting properly the cluster field in  the projected space.

Then, Analogously to what done for each of the 50 regions (see Eq.~\ref{eq:prob_poisson2bis}) we estimate 
the probability of the null hypothesis (i.e. no
clustering) for the new area as
\begin{equation}
 \label{eq:prob_poisson3}
  \mathcal{P}_{Poisson}^{i}({n'}_i\geq n_i) = \sum_{k= N_i}^{\infty}\frac{(\langle n\rangle_i\times(1+h_i)\Delta\Omega_t)^k}{k!}
e^{-\langle n\rangle_i\times(1+h_i)\Delta\Omega_t}\hspace{0.1cm}\rm{,}
\end{equation}
that is the probability to have a number density ${n'}_i$ higher or equal than that observed, $n_i$, according to the Poisson statistics.
For the sake of clarity, hereafter we omit the argument of $\mathcal{P}_{Poisson}^{i}$.
We also define as $r_{\rm min}^i$ and $r_{\rm max}^i$ the minimum and maximum projected distances from the coordinates of the beacon, 
(in our case it is the radio galaxy)
within which the number count excess is detected, respectively.
These radii are equal to $r_{min}^i=\arccos[1+(\overline{t}_i-1)\times(-1+\cos50~{\rm arcsec})]\simeq\sqrt{\overline{t}_i-1}\times50$~arcsec and
$r_{max}^i=\arccos[1+(\overline{t}_i+h_i)\times(-1+\cos50~{\rm arcsec})]
\simeq\sqrt{\overline{t}_i+h_i}\times50$~arcsec,  because of the specific tessellation,
consistently with the adopted procedure.

Accordingly to  what discussed in Sect.~\ref{sec:theoretical_framework}, the null hypothesis is rejected with a 
probability $\mathcal{P}^i = 1-  \mathcal{P}_{Poisson}^{i}$.  
We set $\mathcal{P}^i\equiv0$ if the overdensity starts to be detected from $r_{\rm min}\gtrsim132$~arcsec, 
i.e. if  $\overline{t}_i\geq8$. This is done to reject those 
overdensities that are detected only at large angular separation from the location of the source. 
In fact, since 132~arcsec correspond to $\sim$1.1~Mpc at $z=1.5$, these overdensities might not be associated with our beacon.
However, note that such a constraint does not exclude the possibility to detect structures 
that are extended up 132~arcsec or even higher. These extended structures are detected
if they start at a distance lower than 132~arcsec.

The specific projected distance of 132~arcsec corresponds to 0.8~$h^{-1}$~Mpc ($h=0.71$), that is the scale where
the amplitude of the correlation function between Radio Loud AGN (RLAGN) and Luminous Red Galaxies (LRGs)
is reduced to a few percent ($\sim4\%$) of the value at its maximum, up to ${\rm z}\simeq0.8$
 \citep[e.g.,][]{donoso2010,worpel2013}.

Limiting the number counts to the $h_i+1$ overdense regions between the radii 
$r_{\rm min}^i$ and $r_{\rm max}^i$ does not bias our results
towards overestimating the number count excess because we are allowed to select slightly overdense regions, down to 
$\sim1\sigma$ number count excess. 

Similarly, the probability to detect those fields that show low number count excess or shot-noise fluctuations is negligible.
This is because the number of selected galaxies does not decrease when the $h_i+1$ regions are merged
and then the number count excess probability 
is re-estimated for the new (larger) region delimited by the radii $r_{\rm min}^i$ and $r_{\rm max}^i$.
We will describe in the following sections the noise mitigation procedure and the peak finding algorithm 
adopted to detect overdensities.
As will be clarified below, both of these procedures further suppress 
the probability to detect as overdensity any number count excess that is simply due to shot noise fluctuations.

\subsection{Noise mitigation}\label{sec:noise_subtraction}
%We translate  $\mathcal{P}^i$ in terms of a significance level $\mathcal{S}^i$ that is defined through 
%$\mathcal{P}^i = \textsf{erf}(\mathcal{S}^i/\sqrt{2})$,
%where $\textsf{erf}(x)= \frac{2}{\sqrt{\pi}}\int_0^x e^{-t^2}{\rm dt}$ is the error function.

\begin{figure*} \centering
\subfigure[]{\includegraphics[width=0.45\textwidth,natwidth=610,natheight=642]{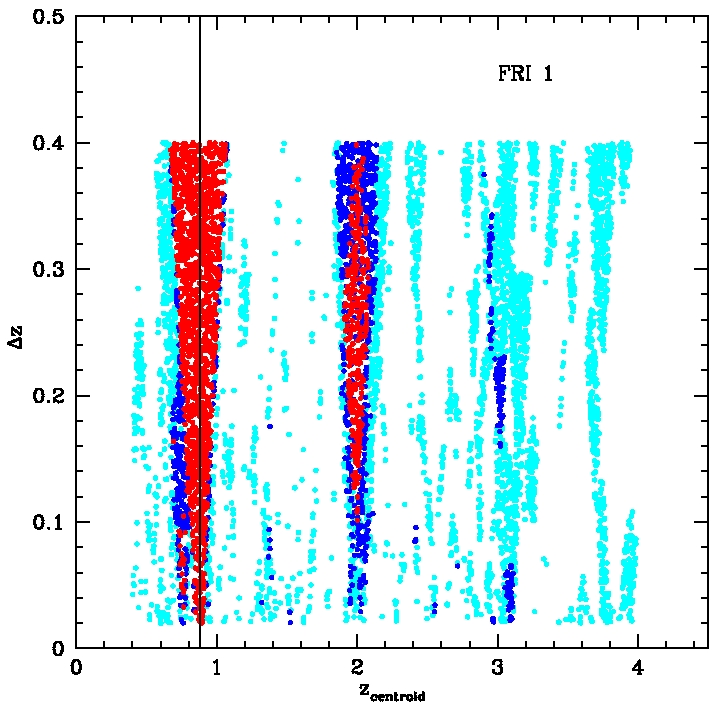}}\qquad
\subfigure[]{\includegraphics[width=0.45\textwidth,natwidth=610,natheight=642]{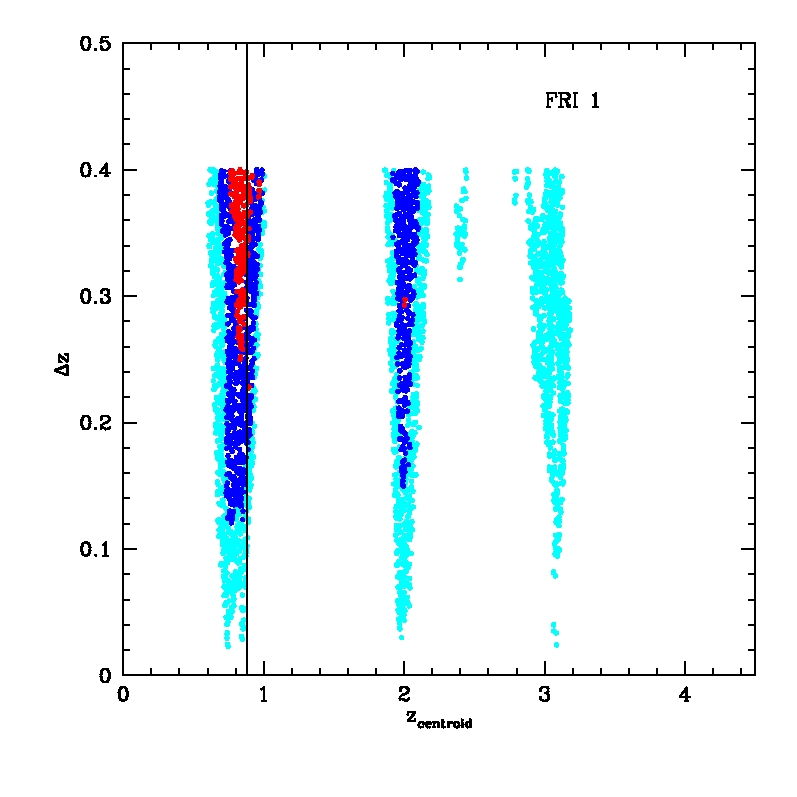}}\qquad
\caption{PPM plots for source 01. Left: no filter is applied. 
Right: the Gaussian filter which eliminates high frequency noisy patterns is applied.
The abscissa of the vertical solid line is at the spectroscopic redshift of the source. 
We plot only the points corresponding to detected overdensities for 
different values of $\Delta z$ and $z_{\rm centroid}$. Color code: $\geq2\sigma$ (cyan points), $\geq3\sigma$ (blue points),
$\geq4\sigma$ (red points). }
\label{fig:PPMplots_filter_before_and_after}
\end{figure*}

In Figure~\ref{fig:PPMplots_filter_before_and_after}, left panel,
we plot $\mathcal{P}$ as a function of 
the redshift bin  $\Delta z$ and its centroid $z_{\rm centroid}$.
We omit the index $i$ corresponding to the specific redshift interval.  We will introduce such an index only where necessary.
%, through the ''erf'' error function.
Red, blue, and cyan colors refer to points with significances $\geq4\sigma$, $\geq3\sigma$,  and $\geq2\sigma$, respectively.
We plot as a vertical solid line the spectroscopic redshift of the source.

Isolated high significance spiky patterns are clearly visible in the plot.
They occur because of 
the presence of a significant source number excess at specific redshifts 
and redshift bins.
However, the fact that such patterns are spiky and extended over scales that are smaller than the typical 
statistical photometric redshift uncertainties ($\sigma_z\sim0.1-0.2$)
suggests that they are not physical and are ultimately due to noise fluctuations.

In order to eliminate such high frequency noisy patterns we apply a Gaussian filter to the function $\mathcal{P}^i$ as follows:
\begin{equation}
  \label{eq:filtering}
  \overline{\mathcal{P}}^i = \frac{\displaystyle\sum_{j}\mathcal{W}^{ij}\mathcal{P}^i}{\displaystyle\sum_j \mathcal{W}^{ij}}\hspace{0.1cm}\rm{,}
\end{equation}
where the kernel
\begin{equation}
 \label{eq:fit}
 \mathcal{W}^{ij}= \begin{cases}
                e^{-\frac{{\zeta_{ij}}^2}{2\sigma_w^2}}         &{\rm if\hspace{0.2cm}\zeta_{ij}}\leq 7.5~\sigma_w~\\
               & \hspace{2.5cm},\\
               0             & {\rm otherwise}
           \end{cases}
\end{equation}
and
\begin{equation}
{\zeta_{ij}}^2 =(z_{\rm centroid}^i-z_{\rm centroid}^j)^2+ ({\Delta z}^i - {\Delta z}^j)^2 \hspace{0.1cm}.
\end{equation}

For practical reasons,
the sum over $j$ is extended only to those points that are at most 7.5-$\sigma_w$
from ($z_{\rm centroid}^i$; ${\Delta z}^i$), so that the kernel has a compact domain and acts as a weighted
local average.
The newly defined function, $\overline{\mathcal{P}}$, is simply the convolution between the Gaussian filter and the 
previously defined probability $\mathcal{P}$.

In practice, patterns that are extended over scales of the order of $\lesssim\sigma_w$ 
with respect to $z_{\rm centroid}$ or $\Delta z$ are removed from the plots.
We choose $\sigma_w=0.02$, that is much lower than the typical statistical photometric redshift uncertainty $\sigma_z\sim0.1-0.2$
of the \citet{ilbert2009} catalog. Therefore, our choice  conservatively removes those patterns that are clearly due to noise.

Isolated local maxima are removed from the plots 
since they are suppressed by the surrounding low 
significance points. Conversely, local maxima that belong to extended high significance patterns still remain associated with
high significance patterns, even if their significance is decreased because of the average procedure. 

A heuristic physical interpretation may be provided. 
Averaging $\mathcal{P}$ among those points that belong to a neighborhood 
of $z_{\rm centroid}$ and $\Delta z$ mimics the presence of the redshift uncertainties at fixed $z_{\rm centroid}$ and $\Delta z$.
This is because changing $z_{\rm centroid}$ and $\Delta z$
by small amounts  has the net effect of including some sources 
and excluding others when the redshift interval is changed.
To understand better such a heuristic  equivalence, in the following we will adopt a variational approach.

We note that $\overline{\mathcal{P}}$ can be naturally interpreted as an effective mean field defined on the space
 of ($z_{\rm centroid}^i$; $\Delta z^i$).
This is because the incoherent fluctuations of $\mathcal{P}$ due to noise and small scales are locally suppressed
by the filter $\mathcal{W}$. 

 There is no theoretical and observational reason to prefer one specific form
for the kernel $\mathcal{W}$. 
In the effective field theory context, since the width of the filter is related to the statistical photometric redshift uncertainties,
it is more relevant than the specific shape of the filter $\mathcal{W}$.
In fact, the Gaussian filter is chosen because its exponential declining 
assures that noisy features associated to scales (that are smaller than
the typical statistical photometric redshift uncertainty) are conservatively removed.

A more physical interpretation of the filtering procedure described in this section may be provided by using a variational approach.
Several methods of analysis of functionals of Poisson processes, including 
variational calculus, 
have been developed in the context of stochastic analysis \citep[e.g.,][]{privault1994,albeverio1996,molchanov_zuyev2000}.
We stress that the following discussion does not intend to be a formal proof, but simply an argument that shows 
how we can assign a physical meaning to the $\overline{\mathcal{P}}$ values and, thus, better explaining the above mentioned 
heuristic equivalence that arises when the filtering procedure is reconsidered. 
A more detailed analysis will be performed in a forthcoming paper.

We adopt a compact notation and we 
define the vector $\overrightarrow{x}=(z_{\rm centroid}$; $\Delta z)$ whose $i$-th component is   
given by ($z_{\rm centroid}^i$; $\Delta z^i$). We consider the galaxies that are in the field of our beacon 
(i.e. the radio galaxy in our case) and whose redshifts belong to a 
neighborhood of the redshift $z_{\rm centroid}$ corresponding to the specific $\overrightarrow{x}$.
The values of $\Delta z$ and $z_{\rm centroid}$ can be expressed as the 
redshift range spanned by the selected galaxies 
and the redshift centroid of that redshift range, respectively.
This implies that the value of $\mathcal{P}$ at $\overrightarrow{x}$ ultimately depends on the set $\{z_j\}$ of the redshifts 
of the galaxies in the field of the beacon, where each set  $\{z_j\}$ is selected in such a way that the corresponding
($z_{\rm centroid};\Delta z$) belongs to a neighborhood of $\overrightarrow{x}$.
This argument shows that $\overrightarrow{x}$ is a function of $\{z_j\}$ and, therefore, $\mathcal{P}$ is 
a function of $\{z_j\}$, i.e. $\mathcal{P}(\overrightarrow{x})=\mathcal{P}(\overrightarrow{x}(\{z_j\}))=\mathcal{P}(\{z_j\})$.

Each set $\{z_j\}$ is a specific realization of the photometric redshifts. Each set belongs to the 
ensemble constituted by all the possible redshift realizations.
These realizations are ideally drawn from the redshift probability distributions of
the galaxies in the field of the beacon.

Since $\mathcal{P}$ is not an analytic function, first we assume that $\mathcal{P}$ is defined on a discrete domain, given by the points 
in Figure~\ref{fig:PPMplots_filter_before_and_after} (left panel), then we consider a local
analytic first-order approximation of $\mathcal{P}$ at $\overrightarrow{x}$.
By construction, the analytic approximation has a local continuous domain. 
For the sake of simplicity, in the following we do not distinguish $\mathcal{P}$ from its analytic approximation.
Adopting the analytic approximation allows us to use a variational approach and expand $\mathcal{P}$ in Taylor series as follows:
\begin{equation}
 \label{eq:var_P_x}
 \delta\mathcal{P} = \overrightarrow{\nabla} \mathcal{P} \cdotp \delta\overrightarrow{x} + 
o\left(\frac{|\delta \overrightarrow{x}|}{\Delta z} \right)\; ,
\end{equation}
where $\delta\mathcal{P}$ is the variation of $\mathcal{P}$ induced by
the variation $\delta\overrightarrow{x}$.

The first term in the right hand side (rhs) of the equation is the scalar product of the variation $\delta\overrightarrow{x}$ and 
the gradient $\overrightarrow{\nabla}$ of $\mathcal{P}$ with 
respect to $\overrightarrow{x}$.

We stress that it is not straightforward to 
provide an explicit expression for all the terms in the Taylor series.\footnote{Let $f$ and $g$ be two 
functions defined on some subset of the real numbers. 
$f(x) = o(g(x))$ as $x\rightarrow x_0$ if and only if for all $K>0$ there exists $\delta>0$ such that $|f(x)|\leq K |g(x)|$, 
for each $x$: $|x-x_0|<\delta$.}
In fact, the explicit expression  depends on the specific number counts in the field, the specific redshift $z_{\rm centroid}$
and redshift bin $\Delta z$. We checked that finite differences $\Delta\mathcal{P}$ of $\mathcal{P}$ corresponding to
small displacements $|\Delta \overrightarrow{x}|$ in the PPM plots (e.g. in Figure~\ref{fig:PPMplots_filter_before_and_after}, left panel)  
satisfy the relation $|\Delta\mathcal{P}|/|\Delta \overrightarrow{x}|\lesssim1 /\Delta z$ for almost all of the $\geq2\sigma$ points 
in the plot (except for a negligible set).
This argument leads to the term
$o\left(|\delta \overrightarrow{x}|/\Delta z \right)$ in Eq.~\ref{eq:var_P_x} and implicitly implies that 
only those points associated with high significance (i.e.  $\gtrsim2\sigma$) patterns are considered.
This constraint does not affect our analysis. In fact, as outlined below, only 
these patterns may be ultimately associated with cluster detections.

The variation $\delta\mathcal{P}$ at $\overrightarrow{x}$ can be alternatively estimated as follows. 
Limiting our analysis to a neighborhood of $\overrightarrow{x}$ and to the field of the beacon, analogously to what done in
Eq.~\ref{eq:var_P_x}, we estimate the variation $\delta_{ph}\mathcal{P}$ of $\mathcal{P}$ induced by the variations 
of $\{z_j\}$ within the ensemble as follows:
%--------------------------------------------------------------------------------------
\begin{equation}
 \label{eq:var_P_ph}
 \delta_{ph}\mathcal{P} = {\sum_j} \frac{\partial\mathcal{P}}{\partial z_j}\delta z_j + 
o\left(\frac{\sqrt{\sum_j \delta z_j^2 / \sum_j 1}}{\Delta z}\right)\; ,
\end{equation}
%-----------------------------------------------------------------------------------------
where the sum over $j$ is restricted to a neighborhood of $\overrightarrow{x}$ as specified above
and to the galaxies in the field of the beacon.
The subscript $ph$ stands for photometric redshifts and it is introduced to distinguish the variation 
$\delta_{ph}\mathcal{P}$ from $\delta\mathcal{P}$.
The first and second terms in the rhs of the equation are the first term and the higher order terms of the Taylor expansion, respectively. 
The second term reported in Eq.~\ref{eq:var_P_ph} is estimated similarly to what done in Eq.~\ref{eq:var_P_x}. 
The chain rule is also applied to express the derivatives with respect to $z_j$ into
 derivatives with respect to $\Delta z$ and $z_{\rm centroid}$.

We stress that the variations $\delta\mathcal{P}$ and $\delta_{ph}\mathcal{P}$ are different;  $\delta\mathcal{P}$ is the variation
of $\mathcal{P}$ due small variations $\delta(\Delta z)$ and $\delta z_{\rm centroid}$ of the redshift bin and the redshift
centroid, respectively. Conversely $\delta_{ph}\mathcal{P}$ is the variation of $\mathcal{P}$ due to the photometric redshift
uncertainties of the redshifts 
of the galaxies in the field of the beacon. 
Equivalently, $\delta_{ph}\mathcal{P}$ is the variation of $\mathcal{P}$
in the ensemble of all the possible redshift realizations of the galaxies in the field
of the beacon.
Combining Eq.~\ref{eq:var_P_x} and Eq.~\ref{eq:var_P_ph} we estimate the difference between the two variations of $\mathcal{P}$ as 
follows:

%--------------------------------------------------------------------------------------
\begin{equation}
 \label{eq:var_P}
 \delta\mathcal{P} - \delta_{ph}\mathcal{P} = o\left(\frac{|\delta \overrightarrow{x}| +
\sqrt{\sum_j \delta z_j^2 / \sum_j 1}}{\Delta z }\right)\; ,
\end{equation}
%-----------------------------------------------------------------------------------------
where the same notation and the chain rule adopted above is used.
The equation, combined with the Eq.~\ref{eq:var_P_x} and Eq.~\ref{eq:var_P_ph}, shows that the two variations
$\delta\mathcal{P}$ and $\delta_{ph}\mathcal{P}$ are equal up to the first order in the perturbations.

By construction, the filter $\mathcal{W}$ removes the fluctuations of $\mathcal{P}$ on small scales.
Because of the exponential declining of the Gaussian filter,  it is effective on scales 
$|\delta \overrightarrow{x}|\lesssim 3\sigma_w=0.06$.
As will be explained below, cluster candidates are selected at the fixed
redshift bin $\Delta z = 0.28$.
This redshift bin corresponds to the estimated statistical 2-$\sigma$ photometric redshift uncertainty 
at $z\sim1.5$ for dim galaxies \citep[i.e. with AB magnitude ${\rm i}^{+}\sim24$,][]{ilbert2009}.

Therefore, $|\delta \overrightarrow{x}| /\Delta z\simeq2\%\ll1$. This inequality implies that the scales are sufficiently 
small to assure that the filter $\mathcal{W}$ suppresses the variation  $\delta\mathcal{P}$ up the first order, as in Eq.~\ref{eq:var_P_x},
i.e.  $\delta\mathcal{P}=0$.

Similarly, the condition $\sqrt{\sum_j \delta z_j^2/ \sum_j 1}\lesssim\Delta z$ is reasonably satisfied, because 
$\sqrt{\sum_j \delta z_j^2/ \sum_j 1}$  approximately corresponds to the quadratic average of the photometric redshift uncertainties
of the galaxies in the field. Such an average is reasonably smaller than the  selected redshift bin $\Delta z = 0.28$.
Therefore, the variation $\delta_{ph}\mathcal{P}$ in Eq.~\ref{eq:var_P_ph} is well approximated, for our purposes, by the linear expansion.

Resuming, Eq.~\ref{eq:var_P_x} and Eq.~\ref{eq:var_P_ph}, combined with 
the two reported inequalities, suggest that in our case both $\delta\mathcal{P}$ and $\delta_{ph}\mathcal{P}$ can be 
expressed in linear theory.
Similarly, Eq.~\ref{eq:var_P}, combined with the two inequalities, tells that  the two variations are equal in first-order approximation.
Therefore this argument suggests that the filter $\mathcal{W}$ simultaneously suppresses (in linear approximation) 
both the variation  $\delta\mathcal{P}$ 
and variation $\delta_{ph}\mathcal{P}$ due to the photometric redshift uncertainties.

This discussion suggests that $\overline{\mathcal{P}}$ is a good estimate of the number count excess probability.
In fact,  as discussed in Sect.~\ref{sec:theoretical_framework}, since the photometric redshift uncertainties add a significant contribution
to the total number count variance, $\mathcal{P}$ represents an overestimate of the true detection significance.
Our procedure takes into account {\it - a posteriori -} the initial overestimation: the significance of the local maxima is 
decreased when the filter is applied. Equivalently, the procedure reasonably
removes, as required, the excess of $\mathcal{P}$ that is ultimately due to the photometric redshift uncertainties and
the corresponding number count variance expressed in Eq.~\ref{eq:clustering_variance}. 

 Note that the parameters adopted here for the Gaussian kernel and those used in the following for the cluster 
detection procedure are chosen because of the properties of the photometric redshifts of our sample and of the \citet{ilbert2009} catalog.
In particular, the parameters $\Delta z$ (= 0.28) and $\sigma_w$ (=0.02) are fine tuned in such a way that the 
linear perturbation theory (see Eq.~\ref{eq:var_P_x} and Eq.~\ref{eq:var_P_ph}) is reasonably correct in both 
the two spaces (that of $\overrightarrow{x}$ and the ensemble of all the redshift realizations).  
 In general, all the parameters are adapted to the specific dataset used.
We verified that our results are independent of a slightly different choice of all these parameters.
This is ultimately due to the fact that the procedure is not performed on physical observables (e.g. on the density field), 
but acts directly on the PPM plots  as those reported in Figure~\ref{fig:PPMplots_filter_before_and_after}.

%
%-------------------- Figure: PPM plots for COSMOS-FRI 01 (Chiaberge et al. 2009)----------
%-------------------------------------------------------------------------------------------

%-------------------- Figure: radial PPM plots for COSMOS-FRI 01 (Chiaberge et al. 2009)----------
\begin{figure*} \centering
\subfigure[]{\includegraphics[width=0.45\textwidth,natwidth=610,natheight=642]{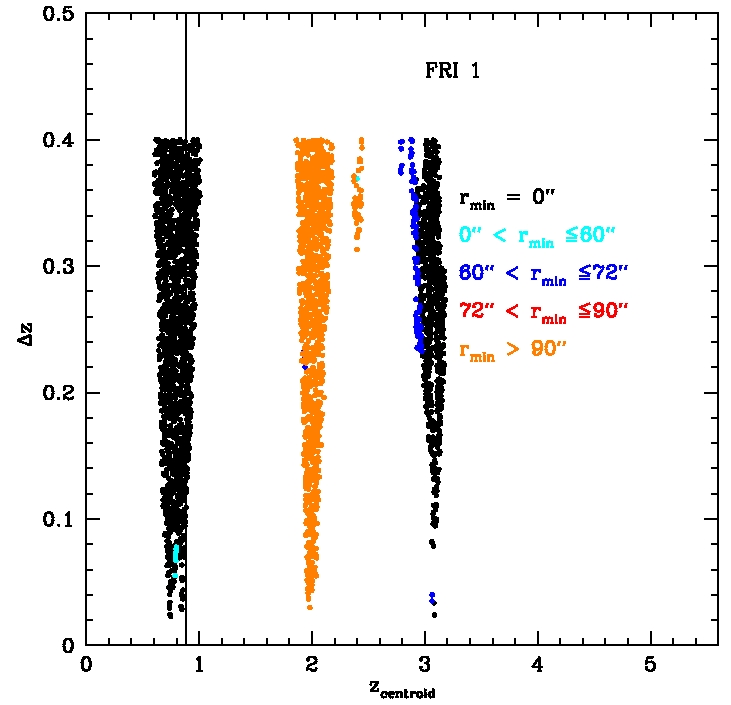}}\qquad
\subfigure[]{\includegraphics[width=0.45\textwidth,natwidth=610,natheight=642]{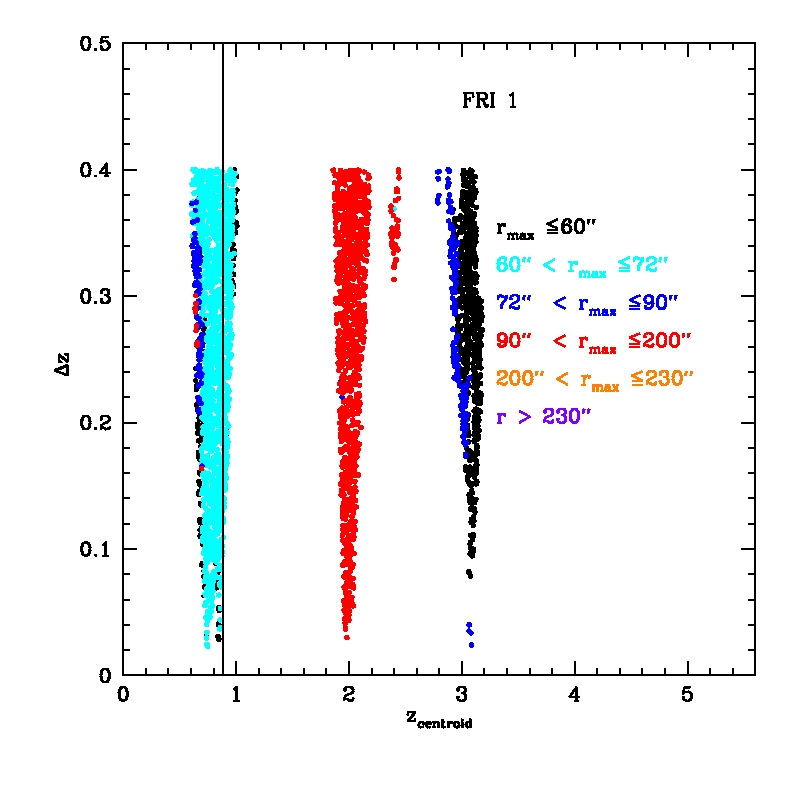}}\qquad
\caption{PPM plots for source 01, $\geq2$-$\sigma$ points are plotted. Radial cluster information concerning
$r_{min}$ (left panel) and $r_{max}$ (right panel).
See legend in the panels for information about the color code adopted.
In each panel the abscissa of the vertical solid line is at the spectroscopic redshift of the source.}
\label{fig:PPMplots_01_rmin_rmax}
\end{figure*}
%-------------------------------------------------------------------------------------------

In Figure~\ref{fig:PPMplots_filter_before_and_after}, right panel,
we plot the function values of  $\overline{\mathcal{P}}$ as a function of $z_{\rm centroid}$ and $\Delta z$.
The same color code for the left panel is adopted.
When the Gaussian filter is applied, isolated noisy patterns are substantially removed from the plot, as clear
from visual comparison of the left panel with the right panel in the figure.
Furthermore, we observe that triangular shape high significance (i.e. $\gtrsim2$-$\sigma$) patterns are still clearly present in the plot.
They are stable with respect to different values of $\Delta z$, i.e. the patterns are extended along  the $\Delta z$ axis.
In particular, they tend to increase their width in the $z_{\rm centroid}$ direction for increasing $\Delta z$. This is due to the fact
that for increasing $\Delta z$ we are including more and more objects that are far from $z_{\rm centroid}$. Therefore,  
we still detect
a number count excess at values of $z_{\rm centroid}$ that are increasingly far from the true redshift of the overdensity,
even if with lower significance.
In fact, lower significance is associated with 
the boundaries of the triangular shape patterns than that related to the central regions of these patterns.

\subsection{Peak finding algorithm}\label{sec:peak_finding_alg}
%Analogously to what done for $\mathcal{P}$, we translate $\overline{\mathcal{P}}$ into the significance $\overline{\mathcal{S}}$, through
%the $\textsf{erf}$ function. 

In this section we will describe our procedure to detect and characterize the overdensities we find in the considered field
by using the PPM plots.
These goals will be achieved by finding the  local maxima of the function $\overline{\mathcal{P}}$.
%Since the pairs ($z_{\rm centroid}^i$;$\Delta z^i$) constitute a discrete set of values, $\overline{\mathcal{S}}$
%is a discrete function.
The peak finding algorithm we will describe is a specific procedure we developed for our discrete case and it belongs to 
a more general context known as Morse theory. Such a theory can be used to find and characterize the
critical points of differentiable functions defined on a manifold \citep[see e.g.][for a review]{guest2001}.
Notably, there are many applications of Morse theory in the context of differential topology \citep{bott1960,milnor1963},
and in quantum field theory \citep[][and following work]{witten1982}.

Firstly, since the high significance patterns in the PPM plots 
are stable with respect to the $\Delta z$ axis we simplify the problem to a 1-d case as follows.
We consider only those points $p_k = (z_{\rm centroid}^k;\Delta z^k)$
such that the redshift bins $\Delta z^k$
fall within a tiny $\delta(\Delta z) = 0.01$ wide interval 
centered at $\Delta  z=0.28$. 
This redshift bin corresponds to the estimated statistical 2-$\sigma$ photometric redshift uncertainty 
at $z\sim1.5$ for dim galaxies \citep[i.e. with AB magnitude ${\rm i}^{+}\sim24$,][]{ilbert2009}.
These magnitudes are typical of the galaxies we expect to find in clusters in the redshift range of our interest.
We verified that our results are stable with respect of a sightly different choice of the redshift bin $\Delta z$.

Then, we sort the points $\{p_k\}$ in increasing order of $z^k_{\rm centroid}$, and we redefine the ordering of the 
points in such a way that
$z_{\rm centroid}^{k+1}\geq z_{\rm centroid}^{k}$.
Hence, our problem is reduced into finding the maxima of $\overline{\mathcal{P}}$ defined on the 1-d discrete
domain $\{z_{centroid}^k\}$. 
Having reduced the dimensionality of the domain is a great simplification, since saddle points are not present
in the 1-d case, where critical points are of only two types: maxima and minima.
%Having reduced by one unit the dimensionaly of the domain is a great semplification. This is because nondegenerate
%critical points of a smooth (i.e. infinite differentiable) real function defined on a 1-D manifold can be of only two types:
%maxima and minima.\footnote{A critical point is a point of the manifold at which the derivative of that function is null. The 
%point is nondegenerate iff the dimension of the 0-eigenspace of the Hessian of the function is zero, at that point.}
%This means that we do not have saddle points, which are instead present in the case of 2-D domains.  

Starting from the significance $s = 2\sigma$, we select those intervals of consecutive points that have significances $\geq s$.
We merge consecutive intervals that are separated by $\delta z_{\rm centroid} = 0.02$ or less 
along the $z_{\rm centroid}$ axis.
We also reject those intervals that are shorter than $\delta z_{\rm centroid} = 0.1$ along the $z_{\rm centroid}$ axis.
%Given these considerations, we fix a significance $s$ and we select the points $p_{k'}\in\{p_k\}$
%such that $\overline{\mathcal{S}}(p_{k'})\geq s$
%We say that $p_{k'}$ is a lower bound if $\overline{\mathcal{S}}(p_{k'+1})\geq s$ and $\overline{\mathcal{S}}(p_{k'-1})< s$.
%The latter condition is not required if $p_{k'}$ is the first element of  $\{p_k\}$.
%Conversely, we say that $p_{k'}$ is an upper bound if 
%$\overline{\mathcal{S}}(p_{k'-1})\geq s$ and $\overline{\mathcal{S}}(p_{k'+1})< s$
%The latter condition is not required if $p_{k'}$ is the last element of  $\{p_k\}$.
%We reconsider the previous definition by requiring that i) each lower bound and the following upper bound
%are separated by at least 0.1 units in the redshift $z_{\rm centroid}$, and ii) each lower bound and the
%previous upper bound are separated by at least 0.02 units in the redshift $z_{\rm centroid}$.
%The lower bounds and upper bounds that do not satisfy the conditions i) and ii) are not considered any more as bounds.
%We note that, by construction, the subset of upper and lower bounds is ordered. 
%In particular, a lower bound always preceeds an upper bound, 
%and the number of upper bounds is the same as that of lower bounds. 
The first condition merges
those intervals that are separated by a tiny separation along the $z_{\rm centroid}$ 
axis. The minimum allowed separation between two consecutive intervals
is set equal to the dispersion $\sigma_w$ adopted for the filtering procedure.
In fact, fluctuations that occur on these scales may not be physical since they occur for redshift separations that
are well below those of the typical statical redshift uncertainties.
Similarly, the second condition is applied in order to detect only high significance features whose length along the $z_{\rm centroid}$ 
axis is at least comparable with the typical statistical photometric redshift uncertainty of the \citet{ilbert2009} sample.
Given the significance $s$, this procedure gives a set of intervals that we define as $s$-intervals.

Then, we increase the significance threshold by a tiny amount $ds=0.1\sigma$ and we repeat the above oulined procedure.
We note that each $(s+ds)$-interval is entirely included within a $s$-interval.
We retain those $s$-intervals that do not cointain any $(s+ds)$-interval, whereas we reject all the other $s$-intervals.

\begin{figure*} \centering
\includegraphics[width=0.45\textwidth,natwidth=610,natheight=642]{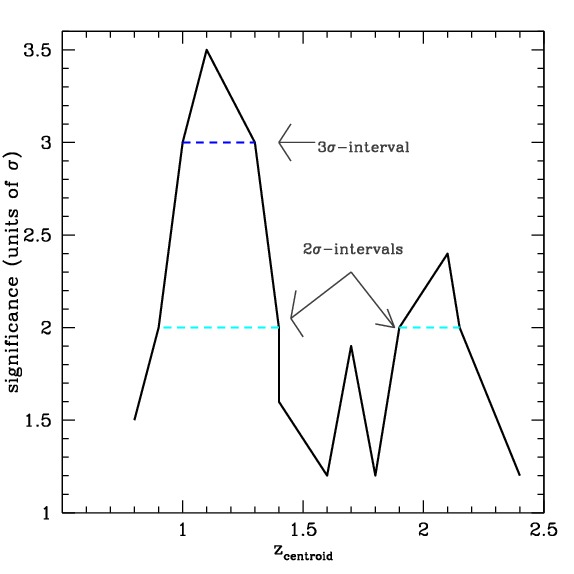}
\caption{ Visual representation of the peak finding algorithm. The centroid of the redshift bin $\Delta z \simeq 0.28$ is plotted 
in the x-axis. The values of $\overline{\mathcal{P}}$ in units of $\sigma$ are reported in the y-axis.
Solid black line: significance as a function of the centroid redshift $z_{\rm centroid}$. 
Horizontal dashed lines: 2$\sigma$- and 3$\sigma$-intervals. 
According to the peak finding procedure, the 2$\sigma$-interval in the plot 
associated with the peak at $z_{\rm centroid} \sim 1.1$
is rejected, since it entirely contains the higher significance  3$\sigma$-interval shown in the figure.
No $s$-interval is associated with the peak at  $z_{\rm centroid} \sim 1.7$. This is because its significance is
less than the 2$\sigma$ threshold.}
\label{fig:morse_plot}
\end{figure*}

The significance $s$ is increased and the procedure is repeated until no $s$-interval is found.
The final set of $s$-intervals represents the local maxima of $\overline{\mathcal{P}}$.
These intervals have different significances $s$ and they are centered at different redshifts $z_{\rm centroid}$.
Each $s$-interval corresponds to a  cluster candidate detection and is associated with
a number count excess found in the given field and around a specific redshift.
In Figure~\ref{fig:morse_plot} we show a visual representation of the peak finding algorithm adopted.
In the following section we will describe how the method estimates cluster properties such as the redshift and 
size.

%-------------------- Figure: Peak finding algorithm ----------
%-------------------------------------------------------------------------------------------

\subsection{Cluster candidates selection} \label{sec:cluster_candidates_selection}
The significance $s$ of a given $s$-interval  
is interpreted as the detection significance of the corresponding cluster candidate.
In this section we describe our procedure that provides (i) an estimate for the redshift of the overdensity, 
(ii) an estimate for the cluster core size, and (iii) a rough estimate for the cluster richness.

We estimate the size of each cluster candidate in terms of the minimum and maximum distances 
from the beacon (in our case the FR~I) at which the overdensity is detected. 
According to our procedure, the points of the $s$-interval are associated with different values of $r_{\rm min}^i$ and $r_{\rm max}^i$.
We estimate a minimum and a maximum projected radius of the overdensity as the average (and the median) 
of the minimum ($r_{\rm min}^i$) 
and maximum distances ($r_{\rm max}^i$) associated with all of the points of the $s$-interval, respectively. 
The uncertainty is estimated with the rms dispersion around the average.
The maximum projected radius provides also an estimate for the cluster core size, as further tested in this work and in Paper~II.

In Figure~\ref{fig:PPMplots_01_rmin_rmax} we show the PPM plots of source 01, 
(after having applied the Gaussian filter) where the radial information 
concerning $r_{min}$ (left panel) and $r_{max}$ (right panel) is considered.
Analogously to what done for Figure~\ref{fig:PPMplots_filter_before_and_after},
we only plot points that are associated with $\geq2$-$\sigma$ overdensities 
(see the legend in the panels for the color code adopted). 
The vertical solid line in each panel is located at the redshift of source 01. 
The values of  $r_{\rm max}^i$ and $r_{\rm min}^i$ associated with 
the high significance patterns of our plots are very stable with respect of the $\Delta z$ 
(see Figure~\ref{fig:PPMplots_01_rmin_rmax}).
Therefore, the particular choice $\Delta z =0.28$ does not affect the results concerning the projected space analysis.

We also estimate the redshift of the cluster as the middle point of the $s$-interval.
To estimate the fiducial uncertainty of the cluster redshift we consider all the sources located within the 
median value of minimum distance and the median value of the maximum
distance from the coordinates of the FR~I beacon within which the overdensity is detected in 
  the projected space. We also limit to the sources that have photometric redshifts within a redshift 
  bin $\Delta z = 0.28$ centered at the estimated cluster redshift. This is done consistently with our detection procedure.
The cluster richness is roughly estimated by the number of the selected sources. 
Then, the cluster redshift is estimated at 1-$\sigma$ level by the rms dispersion around the average of the 
redshifts of the selected sources.
  
In particular, if $N\gg1$ sources were uniformly distributed within the redshift bin $\Delta{\rm z}=0.28$ 
we would obtain a rms dispersion of 0.08. 
We expect the estimated redshift uncertainty to be around this value.

\subsection{Cluster candidate - FR~I association}\label{sec:cl_ff1_ass}
The method described here detects Mpc-scale overdensities
within the entire redshift range spanned by the $z_{\rm centroid}$ values.
Then, we associate with the radio galaxy, only those overdensities 
that are detected in the field, at a redshift compatible with that of the source, i.e.
when the interval centered at the redshift estimated
for the overdensity and with a half-width equal to 2
times the fiducial redshift error intersects the redshift range defined within the radio galaxy redshift
uncertainties. Multiple overdensity associations are not excluded.

\subsection{Considerations about the redshift information}
We point out that the redshift of the FR~I beacon is considered only during the last step of the method procedure, when we
perform the association between the detected overdensities and the radio galaxy.
This is primarily motivated by the fact that we do not have spectroscopic information for most of our FR~Is.
Therefore, our approach is necessarily different from previous studies which select
cluster members using photometric redshifts for the majority of the galaxies 
in the field, but also knowing the spectroscopic redshifts of some of the cluster members \citep[e.g.][]{papovich2010}.

Furthermore, our choice implies that the high significance patterns in our plots (see e.g. Figure~\ref{fig:PPMplots_filter_before_and_after})
have a typical width along the $z_{\rm centroid}$ axis comparable to the statistical photometric redshift uncertainty
$\sigma_z\sim0.054(1+z_{\rm centroid})$.
Such an uncertainty corresponds to sources with $i^+\sim24$ and $1.5<z<3$ sources \citep[see Table~3 of][]{ilbert2009},
which we expect to find in our clusters.  
Therefore, our method estimates the cluster redshifts with similar accuracy.
On the other hand, including the (photometric) redshift information of the FR~I beacon from the beginning of our procedure would imply an increasing of the intrinsic 
scatter due to the FR~I redshift uncertainty.
If we sum in quadrature the redshift uncertainty associated with each FR~I to the statistical photometric redshift uncertainty $\sigma_z\sim0.1-0.2$, the uncertainty increases up to $\sim0.2-0.5$, depending 
on the redshift of the radio galaxy and its uncertainty.
This effect would make our method ineffective to search for high-$z$ cluster candidates around radio galaxies. 

\subsection{Generalization to other datasets}\label{sec:generalization}

 In this section we describe how the PPM can be generalized for its application to other datasets and photometric redshift catalogs,
whose statistically redshift accuracy is possibly comparable or better than that of the \citet{ilbert2009} catalog. 
Surveys with a similar or higher depth than that of COSMOS are preferred.
In particular, the PPM might be applied to both present and future wide field surveys such as SDSS Stripe 82, LSST, and Euclid.

The parameters should be adapted to take into account the different mean field number density of galaxies in the survey
and the statistical photometric redshift uncertainties of the redshift catalog. 
Therefore, it is not straightforward to provide precise rules.

In general, the redshift bin $\Delta z$ at which the overdensities are evaluated in the PPM plots should be set
equal to the 2-$\sigma$ statistical photometric redshift uncertainty of the galaxies at the redshifts of our 
interest and magnitudes typical of the sources we expect to find in clusters at high redshifts.

Consequently the length of the $s$-intervals should be at least about one-third of the specific  
redshift bin $\Delta z$ adopted to select the overdensities in the PPM plots. 
We remind that the $s$-intervals are the high significance intervals in the PPM plots (at the specific $\Delta z$) 
that are associated with overdense regions.
Similarly, the Gaussian dispersion $\sigma_w$ used to remove the noise from the PPM plots should be equal to one-fifth of the above mentioned
minimum  length for the $s$-interval.
As described and motivated in the procedure description 
(see Sect.~\ref{sec:peak_finding_alg}) when defining the $s$-intervals, consecutive intervals that are
separated by an amount $\sigma_w$ ($= 0.02$ in this paper) or less along the  $z_{\rm centroid}$ axis should be always merged.
Such scaling relations are motivated by the fact that the adopted parameters should ultimately rescale linearly with 
the typical statistical photometric redshift accuracy. 

Furthermore, according to the procedure description we do not change the PPM tessellation with increasing cluster redshift. 
This is mainly because we are looking for
overdense regions with sizes 
typical of those of the cluster cores and the angular distance $D_A(z)$ at the redshift $z$ is 
fairly constant between $z\sim1-2$. 
However, it might be appropriate to rescale linearly 
the size of the PPM tessellation by a factor of $\sim D_A(z=1.5)/D_A(z)$ in the case where the PPM is used 
to search for diffuse proto-clusters 
at redshifts significantly higher than $z\sim1-2$ \citep[e.g., at $z\simeq8$,][]{trenti2012}.
This leads to a correction $\sim21\%$, $46\%$, and $51\%$  at redshifts $z\sim4$, $z\sim6$, and $z\sim8$, respectively.

\begin{acknowledgements}
We are grateful for useful comments from an anonymous referee.
We  thank the Space Telescope Science Institute, where part of this work was developed.
We also thank Gianfranco De Zotti, Roberto Gilli, Piero Rosati, and Paolo Tozzi for fruitful discussion.
This work was partially supported by the STScI JDF account D0101.90157, and (G.C.) by both the Internship Program ISSNAF-INAF 2010 and
one of the Foundation Angelo Della Riccia fellowships both in 2012 and in 2013.
\end{acknowledgements}

\end{document}